\documentclass[twocolumn,tighten]{aastex63}
\usepackage{amsmath}
\usepackage{threeparttable}
\usepackage{float}
\usepackage{graphicx}
\usepackage{CJK}
\usepackage{xspace}

\submitjournal{ApJ}

\shorttitle{Faint AGN in NIRCam grism surveys}
\shortauthors{Matthee et al.}

\begin{document}
\begin{CJK*}{UTF8}{gbsn}

\title{Little Red Dots: an abundant population of faint AGN at $z\sim5$ revealed by the EIGER and FRESCO JWST surveys}

\correspondingauthor{Jorryt Matthee}
\email{jorryt.matthee@ist.ac.at}
\author[0000-0003-2871-127X]{Jorryt Matthee}
\affiliation{Department of Physics, ETH Z{\"u}rich, Wolfgang-Pauli-Strasse 27, Z{\"u}rich, 8093, Switzerland}

\affiliation{Institute of Science and Technology Austria (ISTA), Am Campus 1, 3400 Klosterneuburg, Austria}

\author[0000-0003-3997-5705]{Rohan P. Naidu}
\thanks{NASA Hubble Fellow}
\affiliation{MIT Kavli Institute for Astrophysics and Space Research, 77 Massachusetts Avenue, Cambridge, 02139, Massachusetts, USA}

\author[0000-0003-2680-005X]{Gabriel Brammer}
\affiliation{Cosmic Dawn Center (DAWN), Niels Bohr Institute, University of Copenhagen, Jagtvej 128, K\o benhavn N, DK-2200, Denmark}

\author[0000-0002-0302-2577]{John Chisholm}
\affiliation{Department of Astronomy, The University of Texas at Austin, 2515 Speedway, Stop C1400, Austin, TX 78712-1205, USA}

\author[0000-0003-2895-6218]{Anna-Christina Eilers}
\affiliation{MIT Kavli Institute for Astrophysics and Space Research, 77 Massachusetts Avenue, Cambridge, 02139, Massachusetts, USA}

\author[0000-0003-4700-663X]{Andy Goulding}
\affiliation{Department of Astrophysical Sciences, Princeton University,Princeton, NJ 08544, USA}

\author[0000-0002-5612-3427]{Jenny Greene}
\affiliation{Department of Astrophysical Sciences, Princeton University,Princeton, NJ 08544, USA}

\author[0000-0001-9044-1747]{Daichi Kashino}
\affiliation{National Astronomical Observatory of Japan, 2-21-1 Osawa, Mitaka, Tokyo 181-8588, Japan}
\affiliation{Institute for Advanced Research, Nagoya University, Nagoya 464-8601, Japan}

\author[0000-0002-2057-5376]{Ivo Labbe} 
\affiliation{Centre for Astrophysics and Supercomputing, Swinburne University of Technology, Melbourne, VIC 3122, Australia}

\author[0000-0002-6423-3597]{Simon J.~Lilly}
\affiliation{Department of Physics, ETH Z{\"u}rich, Wolfgang-Pauli-Strasse 27, Z{\"u}rich, 8093, Switzerland}

\author[0000-0003-0417-385X]{Ruari Mackenzie}
\affiliation{Department of Physics, ETH Z{\"u}rich, Wolfgang-Pauli-Strasse 27, Z{\"u}rich, 8093, Switzerland}

\author[0000-0001-5851-6649]{Pascal A. Oesch}
\affiliation{Department of Astronomy, University of Geneva, Chemin Pegasi 51, 1290 Versoix, Switzerland}
\affiliation{Cosmic Dawn Center (DAWN), Niels Bohr Institute, University of Copenhagen, Jagtvej 128, K\o benhavn N, DK-2200, Denmark}

\author[0000-0001-8928-4465]{Andrea Weibel}
\affiliation{Department of Astronomy, University of Geneva, Chemin Pegasi 51, 1290 Versoix, Switzerland}

\author[0000-0003-3735-1931]{Stijn Wuyts}
\affil{Department of Physics, University of Bath, Claverton Down, Bath, BA2 7AY, UK}

\author[0000-0003-1207-5344]{Mengyuan Xiao}
\affiliation{Department of Astronomy, University of Geneva, Chemin Pegasi 51, 1290 Versoix, 
Switzerland}

\author[0000-0002-3120-7173]{Rongmon Bordoloi}
\affiliation{Department of Physics, North Carolina State University, Raleigh, 27695, North Carolina, USA}

\author[0000-0002-4989-2471]{Rychard Bouwens}
\affiliation{Leiden Observatory, Leiden University, NL-2300 RA Leiden, Netherlands}

\author[0000-0002-8282-9888]{Pieter van Dokkum}
\affiliation{Astronomy Department, Yale University, 52 Hillhouse Ave, New Haven, CT 06511, USA}

\author[0000-0002-8096-2837]{Garth Illingworth}
\affiliation{Department of Astronomy and Astrophysics, University of California, Santa Cruz, CA 95064, USA}

\author[0000-0001-5346-6048]{Ivan Kramarenko}
\affiliation{Department of Astronomy, University of Geneva, Chemin Pegasi 51, 1290 Versoix, Switzerland}

\author[0000-0003-0695-4414]{Michael V. Maseda}
\affiliation{Department of Astronomy, University of Wisconsin-Madison, 475 N. Charter St., Madison, WI 53706 USA}

\author[0000-0002-3407-1785]{Charlotte Mason}
\affiliation{Cosmic Dawn Center (DAWN), Niels Bohr Institute, University of Copenhagen, Jagtvej 128, K\o benhavn N, DK-2200, Denmark}
\affiliation{Niels Bohr Institute, University of Copenhagen, Jagtvej 128, DK-2200 Copenhagen N, Denmark}

\author[0000-0001-5492-4522]{Romain A. Meyer}
\affiliation{Max Planck Institut f\"ur Astronomie, K\"onigstuhl 17, D-69117, Heidelberg, Germany}
\affiliation{Department of Astronomy, University of Geneva, Chemin Pegasi 51, 1290 Versoix, Switzerland}

\author[0000-0002-7524-374X]{Erica J. Nelson}
\affiliation{Department for Astrophysical and Planetary Science, University of Colorado, Boulder, CO 80309, USA}

\author[0000-0001-9687-4973]{Naveen A. Reddy}
\affiliation{Department of Physics and Astronomy, University of California, Riverside, 900 University Avenue, Riverside, CA 92521, USA}

\author[0000-0003-4702-7561]{Irene Shivaei}
\affiliation{Steward Observatory, University of Arizona, Tucson, AZ 85721, USA}
\affiliation{Centro de Astrobiolog\'{i}a (CAB), CSIC-INTA, Ctra. de Ajalvir km 4, Torrej\'{o}n de Ardoz, E-28850, Madrid, Spain}

\author[0000-0003-3769-9559]{Robert A.~Simcoe}
\affiliation{MIT Kavli Institute for Astrophysics and Space Research, 77 Massachusetts Avenue, Cambridge, 02139, Massachusetts, USA}

\author[0000-0002-5367-8021]{Minghao Yue}
\affiliation{MIT Kavli Institute for Astrophysics and Space Research, 77 Massachusetts Avenue, Cambridge, 02139, Massachusetts, USA}

\begin{abstract}  
Characterising the prevalence and properties of faint active galactic nuclei (AGN) in the early Universe is key for understanding the formation of supermassive black holes (SMBHs) and determining their role in cosmic reionization. We perform a spectroscopic search for broad H$\alpha$ emitters at $z\approx4-6$ using deep {\it JWST}/NIRCam imaging and wide field slitless spectroscopy from the EIGER and FRESCO surveys. We identify 20 H$\alpha$ lines at $z=4.2-5.5$ that have broad components with line widths from $\sim1200-3700$ km s$^{-1}$, contributing $\sim30-90$ \% of the total line flux. We interpret these broad components as being powered by accretion onto SMBHs with implied masses $\sim10^{7-8}$ M$_{\odot}$. In the UV luminosity range M$_{\rm UV, AGN+host}=-21$ to $-18$, we measure number densities of $\approx10^{-5}$ cMpc$^{-3}$. This is an order of magnitude higher than expected from extrapolating quasar UV luminosity functions. Yet, such AGN are found in only $<1$ \% of star-forming galaxies at $z\sim5$. The number density discrepancy is much lower when compared to the broad H$\alpha$ luminosity function. The SMBH mass function agrees with large cosmological simulations. In two objects we detect complex H$\alpha$ profiles that we tentatively interpret as caused by absorption signatures from dense gas fueling SMBH growth and outflows. We may be witnessing early AGN feedback that will clear dust-free pathways through which more massive blue quasars are seen. We uncover a strong correlation between reddening and the fraction of total galaxy luminosity arising from faint AGN. This implies that early SMBH growth is highly obscured and that faint AGN are only minor contributors to cosmic reionization. 
\end{abstract}


\section{Introduction}
\label{sec:introduction}

With its unprecedented infrared capabilities, {\it JWST} has opened new opportunities to study the distant Universe. Various recent studies have exemplified {\it JWST}'s ability to identify relatively faint active galactic nuclei (AGN) in the early Universe ($z>3$) by means of spectroscopy \citep{Carnall23,Harikane23,Kocevski23,Larson23,Ubler23,Maiolino23} as well as high resolution imaging and the modeling of spectral energy distributions (SEDs; e.g. \citealt{Endsley22b,Furtak22,Bogdan23,Labbe23,Onoue23}). Prior to {\it JWST}, AGN samples at these redshifts were mostly limited to relatively bright, $\gtrsim 5\times L^{*}$ systems ($M_{\rm{UV}}\lesssim-22$) for which ground-based rest-UV spectroscopy was feasible \citep[e.g.,][]{Kulkarni19, Niida20, Shin22}. Towards fainter magnitudes, the number density of faint AGN is uncertain by more than two orders of magnitude \citep[e.g.,][]{Parsa18, Giallongo19, Morishita20,Shen20, FinkelsteinBagley22}. 

Constraining the abundance and properties of these faint AGN has wide-ranging implications for a number of frontiers in extragalactic astronomy. These sources may play a significant role in the final stages of the hydrogen reionization of the Universe \citep[e.g.,][]{Madau15,Finkelstein19}, provided they reside in environments conducive to ionizing photon escape. The number density and properties of black holes with masses M$_{\rm BH}\sim 10^{6-7}$ M$_{\rm{\odot}}$ may test scenarios of black hole seeding and growth that explain the presence of extreme supermassive black holes at $z\gtrsim6$ (with masses up to $\approx10^{10} M_{\rm{\odot}}$; \citealt{Volonteri10,Eilers22}) that have formed in less than a billion years \citep[e.g.,][]{Ricarte18,Greene20,Li22,Trinca22,Bogdan23,Kokorev23,Goulding23}.

{\it JWST}'s infrared spectroscopic capabilities enable the systematic identification of faint AGN at high-redshift through broad wing Balmer emission-lines similar to studies in the local Universe \citep[e.g.][]{SternLaor12,ReinesVolonteri15}. So far, NIRSpec spectroscopy has confirmed $\approx20$ faint AGN with such broad Balmer lines \citep[][]{Carnall23,Furtak23,Greene23,Harikane23,Kocevski23,Kokorev23,Larson23,Maiolino23b,Ubler23}. These sources have inferred black hole masses $\approx10^{6}-10^{8} M_{\rm{\odot}}$, and are UV faint ($M_{\rm{UV}}\lesssim-18$). However, as discussed in \citet{Harikane23}, the selection function of NIRSpec multi-object spectroscopy is complicated by diverse targeting choices (e.g., prioritizing the highest redshift galaxies, selecting sources from a mixture of {\it HST}/{\it JWST} photometry) as well as challenges with the observations themselves. This has kept us from answering the most fundamental questions about this population -- e.g., how common are these sources \citep[e.g.][]{Giallongo19}? What stage of supermassive BH growth do they correspond to \citep[e.g.][]{Kocevski23}? Can ionizing photons escape from faint AGN, or are they obscured \citep[e.g.][]{Fujimoto22,Endsley23}?

Blind, wide-area surveys with simple selection functions are required to measure the numbers and nature of faint AGN. NIRCam grism spectroscopy has emerged as a powerful, complementary mode to slit-based spectroscopy. By acquiring high-resolution ($R\approx1600$) spectra for \textit{every} galaxy in the field of view, grism surveys are delivering the largest spectroscopic samples of $z\gtrsim4$ sources in {\it JWST}'s first year of operations \citep[e.g.,][]{Kashino23, Matthee23,Oesch23,Helton23}. The much higher spectral resolution compared to the {\it HST} grisms that were already used to identify AGN at $z\approx1$ ($R\approx100$; e.g. \citealt{Nelson12,Brammer12,Momcheva16}) has allowed for relatively straightforward disentangling of overlapping spectra to extract emission lines. 

Here we perform a dedicated search for broad line (BL) H$\alpha$ emitters in two of the largest Cycle 1 grism programs -- EIGER \citep[][]{Kashino23} and FRESCO \citep[][]{Oesch23} -- currently totalling $\approx70$ hours of {\it JWST} observing time over $\approx230$ arcmin$^2$. We aim to measure the faint AGN number density and characterise the properties of this population using NIRCam imaging and spectroscopy of a well-defined, flux-limited H$\alpha$ sample in a volume $6\times10^5$ cMpc$^{3}$ over $z=4.0-6.0$. A plan for this paper follows: we briefly present the observations and data reduction in \S \ref{sec:data}. We discuss the selection and identification of the BL H$\alpha$ emitters, and their emission-line measurements in the grism data in \S $\ref{sec:identification}$. In \S $\ref{sec:properties}$ we present the properties of the BL H$\alpha$ emitters, where we make the case for an AGN origin of the broad H$\alpha$ components based on the imaging and spectroscopic data, and we present the properties of the galaxies and their supermassive black holes. In \S $\ref{sec:numberdensity}$ we present the measured number density of faint AGN. We discuss and interpret our results in the context of earlier results, supermassive black hole formation and growth scenarios and the role of faint AGN in the reionization of the Universe in \S $\ref{sec:discussion}$. We summarise our results and their interpretation in \S $\ref{sec:summary}$.

Throughout this work we assume a flat $\Lambda$CDM cosmology with $H_0=70$ km s$^{-1}$ Mpc$^{-1}$ and $\Omega_M=0.3$. Magnitudes are listed in the AB system.

\section{Data}
\label{sec:data}
In order to cover a large cosmic volume in various independent sight-lines and span a significant redshift range, we combine the {\it JWST}/NIRCam \citep{Rieke23} imaging and wide field slitless spectroscopic (WFSS) data from the EIGER \citep{Kashino23} and FRESCO \citep{Oesch23} surveys. EIGER (PID: 1243, PI: S. Lilly) is a large program around six high-redshift $z=6-7$ quasars with WFSS in the F356W filter. FRESCO (PID: 1895, PI: P. Oesch) is a medium-sized program in the GOODS-N and GOODS-S fields with WFSS in the F444W filter. In total, the spectroscopic data used in this paper amounts to 70 hours of exposure time (38.8h from EIGER, 31.2h from FRESCO).

\begin{figure*}
    \centering
    \includegraphics[width=18cm]{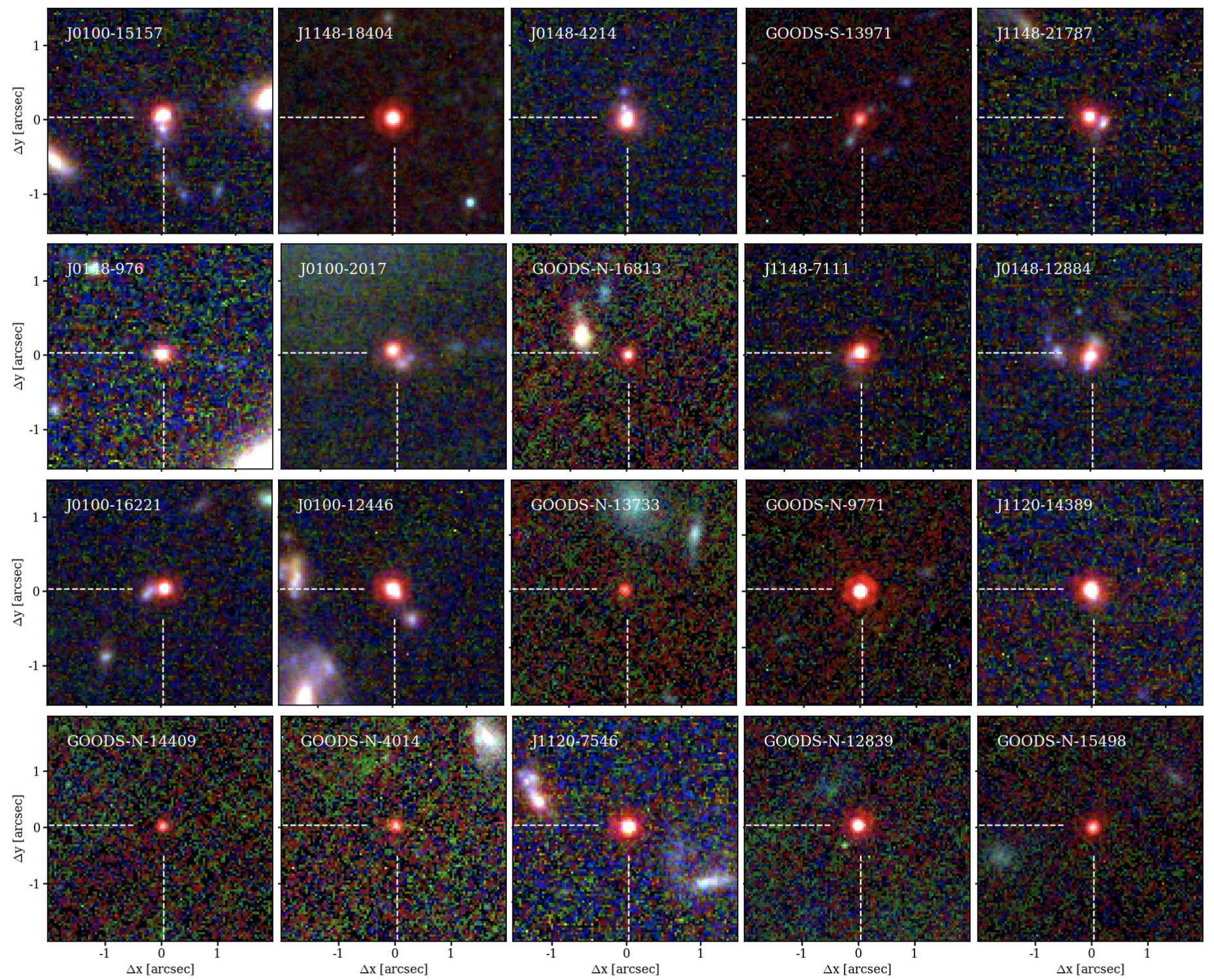}
    \caption{{\bf False color stamps of the 20 BL H$\alpha$ emitters at $\mathbf{z=4.2-5.5}$ identified in this work ordered by their broad-to-total H$\alpha$ luminosity ratio, from top-left to bottom-right.} For the EIGER sample (J*) we used {\it JWST} imaging data in the F115W/F200W/F356W filters in their native resolution, while F182M/F210M/F444W was used in the FRESCO sample (GOODS-*). We use a high stretch to highlight colour differences between various components. BL H$\alpha$ emitters stand out as red point sources. Blue companion galaxies can be identified in a large fraction of the EIGER sample with the deepest imaging data.
    }
    \label{fig:RGBstamps}
\end{figure*}

\subsection{Observational setup}
We use data from the four quasar fields that are part of the EIGER survey that have been observed before February 2023 (J0100+2802, J1148+5251, J1120+0641 and J0148+0600). The quasars are at $z\sim6-7$ and therefore located far behind the objects of study in this paper in these fields. As detailed in \cite{Kashino23}, EIGER observes with a 2x2 mosaic totalling 26 arcmin$^2$ that centers on the quasar, which is covered by all four visits. EIGER uses the grismR that disperses spectra in the horizontal direction on the camera, in combination with the F356W filter in the long wavelength channel that spans $\lambda\approx3.1-4.0$ $\mu$m. Due to the length of the $R\sim1600$ spectra on the detectors, not all sources in the field of view have full spectral coverage. The effective field of view is maximum around the red end of the wavelength coverage, while it is 20 \% smaller at $\lambda=3.1$ $\mu$m. In each of the four visits, the total spectroscopic exposure time is 8.8ks. An additional 1.6ks are spent on direct and out of field imaging in the F356W filter. Deep imaging data in the short wavelength filters F115W and F200W are taken simultaneously with the grism spectroscopy. Observations in the F200W filter were carried out during direct and out of field imaging and thus received more exposure time than the F115W data.

We also use the FRESCO survey that targets the well known Northern and Southern GOODS fields. While the GOODS fields are covered by extensive deep {\it HST} photometry \citep[e.g.][]{Giavalisco96,Koekemoer11,Illingworth13,Whitaker19}, we focus mostly on the {\it JWST} photometry and spectroscopy. Besides FRESCO data, in the GOODS-S field we also include NIRCam medium-band imaging data in the F182M and F210M filters from the JEMS program \citep{Williams23}. GOODS-S was observed in November 2022 and the observations of the GOODS-N field were taken in February 2023. As detailed in \citet{Oesch23}, the FRESCO footprint is a 2x4 mosaic totalling $\approx65$ arcmin$^2$ that is aimed at maximising the sky area with complete spectral coverage. Similar to EIGER, the grismR is used yielding a single dispersion direction over the majority of the field. The F444W filter covers $\lambda=3.8-4.9\, \mu$m. 85 \% of the field of view has full spectral coverage. The spectroscopic observing time per visit is 7 ks, whereas the direct and out of field imaging in the F444W filter amounts to 0.9 ks. The F182M and F210M medium-bands are used in the SW channel for the grism exposures, where F182M was also observed during the direct imaging.

\subsection{Data reduction and sensitivity}
The NIRCam WFSS data of both fields and the imaging data from EIGER were reduced following \cite{Kashino23}. The FRESCO imaging data were reduced with similar steps as the EIGER imaging data (see \citealt{Oesch23}), but using the \texttt{grizli} \citep{Brammer22} tool. For the WFSS data, we first process the raw exposure files with the \texttt{Detector1} step from the \texttt{jwst} pipeline (version 1.9.4) and use \texttt{Spec2} to assign an astrometric solution. We then flat-field the data using \texttt{Image2} and remove 1/f noise and variations in the sky background using the median counts in each column. As the main goal of our analysis is emission-line science, we subtract the continuum emission (regardless of whether it is contamination or from sources themselves) using a running median filter along each single row as described in \cite{Kashino23}. The continuum-filtering is a two step procedure, where detected emission-lines are identified in the first iteration in order to mask them for the final continuum removal. The standard running median procedure is performed with a kernel (51 pixels) along the dispersion axis that has a hole in the center (9 pixels wide) designed to avoid over-subtracting narrow lines. However, faint extended wings of broad emission-lines may still slightly be over-subtracted. We therefore optimise the kernel used for the continuum subtraction for sources individually after identifying them. In practice, this means that we use a much wider kernel with a larger central hole (151 pixels, 31 pixels, respectively). The downside of such wider kernel are possible residuals from the edges of the continuum trace from neighbouring sources. 

Comparing the data-sets, we find that the zodiacal background level and the exposure time mostly determine the sensitivity of the WFSS data. The EIGER (FRESCO) data have a typical spectral sensitivity of $1 (2)\times10^{-18}$ erg s$^{-1}$ cm$^{-2}$ (3$\sigma$). Aperture-matched photometry in the imaging data as described in \cite{Kashino23} and \cite{Oesch23} shows that the $5\sigma$ sensitivity is $\approx28-28.5$ AB magnitude, with the EIGER data being somewhat more sensitive due to longer exposure times and the use of wider filters.

\section{Identification of broad line H$\alpha$ emitters} \label{sec:identification}
\subsection{Selection criteria} \label{sec:colors}
For a systematic search of broad line H$\alpha$ emitters, we inspected spectra for all sources with at least one emission line in the EIGER and FRESCO data. To identify candidate broad line emitters, we fit their emission-line profiles with a combination of a narrow and a broad emission-line (see \S $\ref{sec:fit1D}$ for details). We then inspect all objects for which the broad component is identified with S/N$>5$ in order to determine whether these are consistent with being H$\alpha$ emitters at $z\approx5$ and whether the broad component is not included to account for the spatial extent of the object along the dispersion direction. In order to facilitate the determination of the origin of the broad component, we limit ourselves to broad components with full width half maximum $v_{\rm FWHM, H\alpha, broad}>1000$ km s$^{-1}$. Finally, in order to mitigate the impact of the wavelength-dependent sensitivity of our data, we impose a conservative lower limit to the H$\alpha$ luminosity of the broad component L$_{\rm H\alpha, broad} > 2\times10^{42}$ erg s$^{-1}$. The selection criteria are summarised in Equation $\ref{eq:selection}$. 

\begin{equation} \label{eq:selection}
\begin{split}
     {\rm S/N_{\rm H\alpha, broad} > 5}, \\
    {\rm L_{\rm H\alpha, broad} > 2\times10^{42} \, erg \, s^{-1} } , \\
    v_{\rm FWHM, H\alpha, broad} > 1000 \, {\rm km \,  s^{-1}. }
\end{split}
\end{equation}

\begin{figure}
    \centering
    \includegraphics[width=8.4cm]{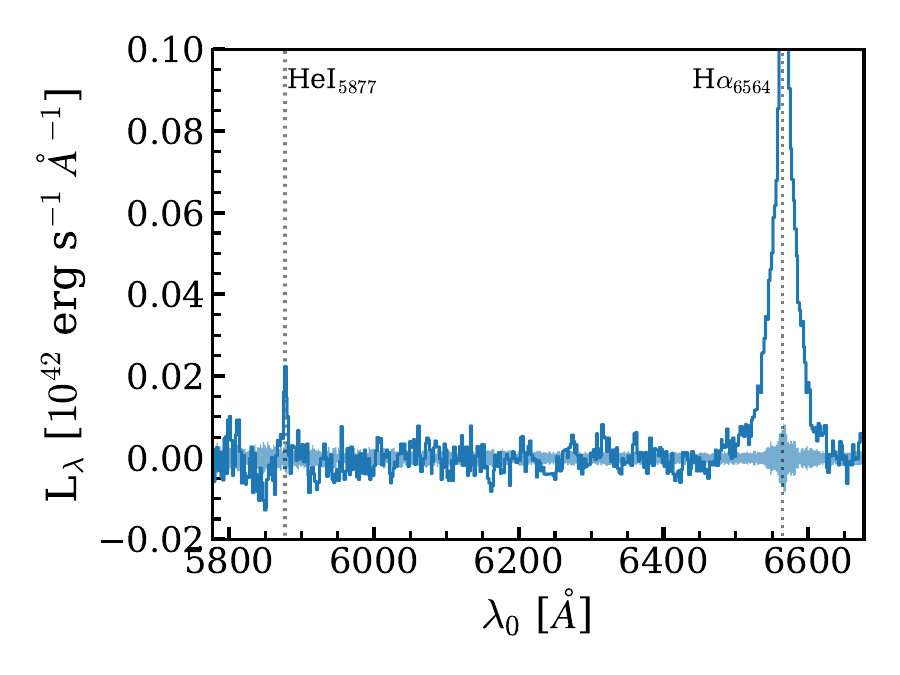}
    \caption{{\bf Median stacked spectrum of our sample assuming the broad line is H$\alpha$ at $z\sim5$.} Our stacked spectrum reveals HeI$_{5877}$ emission at a S/N of $\approx10$ validating that the single bright emission-lines are indeed H$\alpha$. 
     }
    \label{fig:HeI}
\end{figure}

We identify 20 BL H$\alpha$ emitters in this work, whose false-color stamps we show in Fig. $\ref{fig:RGBstamps}$. The majority of objects are characterised by a red, point-like morphology. The IDs, coordinates and redshifts of the sample are listed in Table $\ref{tab:LRDsample}$. Most objects only display a single emission line in our spectra, which we interpret as H$\alpha$, since broad H$\beta$ would be accompanied by [OIII] emission and broad Paschen lines have been identified and removed from the sample because of the detection of other lines as [SIII] or HeI (in particular in the case of Paschen-$\gamma$).The observed equivalent widths of the emission-lines are typically 3000 {\AA}. This strongly suggests that the lines are H$\alpha$ lines (with rest-frame EWs $\sim500$ {\AA}). If the lines would alternatively be Paschen-$\alpha$ or Paschen-$\beta$ at redshifts $z\approx 1.2, 2.2$, respectively, the implied rest-frame EWs would be $\sim1000$ {\AA}. This would be about a hundred times higher than the EWs of those lines in the average quasar spectrum at low-reshift \citep[e.g.][]{Glikman06}, and ten times higher than in a $z=7$ quasar spectrum \citep{Bosman23}. This is challenging to understand given that these lines have fluxes typically $\sim3-10$ \% of H$\alpha$. Indeed, in the \citealt{Glikman06} spectrum, the H$\alpha$ EW is about 10-15 times higheer than the EWs from the Paschen lines. In addition, we detect HeI$_{5877}$ with S/N$\approx10$, see Fig. $\ref{fig:HeI}$, and HeI$_{7065}$ with S/N$\approx5$ in the median stacked spectrum of the full sample corroborating the redshift identification. This emission-line is also detected in the individual spectrum of J0100-12446 with a S/N$\sim5$. While photometric information was not primarily used in the selection of these objects, the photometric redshifts of the objects in the GOODS fields derived from {\it HST}+{\it JWST} photometry and EAZY \citep{brammer08} agree very well with the spectroscopic redshift ($\langle \frac{z_{\rm spec} - z_{\rm phot}}{1+z_{\rm spec}}\rangle = -0.01$, with the largest outlier having a redshift difference of $\Delta z$=0.5). All objects in FRESCO display a colour break consistent with a Lyman-break at $z\sim5$ (Appendix $\ref{appendix:photometry}$). One object in our sample (GOODS-S-13971) shows Lyman-$\alpha$ emission in VLT/MUSE data very close to the H$\alpha$ redshift \citep{Bacon23}.

\subsection{JWST colors} \label{sec:colors}
Here we contextualise the BL H$\alpha$ emitters to the colors of the general source population identified in {\it JWST} photometry, noting that no color selection criteria were applied to our sample.

\begin{figure}
    \centering
    \includegraphics[width=8.1cm]{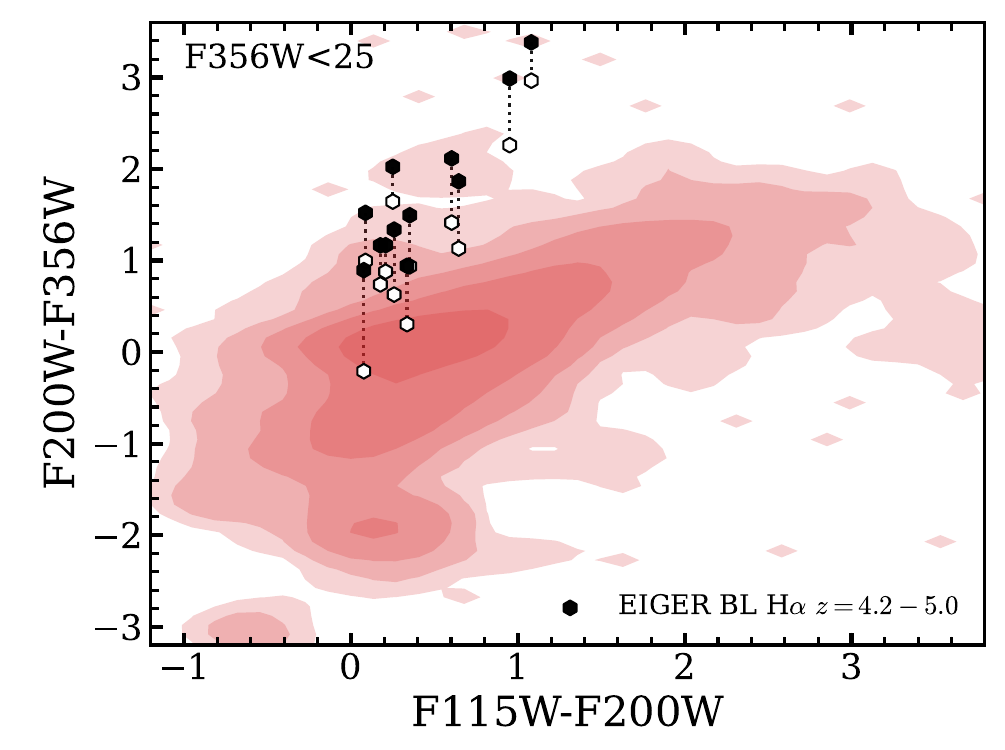} \\
    \includegraphics[width=8.1cm]{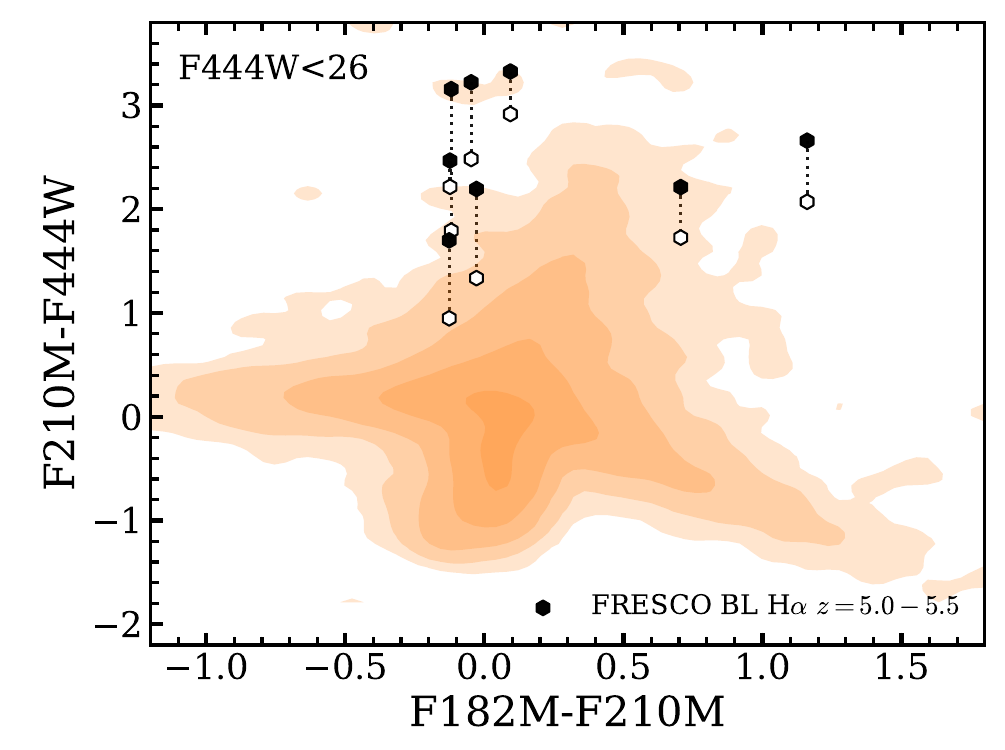} \\
    \caption{{\bf JWST colors of the objects identified in the EIGER (top) and FRESCO surveys (bottom) surveys.} In the top panel, the red contours show the color-color distribution of objects with F356W$<25$ in the EIGER survey, while the orange contours in the bottom panel show the distribution of objects with F444W$<26$ in the FRESCO survey. These magnitude limits correspond to the range probed by the BL H$\alpha$ emitters. Filled hexagons mark the observed locations of BL H$\alpha$ emitters, open hexagons use the H$\alpha$-subtracted F356W or F444W magnitudes. These open data-points illustrate the continuum colors. } 
    \label{fig:colselection}
\end{figure} 

In the EIGER data, we select 12 objects as broad line H$\alpha$ emitters at $z\approx4-5$. While it was not used as a selection criterion, all these sources are spatially very compact. Interestingly, the BL H$\alpha$ emitters are found towards the rarest regions in our color-color diagram in Fig $\ref{fig:colselection}$. They are relatively blue in F115W - F200W while extending to the reddest F200W - F356W colors, even when removing the contribution of the strong emission-line to the F356W photomtery. This is unlike the dominant population of bright galaxies with red F200W - F356W colors that are dusty star-forming galaxies at $z\approx1-4$ \citep[e.g.][]{Bisigello23,Glazebrook23} for which we detect Paschen, HeI and [SIII] lines depending on their redshift. These lower redshift objects typically have much redder F115W - F200W colors. High-redshift $z=5-7$ [OIII] emitters \citep[e.g.][]{Matthee23,Rinaldi23} have relatively similar colours as BL H$\alpha$ emitters, typically being blue in F115W - F200W and red in F200W - F356W. However, they are not as red in the latter color as their redness is caused by line emission on top of a relatively flat continuum. In Table $\ref{tab:linefits}$ we list in addition to the observed magnitudes also the colour excess in the long-wavelength filter that is due to the H$\alpha$ line emission. This excess, $\Delta m_{\rm LW} = m_{\rm obs} - m_{\rm line \, corrected}$, is estimated by subtracting the measured H$\alpha$ line-flux in the grism data from the observed photometry ($m$ either corresponds to F356W or F444W). While the excess is typically relatively high, 0.7 magnitude, we find that the underlying continuum is (very) red in all cases except for J0100-15157 that has a flat color. Therefore, BL H$\alpha$ emitters are characterised by a very red optical continuum. We also find H$\alpha$ emitters at $z\sim5$ that have narrow emission lines and [NII] detections. These objects -- similar to some objects presented in \cite{Arrabal23} -- typically have significantly more extended morphologies than the broad line emitters and often have redder F115W-F200W colors, likely due to strong dust attenuation. There are four objects with very similar colors and morphologies as the broad line H$\alpha$ emitters, but for those our spectral coverage is incomplete as they are located towards the edges of the survey area, preventing us from detecting line-emission.

Eight broad line H$\alpha$ emitters were identified in the FRESCO data. Similar to the BL H$\alpha$ emitters in the EIGER data, we find that these are located in rare locations in the color-color diagram. They have the reddest F210M-F444W colors but atypical F182M-F210M colors (either relatively blue, or red). We note that MgII line-emission may contribute to the F182M photometry in all FRESCO BL H$\alpha$ emitters. One of the objects with red F182M-F210M colours is among the faintest in the sample, such that its colour is consistent with being flat (see Appendix $\ref{appendix:photometry}$).

\begin{table}
    \centering
    \caption{{\bf Coordinates and redshifts of the 20 BL H$\alpha$ emitters identified in this work.} Coordinates are in the J2000 reference frame.} 
    \begin{tabular}{cccc}
    ID & R.A. & Dec. & $z_{\rm spec}$ \\ \hline
    GOODS-N-4014  & 12:37:12.03 & +62:12:43.36 & 5.228  \\
GOODS-N-9771  & 12:37:07.44 & +62:14:50.31 & 5.538 \\ 
GOODS-N-12839 & 12:37:22.63 & +62:15:48.11 & 5.241 \\
GOODS-N-13733 & 12:36:13.70 & +62:16:08.18 & 5.236 \\
GOODS-N-14409 & 12:36:17.30 & +62:16:24.35 & 5.139 \\
GOODS-N-15498 & 12:37:08.53 & +62:16:50.82 & 5.086 \\
GOODS-N-16813 & 12:36:43.03 & +62:17:33.12 & 5.355 \\
GOODS-S-13971 & 3:32:33.26 & --27:47:24.90 & 5.481 \\
J1148-7111 & 11:48:24.41 & +52:54:28.66 & 4.339 \\
J1148-18404 & 11:48:13.91 & +52:51:46.09 & 5.011 \\
J1148-21787 & 11:48:05.14 & +52:50:01.04 & 4.277 \\
J0100-2017 & 01:00:13.93 & +28:04:20.69 & 4.938 \\
J0100-12446 & 01:00:11.58 & +28:00:34.98 & 4.699 \\
J0100-15157 & 01:00:07.26 & +28:03:00.64 & 4.941 \\
J0100-16221 & 01:00:08.17 & +28:03:05.68 & 4.349 \\
J0148-976 & 01:48:35.08 & +05:57:20.97 & 4.163 \\
J0148-4214 & 01:48:33.29 & +05:59:50.04 & 5.019 \\
J0148-12884 & 01:48:41.58 & +06:00:57.30 & 4.602 \\
J1120-7546 & 11:19:59.86 & +06:39:17.01 & 4.967 \\
J1120-14389 & 11:20:00.89 & +06:43:10.42 & 4.897 \\\hline
    \end{tabular}
    \label{tab:LRDsample}
\end{table}

\begin{figure}
    \centering
    \begin{tabular}{cc}
    \hspace{-0.4cm}
 \includegraphics[width=4.5cm]{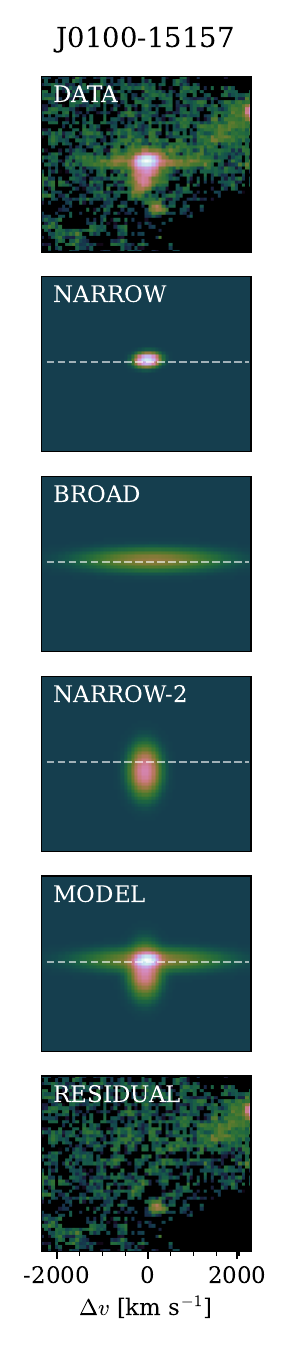} &
     \hspace{-0.8cm}
 \includegraphics[width=4.5cm]{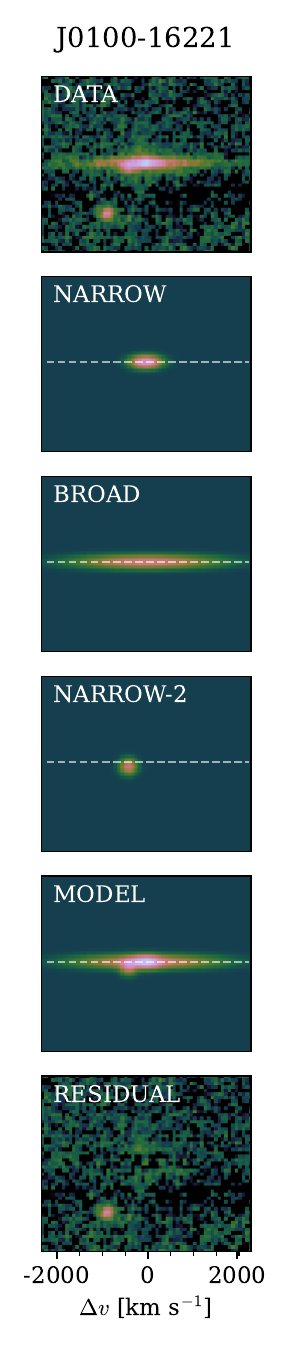} \\
        \end{tabular}\vspace{-0.8cm}
    \caption{{\bf The spatially resolved decomposition of the various H$\alpha$ emitting components in the grism data of two example BL H$\alpha$ emitters.} The color scaling follows a power-law with $\gamma=0.33$ to highlight low surface brightness emission. }
    \label{fig:2D_fits}
\end{figure}

\begin{figure*}
    \centering
    \begin{tabular}{ccccc}
    \hspace{-0.4cm}
 \includegraphics[height=3.45cm]{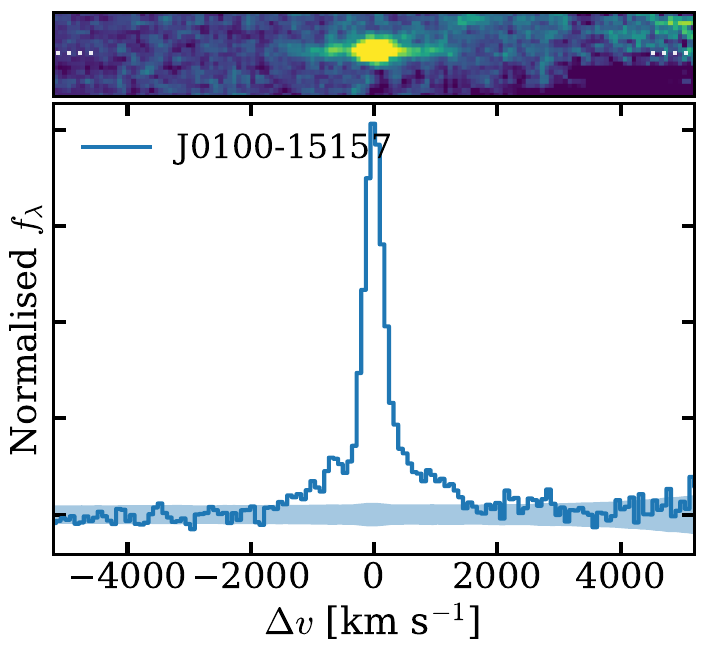} &
 \hspace{-0.45cm}
 \includegraphics[height=3.45cm,trim={0.75cm 0 0 0},clip]{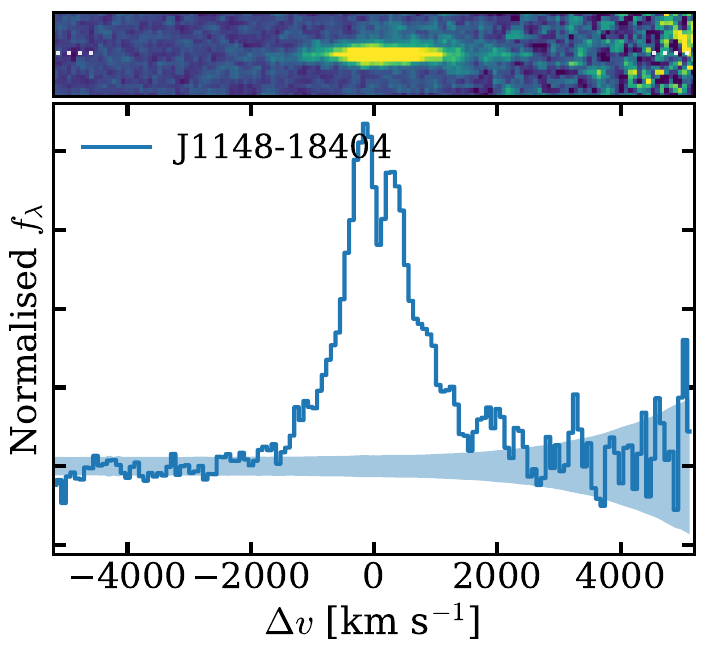} &
 \hspace{-0.45cm}
 \includegraphics[height=3.45cm,trim={0.75cm 0 0 0},clip]{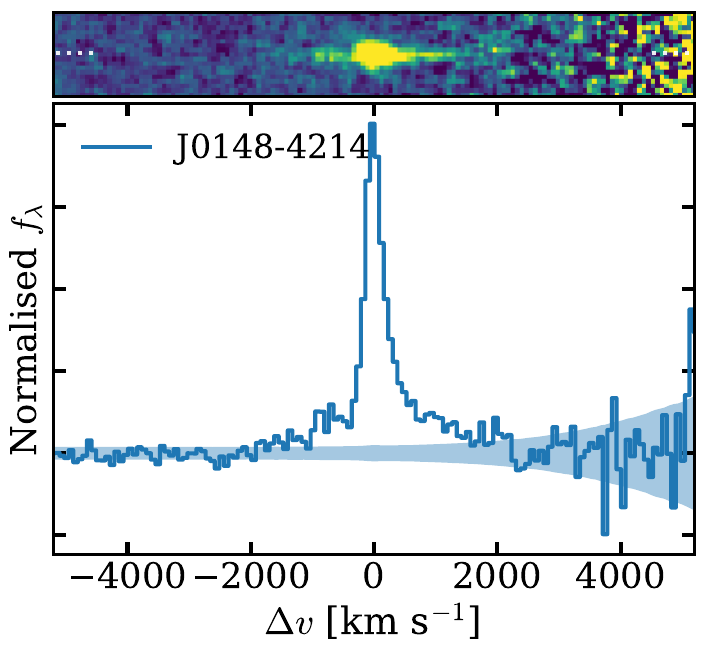} &
 \hspace{-0.45cm}
 \includegraphics[height=3.45cm,trim={0.75cm 0 0 0},clip]{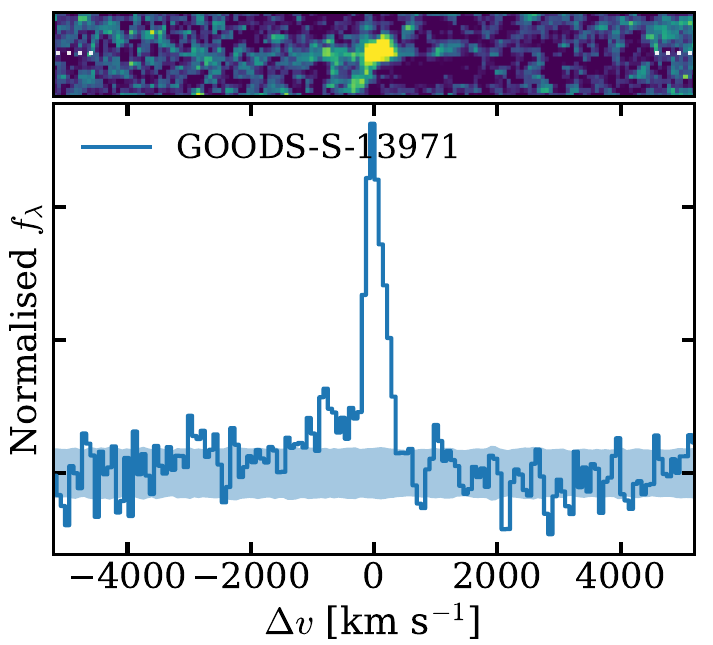} &
 \hspace{-0.45cm}
 \includegraphics[height=3.45cm,trim={0.75cm 0 0 0},clip]{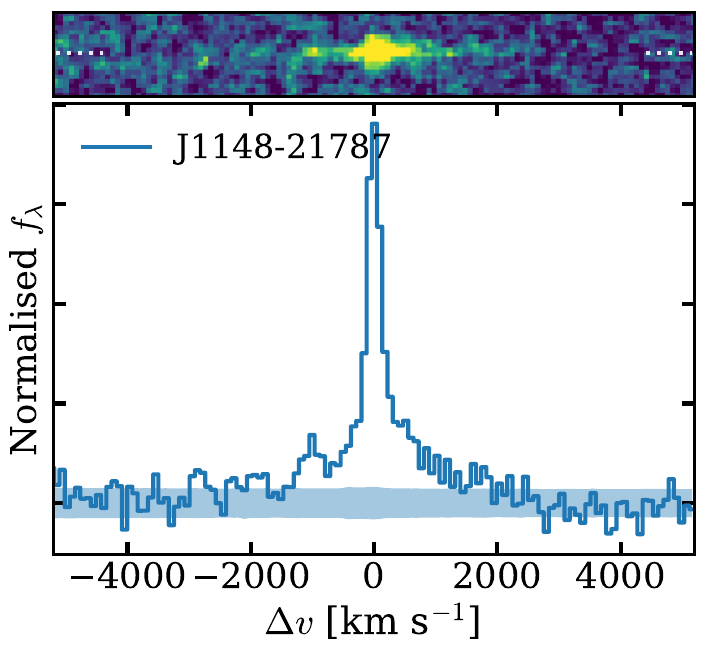} \\
 \hspace{-0.4cm}
 \includegraphics[height=3.45cm]{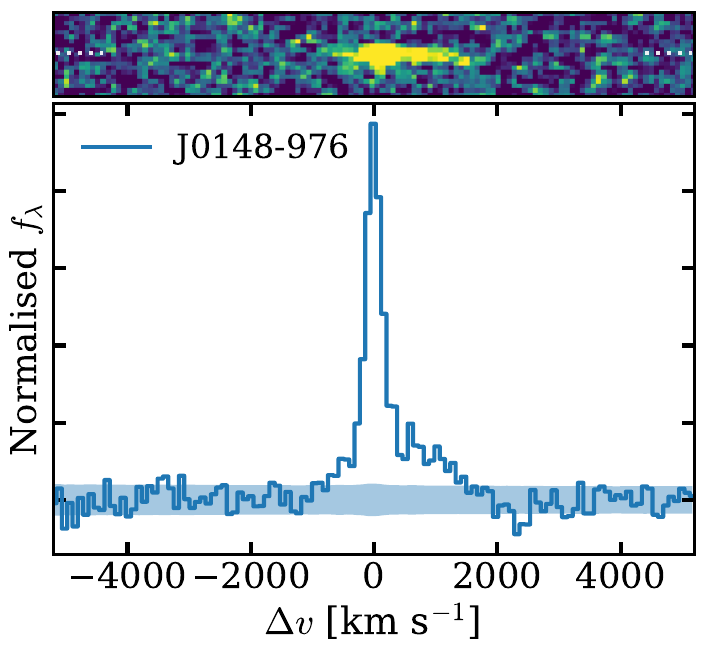} &
 \hspace{-0.45cm}
 \includegraphics[height=3.45cm,trim={0.75cm 0 0 0},clip]{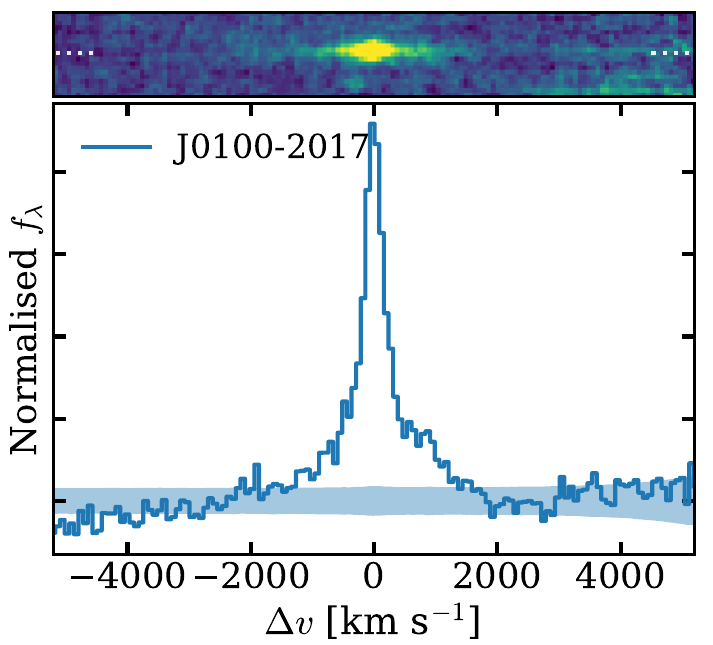} &
 \hspace{-0.45cm}
 \includegraphics[height=3.45cm,trim={0.75cm 0 0 0},clip]{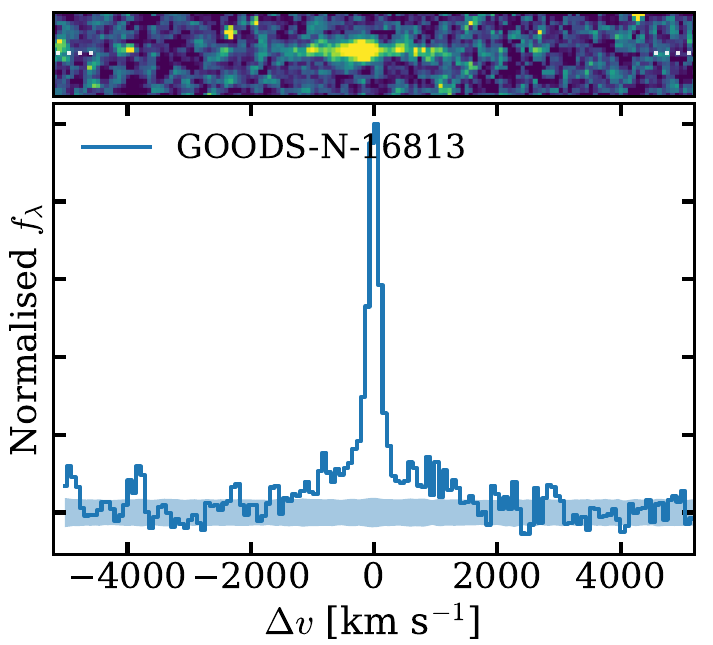} &
 \hspace{-0.45cm}
 \includegraphics[height=3.45cm,trim={0.75cm 0 0 0},clip]{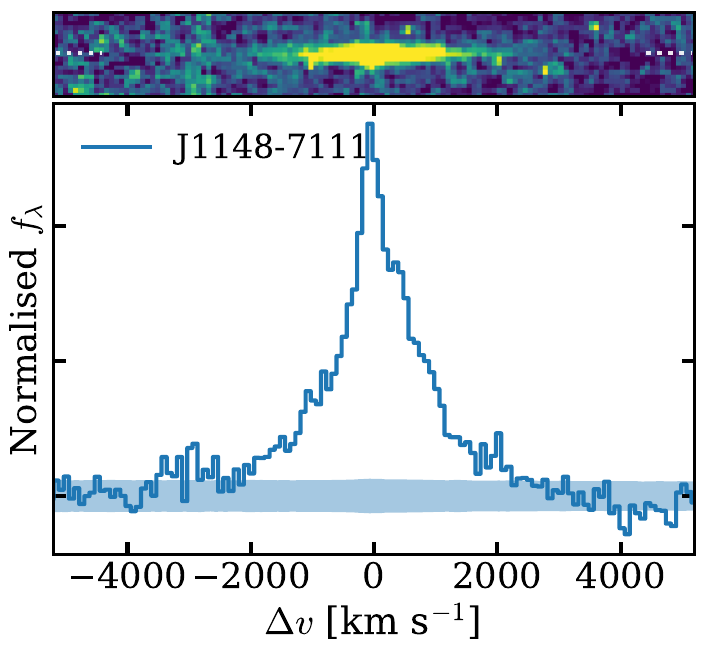} &
 \hspace{-0.45cm}
 \includegraphics[height=3.45cm,trim={0.75cm 0 0 0},clip]{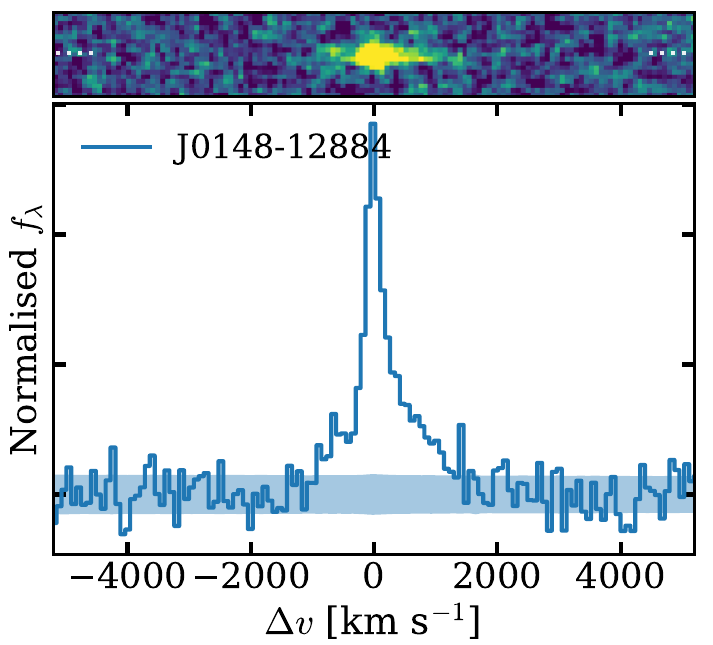} \\
  \hspace{-0.4cm}
  \includegraphics[height=3.45cm]{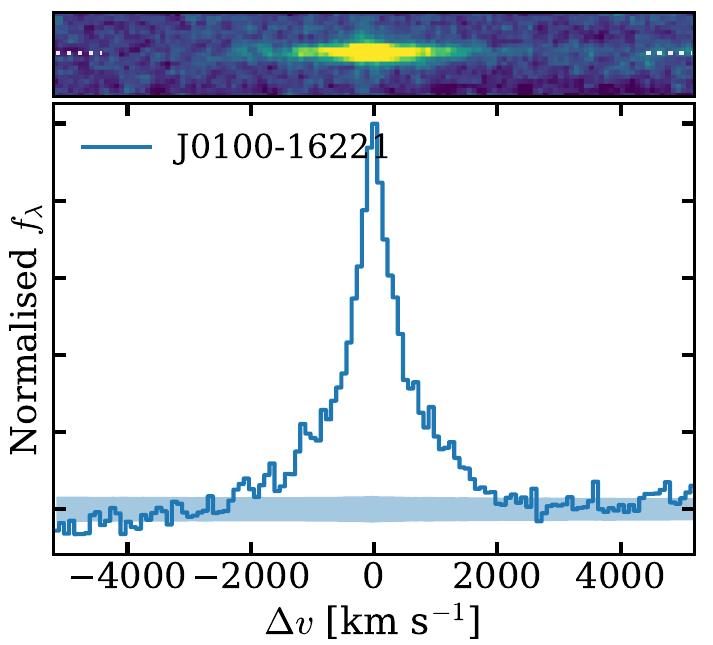} &
 \hspace{-0.45cm}
 \includegraphics[height=3.45cm,trim={0.75cm 0 0 0},clip]{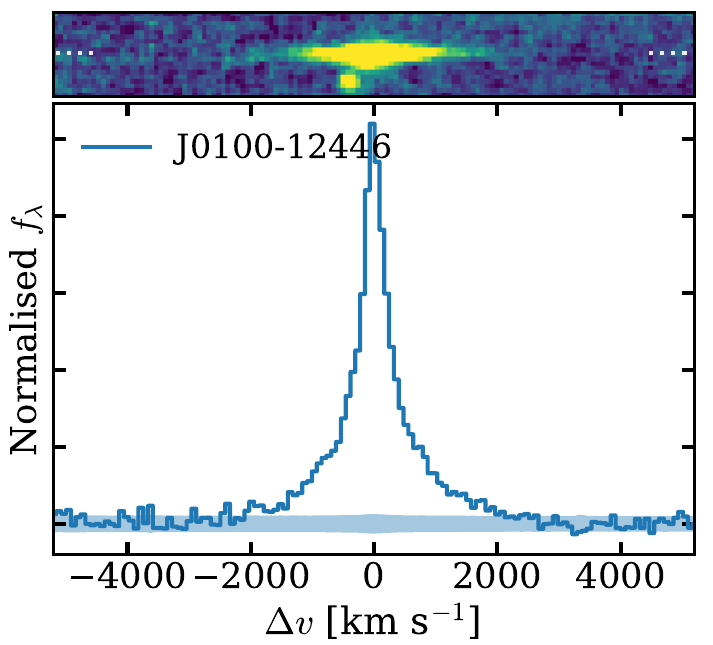} &
 \hspace{-0.45cm}
 \includegraphics[height=3.45cm,trim={0.75cm 0 0 0},clip]{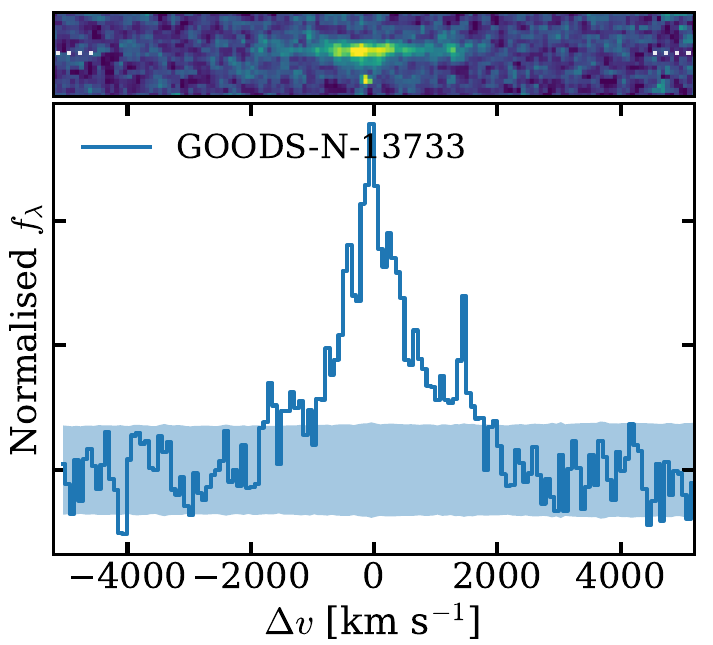} &
 \hspace{-0.45cm}
 \includegraphics[height=3.45cm,trim={0.75cm 0 0 0},clip]{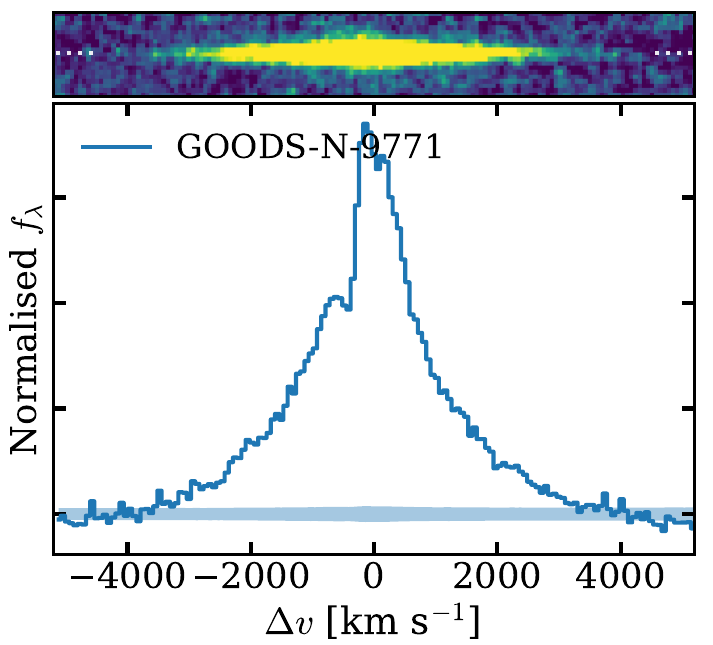} &
 \hspace{-0.45cm}
 \includegraphics[height=3.45cm,trim={0.75cm 0 0 0},clip]{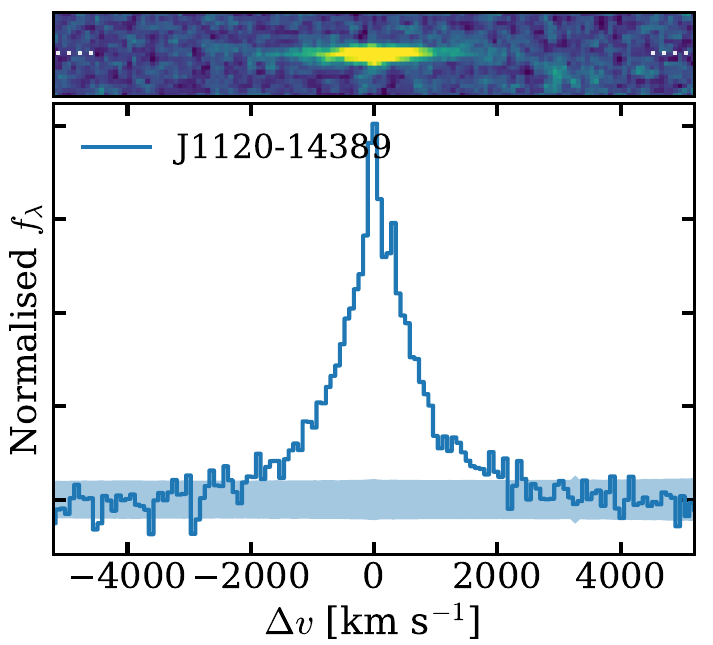} \\
 \hspace{-0.4cm}
  \includegraphics[height=3.45cm]{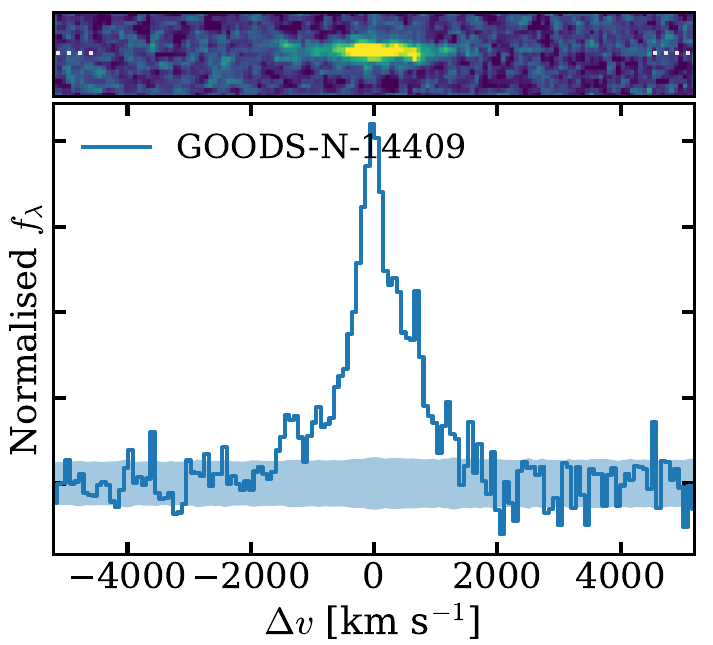} &
 \hspace{-0.45cm}
 \includegraphics[height=3.45cm,trim={0.75cm 0 0 0},clip]{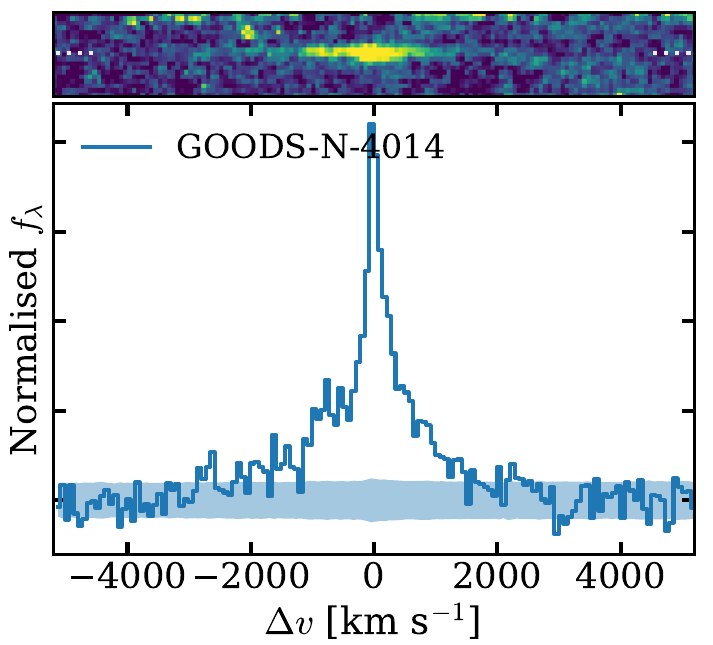} &
 \hspace{-0.45cm}
 \includegraphics[height=3.45cm,trim={0.75cm 0 0 0},clip]{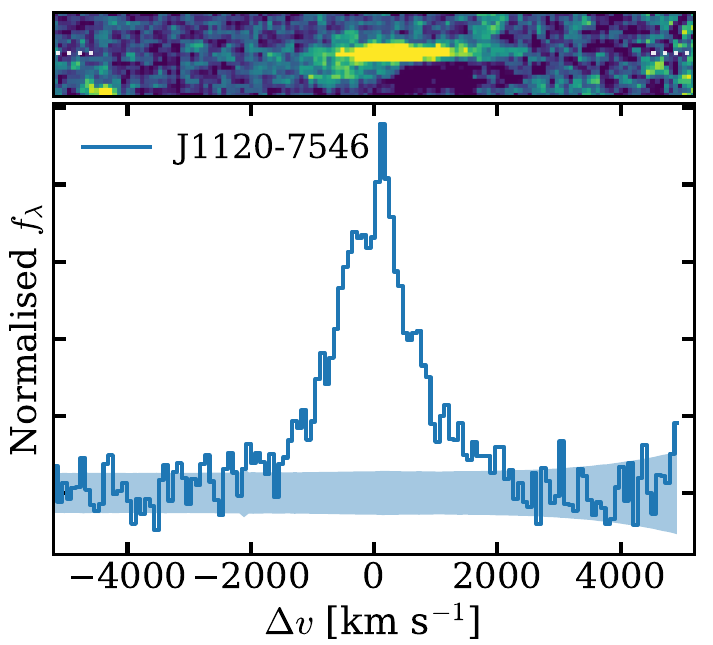} &
 \hspace{-0.45cm}
 \includegraphics[height=3.45cm,trim={0.75cm 0 0 0},clip]{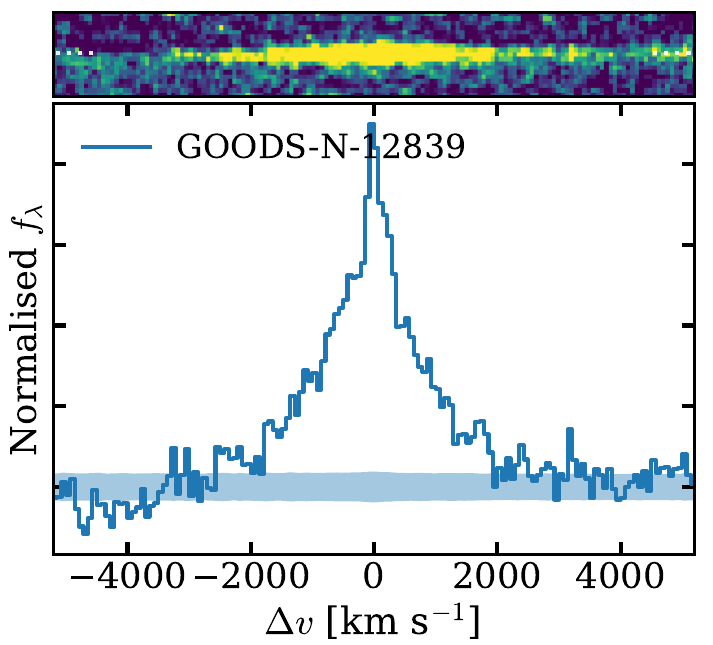} &
 \hspace{-0.45cm}
 \includegraphics[height=3.45cm,trim={0.75cm 0 0 0},clip]{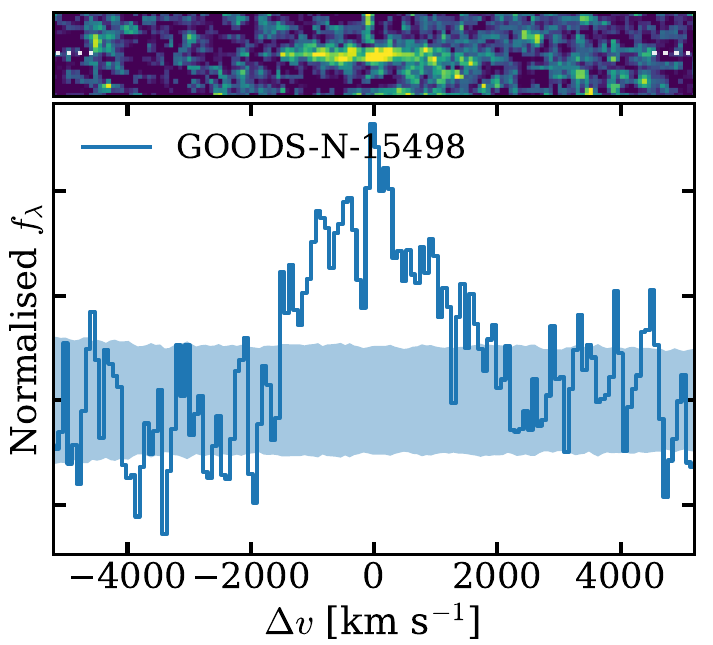} \\
        \end{tabular}
    \caption{{\bf Two dimensional and optimally extracted 1D H$\alpha$ spectra for the 20 BL H$\alpha$ emitters at $z\sim5$ identified in this work.} The objects are ordered by increasing broad-to-total H$\alpha$ flux ratio, as in Fig. $\ref{fig:RGBstamps}$. The spectra are centered on the redshift of the narrow component and normalised to the peak of the line emission. The blue shaded region shows the errors. The red part of the spectrum of GOODS-S-13971 is heavily impacted by contamination from a foreground galaxy.}
    \label{fig:1Dgrid}
\end{figure*}

\subsection{Optimal spectral extraction and cleaning of emission-line contamination} \label{sec:optextract} 
The NIRCam WFSS data on the BL H$\alpha$ emitters yields spatially resolved spectra with a resolution of $R\sim1600$. We extract 2D spectra using \texttt{grismconf}\footnote{https://github.com/npirzkal/GRISMCONF} using the V4 trace models\footnote{https://github.com/npirzkal/GRISM$\_$NIRCAM}. Pixel-level corrections are applied to center the emission-lines in the 2D spectrum. In the EIGER data, a fraction of the objects is covered by observations of both NIRCam modules, yielding two orthogonal dispersion directions (see \citealt{Kashino23}). Based on visual inspection, we either use the combined spectrum, or limit ourselves to spectra from a single dispersion direction in case the other is heavily contaminated. In the FRESCO data only a single dispersion direction is available at the position of the candidates.

We find that five BL H$\alpha$ emitters show one or more narrow emission-lines at a slight spatial offset from the broad component, in addition to a co-spatial narrow component. We show the 2D H$\alpha$ spectra for two of those objects in Fig. $\ref{fig:2D_fits}$. We find that these additional narrow lines originate from closely separated companions that are also visible in the imaging data (see the relatively blue companion objects in various stamps in Fig. $\ref{fig:RGBstamps}$). Before extracting 1D spectra that we use to model the line-profile, we remove such additional narrow companions by fitting the 2D spectrum with a three component model using the \texttt{lmfit} package in \texttt{python}. These models consist of two spatially compact 2D gaussians that are close (i.e. within 1 pixel, 0.06$''$) to the center of the 2D spectrum and have narrow and broad line-widths, respectively. We add a second narrow gaussian whose location is allowed to vary freely. The central velocity of each component is a free parameter. We find the best fitting model by using a least squares $\chi^2$ minimisation. A second narrow component is included if this reduces the $\chi^2_{\rm red}$ by $>1$. Fig. $\ref{fig:2D_fits}$ shows two example fits of objects with a secondary narrow component. The residual maps reveal further narrow components at somewhat larger spatial separations that are also associated to the galaxies, but they do not contaminate the 1D spectrum of the object in the center of the trace and are therefore not removed.

After removing such emission-line contamination to the main central galaxy for five BL H$\alpha$ emitters, we optimally extract a 1D spectrum with a weighting determined by the collapsed sum of the main narrow and broad components in the spectral direction \citep[e.g.][]{Horne86}. We show the cleaned 1D spectra of all 20 BL H$\alpha$ emitters in Fig. $\ref{fig:1Dgrid}$.

\subsection{1D line fitting} \label{sec:fit1D}
We now use multi-component gaussian fitting to characterise the optimally extracted H$\alpha$ emission-line spectra shown in Fig. $\ref{fig:1Dgrid}$. The main aim is to determine the relative luminosity and line-widths of the components. We roughly follow the methodology outlined in \cite{Ubler23} and simultaneously fit narrow and broad components of H$\alpha$ and [NII]$_{6549, 6585}$ where the line-ratio of the latter doublet is fixed to 1:2.94. The line centroids and velocity widths of H$\alpha$ and [NII] are tied to each other, but the [NII]$_{6585}$/H$\alpha$ line-ratio of the broad and narrow components may vary independently (from [NII]$_{6585}$/H$\alpha$ = 0 - 1). After an initial guess of the line center, we include the flux within $\pm$ 5000 km s$^{-1}$ for fitting the line-profile. We fit the spectra with a single and a two component gaussian model using \texttt{lmfit}. We use the Adaptive Memory Programming for Global Optimization method as we find that this method yields the most robust results against the initial parameter guesses. In the two component fits, we force the central velocities of the narrow and broad components to be the same, but all other parameters can vary freely. 

\begin{figure}
    \centering
     \includegraphics[width=8.2cm]{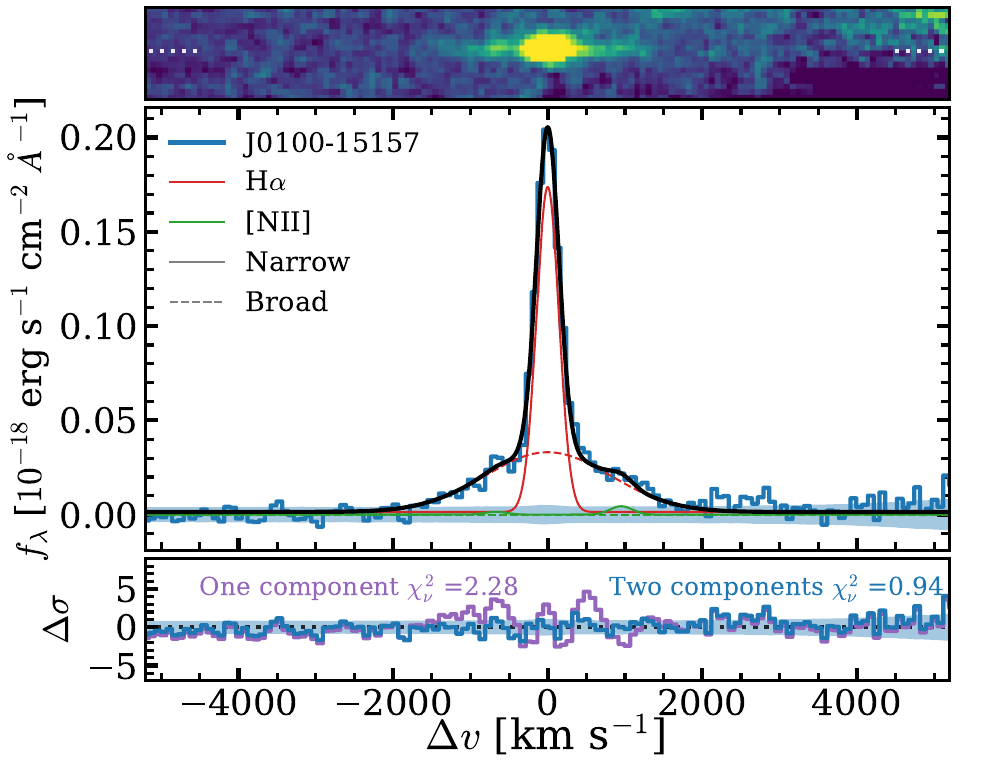} \\  
    \includegraphics[width=8.2cm]{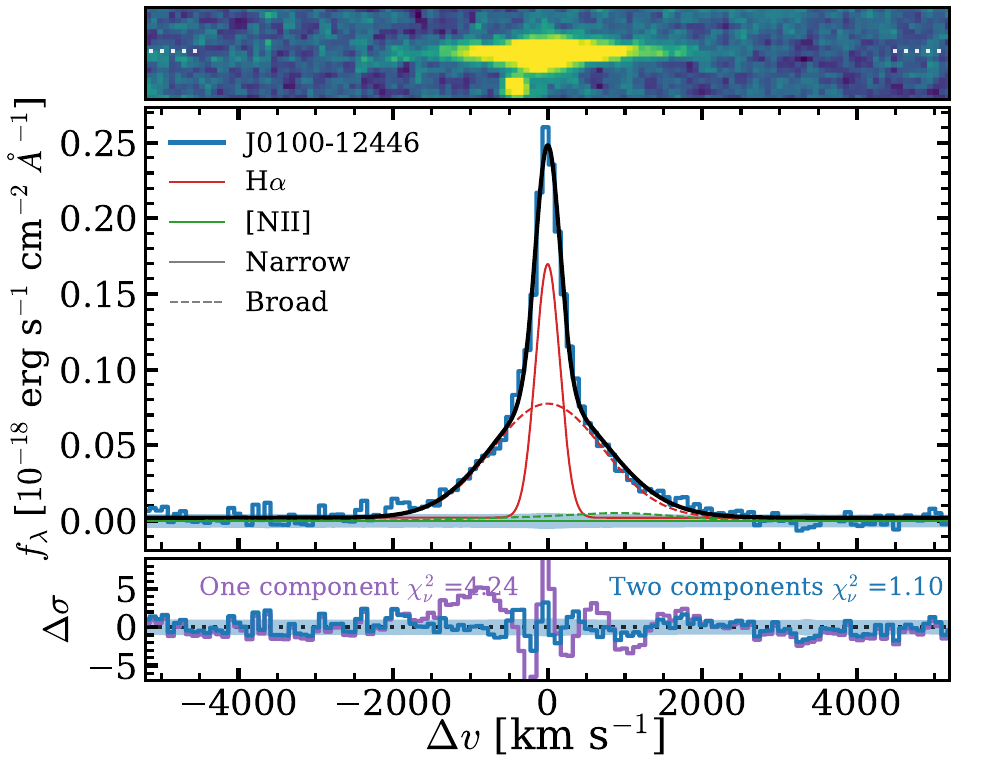} \\
        \includegraphics[width=8.2cm]{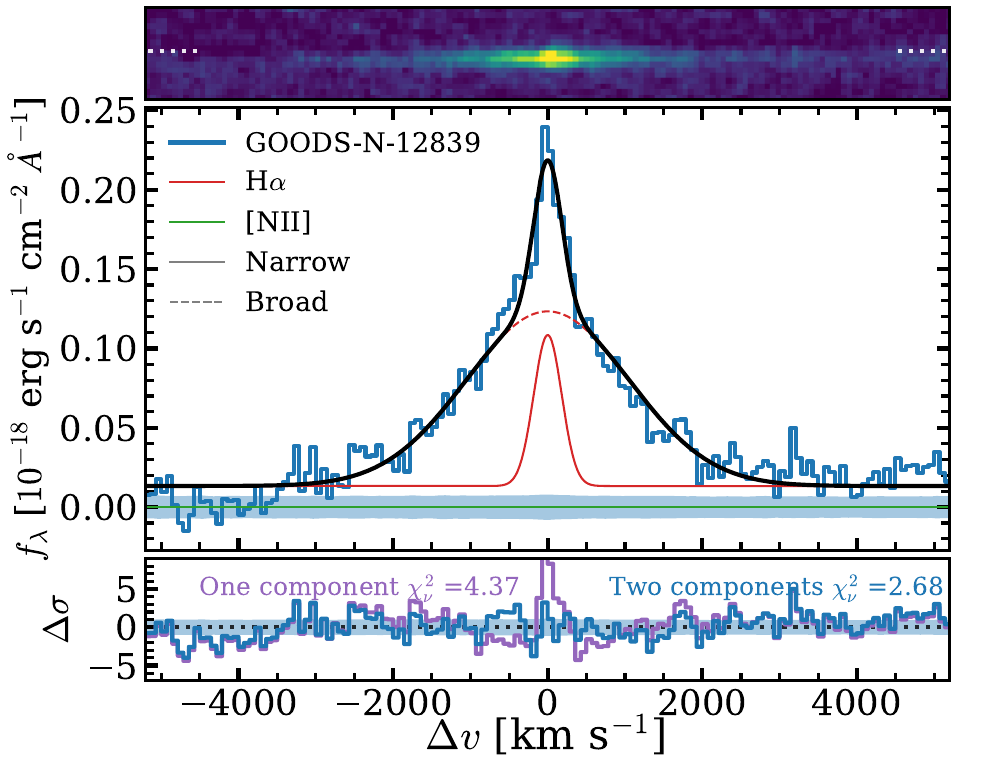} \\ \vspace{-0.2cm}
    \caption{{\bf Example fits of three broad H$\mathbf{\alpha}$ line profiles}. In each of the three panels, we show the 2D emission-line spectrum, the optimally extracted 1D spectrum (blue, errors in blue shades) and the best-fit two component model (black solid line, where the solid red line shows the narrow H$\alpha$ component and the dashed red line the broad H$\alpha$ component, and green shows [NII]) and the residual of the shown two component and the best-fit single component model.
 The examples shown span the range of broad to total flux ratios from L$_{\rm broad}$/L$_{\rm tot}$ = 0.28, 0.66 to 0.87, from top to bottom. Fits for the full sample are shown in Fig. $\ref{fig:profilegrid}$.  } 
    \label{fig:1D_fit}
\end{figure}

Fig. $\ref{fig:1D_fit}$ shows example fitting results for three BL H$\alpha$ emitters that span the range in relative broad-to-total flux ratios. We find that the inclusion of a broad component improves all fits, typically by $\Delta \chi^2_{\rm red}\approx-1$ measured over the $\pm5000$ km s$^{-1}$ window. Additionally, as Fig. $\ref{fig:1D_fit}$ illustrates, the residuals of single component fits typically are not random and show underestimated wings and/or line centers. Broad components are typically detected with S/N=10 (the minimum S/N is 5, for GOODS-S-13971 and the maximum is 40, for GOODS-N-9771). Despite fixing the central velocities of the narrow and broad component to the same value, we find no strong residuals. When allowing the relative velocities to vary, we find consistent results albeit with somewhat larger uncertainties. The full width half maximum (FWHM) of the narrow components are typically 340 km s$^{-1}$ (uncorrected for the line spread function; they are marginally resolved), while broad components typically have FWHMs of $\approx2000$ km s$^{-1}$ (ranging from 1160 - 3700 km s$^{-1}$). [NII] emission is not detected with S/N$>3$ in any of the objects, neither in the broad or narrow components. The broad components typically constitute 65 \% of the total H$\alpha$ flux (the minimum is 27 \%, the maximum is 97 \%, see Fig. $\ref{fig:1D_fit}$). Table $\ref{tab:linefits}$ lists the key fitted properties of the broad components. We note that the red part of the line-profile of GOODS-S-13971 is strongly impacted by the edge of the trace of a contaminating foreground object in the grism data. Therefore, we only include flux bluewards of the line-centroid when fitting this object.

\begin{table*} 
    \centering
    \caption{{\bf Properties of the broad H$\mathbf{\alpha}$ line emitters.} $^{\star}$ The F200W magnitudes from the FRESCO sample in the GOODS fields are computed by averaging the flux in the F182M and F210M filters. $^{\dagger}$ The magnitude in the long-wavelength filter that contains the H$\alpha$ line, which is F444W and F356W for the FRESCO and EIGER sample, respectively. The rest-frame H$\alpha$ equivalent width, EW$_{0, \rm H\alpha}$ corresponds to the sum of the narrow and broad component. $\Delta m_{\rm LW}$ is the magnitude boost in the long-wavelength filter due to the H$\alpha$ line emission as described in \S $\ref{sec:colors}$. }
    \begin{tabular}{cccccccc}
    ID & L$_{\rm broad}$/L$_{\rm tot}$ & L$_{\rm broad}$/$10^{42}$ erg s$^{-1}$ & $v_{\rm FWHM}$/km s$^{-1}$ & F200W$^{\star}$ & F356W or F444W$^{\dagger}$ & EW$_{0, \rm H\alpha}$/{\AA} & $\Delta m_{\rm LW}$ \\ \hline
GOODS-N-4014 & $ 0.79 \pm 0.01 $ & $ 4.6 \pm 0.2 $ & $ 2103 \pm 159 $ & $ 28.08^{+0.76}_{-0.47} $ & $ 24.98^{+0.05}_{-0.05} $ & $ 488^{+64}_{-53} $ & $ -0.6 $ \\
GOODS-N-9771 & $ 0.71 \pm 0.01 $ & $ 44.7 \pm 1.1 $ & $ 3739 \pm 112 $ & $ 26.11^{+0.05}_{-0.05} $ & $ 23.01^{+0.06}_{-0.05} $ & $ 789^{+69}_{-65} $ & $ -1.4 $ \\
GOODS-N-12839 & $ 0.87 \pm 0.01 $ & $ 18.8 \pm 0.7 $ & $ 2482 \pm 147 $ & $ 26.92^{+0.10}_{-0.09} $ & $ 23.73^{+0.06}_{-0.05} $ & $ 571^{+69}_{-65} $ & $ -0.7 $ \\
GOODS-N-13733 & $ 0.70 \pm 0.01 $ & $ 2.4 \pm 0.2 $ & $ 2208 \pm 200 $ & $ 27.91^{+0.22}_{-0.19} $ & $ 25.41^{+0.06}_{-0.05} $ & $ 418^{+64}_{-62} $ & $ -0.5 $ \\
GOODS-N-14409 & $ 0.79 \pm 0.02 $ & $ 3.6 \pm 0.4 $ & $ 1474 \pm 190 $ & $ 27.71^{+0.24}_{-0.19} $ & $ 25.53^{+0.06}_{-0.06} $ & $ 657^{+127}_{-142} $ & $ -0.9 $ \\
GOODS-N-15498 & $ 0.97 \pm 0.01 $ & $ 5.3 \pm 1.0 $ & $ 2360 \pm 214 $ & $ 28.10^{+0.23}_{-0.19} $ & $ 24.71^{+0.06}_{-0.05} $ & $ 376^{+128}_{-106} $ & $ -0.4 $ \\
GOODS-N-16813 & $ 0.57 \pm 0.01 $ & $ 4.5 \pm 0.5 $ & $ 2033 \pm 219 $ & $ 26.47^{+0.09}_{-0.08} $ & $ 24.84^{+0.06}_{-0.05} $ & $ 567^{+83}_{-78} $ & $ -0.8 $ \\
GOODS-S-13971 & $ 0.48 \pm 0.03 $ & $ 2.5 \pm 0.5 $ & $ 2192 \pm 479 $ & $ 26.76^{+0.06}_{-0.06} $ & $ 24.35^{+0.05}_{-0.06} $ & $ 231^{+59}_{-54} $ & $ -0.2 $ \\
J1148-7111 & $ 0.60 \pm 0.01 $ & $ 5.5 \pm 0.4 $ & $ 2967 \pm 259 $ & $ 26.58^{+0.06}_{-0.06} $ & $ 24.47^{+0.04}_{-0.04} $ & $ 645^{+73}_{-70} $ & $ -0.7 $ \\
J1148-18404 & $ 0.43 \pm 0.02 $ & $ 3.3 \pm 0.7 $ & $ 2886 \pm 346 $ & $ 27.74^{+0.06}_{-0.06} $ & $ 24.36^{+0.04}_{-0.04} $ & $ 386^{+63}_{-67} $ & $ -0.4 $ \\
J1148-21787 & $ 0.48 \pm 0.03 $ & $ 3.2 \pm 0.8 $ & $ 2321 \pm 360 $ & $ 26.10^{+0.06}_{-0.05} $ & $ 24.58^{+0.05}_{-0.05} $ & $ 520^{+110}_{-111} $ & $ -0.5 $ \\
J0100-2017 & $ 0.52 \pm 0.02 $ & $ 3.2 \pm 0.4 $ & $ 1953 \pm 196 $ & $ 26.36^{+0.05}_{-0.05} $ & $ 25.02^{+0.05}_{-0.05} $ & $ 584^{+76}_{-78} $ & $ -0.7 $ \\
J0100-12446 & $ 0.67 \pm 0.01 $ & $ 6.8 \pm 0.3 $ & $ 1681 \pm 89 $ & $ 26.33^{+0.04}_{-0.04} $ & $ 24.47^{+0.04}_{-0.04} $ & $ 624^{+47}_{-45} $ & $ -0.7 $ \\
J0100-15157 & $ 0.28 \pm 0.01 $ & $ 3.1 \pm 0.3 $ & $ 1786 \pm 136 $ & $ 25.58^{+0.04}_{-0.04} $ & $ 24.69^{+0.04}_{-0.05} $ & $ 777^{+61}_{-59} $ & $ -1.1 $ \\
J0100-16221 & $ 0.65 \pm 0.01 $ & $ 3.4 \pm 0.3 $ & $ 2145 \pm 127 $ & $ 26.41^{+0.05}_{-0.05} $ & $ 24.92^{+0.04}_{-0.04} $ & $ 543^{+67}_{-67} $ & $ -0.6 $ \\
J0148-976 & $ 0.51 \pm 0.02 $ & $ 2.4 \pm 0.3 $ & $ 1445 \pm 236 $ & $ 26.06^{+0.05}_{-0.05} $ & $ 25.12^{+0.05}_{-0.05} $ & $ 622^{+94}_{-94} $ & $ -0.6 $ \\
J0148-4214 & $ 0.47 \pm 0.01 $ & $ 2.8 \pm 0.2 $ & $ 1768 \pm 166 $ & $ 25.81^{+0.04}_{-0.04} $ & $ 24.65^{+0.04}_{-0.04} $ & $ 390^{+29}_{-27} $ & $ -0.4 $ \\
J0148-12884 & $ 0.65 \pm 0.02 $ & $ 2.3 \pm 0.3 $ & $ 1166 \pm 160 $ & $ 25.97^{+0.05}_{-0.05} $ & $ 24.80^{+0.05}_{-0.05} $ & $ 304^{+58}_{-61} $ & $ -0.3 $ \\
J1120-7546 & $ 0.86 \pm 0.02 $ & $ 7.4 \pm 0.8 $ & $ 1843 \pm 189 $ & $ 27.68^{+0.14}_{-0.13} $ & $ 24.69^{+0.04}_{-0.04} $ & $ 592^{+113}_{-111} $ & $ -0.7 $ \\
J1120-14389 & $ 0.74 \pm 0.01 $ & $ 4.2 \pm 0.4 $ & $ 2342 \pm 149 $ & $ 26.62^{+0.05}_{-0.05} $ & $ 24.59^{+0.04}_{-0.04} $ & $ 362^{+53}_{-47} $ & $ -0.4 $ \\ \hline
    \end{tabular}
    \label{tab:linefits}
\end{table*}

\section{Properties of broad line H$\alpha$ emitters} \label{sec:properties}
\subsection{The case for an AGN origin} \label{sec:AGN}
Having established the methods underlying the selection and the emission-line measurements of the broad line H$\alpha$ emitters in the EIGER and FRESCO surveys at $z\sim5$, we now argue why the most likely origin of the broad line emission is nuclear black hole activity. 
The relatively narrow wavelength coverage of the {\it JWST}/NIRCam grism spectra in a single broadband filter ($\Delta \lambda_0\approx200$ nm) prevents us to use well-known emission-line diagnostics \citep[e.g.][]{BPT81} to identify the excitation source of the broad and narrow components, as our spectra typically only cover the bright H$\alpha$ and [NII] lines.

Despite being covered by deep X-Ray data, we find that none of the BL H$\alpha$ emitters in the GOODS fields is matched to published X-Ray detections \citep{Cappelluti16}. By inspecting the {\it Chandra} data in the GOODS fields (similar to the method employed in \citealt{Bogdan23}), we measure $3\sigma$ upper limits of $\approx1\times10^{-16}$ erg s$^{-1}$ cm$^{-2}$ in the 2-7 keV hard band in GOODS-N, and $3\times10^{-17}$ erg s$^{-1}$ cm$^{-2}$ in GOODS-S. At $z\approx5$, the 2-7 keV band probes 12—50 keV rest-frame, which should basically be obscuration independent and an excellent tracer of the intrinsic X-ray luminosity. For a typical AGN X-ray powerlaw slope of $\Gamma=1.9$ \citep[e.g.][]{Nanni17}, the negative k-correction in the X-rays implies $3\sigma$ upper limits of $L_{X,\rm intrinsic} < 3\times10^{43}$ erg s$^{-1}$. Given these limits, our arguments for an AGN origin are therefore based on spatial information in the grism and imaging data, and the line profiles. 

\subsubsection{Spatially resolved spectroscopy} \label{sec:resolvedspec}
The NIRCam grism data allow us to perform spatially resolved spectroscopy at a resolution of $\approx0.1''$ (about 600 pc at $z\approx5$). As already discussed in Section $\ref{sec:optextract}$, we identified multiple narrow components at close separations to the sources that displays a broad component (see Fig. $\ref{fig:2D_fits}$). Here we investigate differences in the spatial extent of the narrow and broad components with the 2D stacked spectrum of the full sample. We create this stacked 2D spectrum by first shifting each spectrum to the rest-frame wavelength, then correcting for differences in luminosity distance and finally construct the median stacked spectrum and its uncertainty from 100 bootstrap realisations (with replacement) of the sample. Unlike the 1D spectra, we do not remove narrow lines from nearby components from the spectrum (as in Section $\ref{sec:optextract}$) before stacking.

\begin{figure}
    \centering
    \includegraphics[width=8cm]{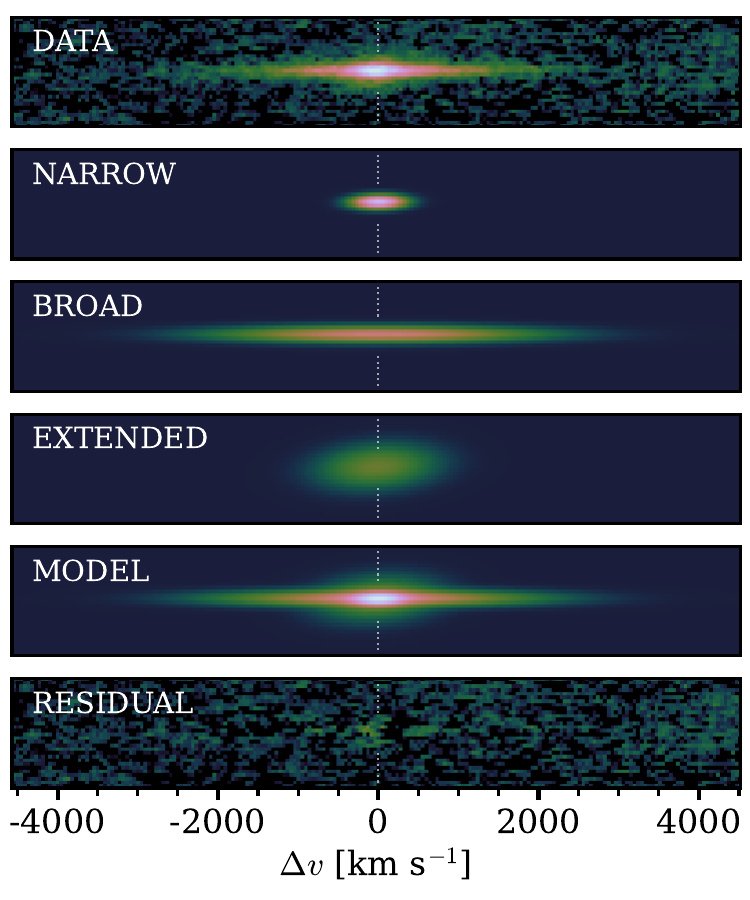}
    \caption{{\bf Median stacked 2D H$\alpha$ spectrum of the 20 BL H$\alpha$ emitters at $\mathbf{z\sim5}$.} The top panel shows the continuum-removed data, the next three panels show the three individual components that are combined in the total model (fifth panel). The residuals from the best-fit model are shown in the bottom panel. A power law scaling with exponent $\gamma=0.33$ is used to highlight low surface brightness emission.}
    \label{fig:2D_stack}
\end{figure}

Fig. $\ref{fig:2D_stack}$ shows the stacked 2D spectrum of the full sample. We require a three component gaussian model to yield a good fit (with $\chi^2_{\rm red}=0.95$) to the data that does not leave strong residual structures. The best-fit three component model to the stacked spectrum consists of a narrow and a broad component that are spatially unresolved in the spatial direction, in addition to a second relatively narrow component (named ``extended" component) that is significantly more extended in the spatial direction (with FWHM of 0.45$''$, about 2.5 kpc). The best-fit line-widths (FWHM) of the components are 430, 2550 and 1000 km s$^{-1}$, for the narrow, broad and third component, respectively. However, we note that any spatial extent in the dispersion direction would lead to artificial line broadening in the grism data. This may plausibly be the case for the third component that is spatially extended in the direction orthogonal to the dispersion. If we assume that this spatial extent is spherically symmetric, the corrected line-width would be $\approx800$ km s$^{-1}$. Despite allowing the spatial and spectral centroids of the third component to vary, we find that all three components are centred on the same spectral and spatial position. The majority, 54 \%, of the flux is in the broad component, while similar fractions of the flux are found in the narrow (24 \%) and extended (22 \%) component. 

The fact that the majority of the H$\alpha$ flux originates from a very broad, spatially unresolved component, is strongly suggestive of a dominant AGN origin powering the broad H$\alpha$ emission. The interpretation of the other narrower components, in particular the spatially extended component, is more ambiguous. Possible explanations include (combinations of) i) emission from nearby companions and/or clumps in the host galaxies that are powered by star formation, ii) diffuse H$\alpha$ emission potentially originating from outflowing (shock ionized) gas (possibly connected to extended Lyman-$\alpha$ emission; e.g. \citealt{Farina19}), or iii) the non gaussian-shape of the point spread function (PSF) and/or non-gaussian spatial shape of the narrow component. Evidence for a contribution from emission from companions has already been identified in Section $\ref{sec:optextract}$. The L$_{\rm broad}$/L$_{\rm tot}$ ratio of 54 \% in the stack, compared to a median of 65 \% for individual measurements (Table $\ref{tab:linefits}$) -- from which the strongest companion contaminants were removed -- suggests that these companions contribute at least 10 \% of the total flux of the stack, and they have a dominant contribution to the flux in the extended component.

\subsubsection{High resolution IR imaging} \label{sec:highresimaging}
The high-resolution NIRCam imaging data in various filters provides more information on the origin of the broad line emission. In Fig. $\ref{fig:stamps}$ we show $1.5\times1.5''$ stamps of example BL H$\alpha$ emitters (the full set is shown in Appendix $\ref{appendix:stamps}$) in a high resolution short-wavelength filter at 2 micron and in the red filter that contains the H$\alpha$ line-emission. For FRESCO, we sum the data in the F182M and F210M medium-bands to increase the sensitivity in the short wavelength. We use a power-law scaling (with $\gamma=0.6$) to highlight low surface brightness emission such as the hexagonal PSF effects. The clear appearance of these features is indicative of a point-source(-like) object in the red filter. In some cases the point-source is subdominant to other nearby components in the blue filter (e.g. GOODS-S-13971). While these are compact objects in {\it JWST}/NIRCam false-color images including the long wavelength channel, they can have different apparent morphologies in bluer wavelength filters. 

We interpret the nearby extended objects as companions or components of the host galaxies of the point-source object. Indeed, in Section $\ref{sec:optextract}$ we show examples where we detect narrow H$\alpha$ line emission from (some of) these objects. These companions span F115W magnitudes from 26 to 29 and have typical separations of 0.15-0.3$''$ to the compact objects, i.e. these are systems at distances of 1-2 kpc which is well within any plausible virial radius, and therefore suggest ongoing interactions.

\begin{figure}
    \centering
\includegraphics[width=8.1cm]{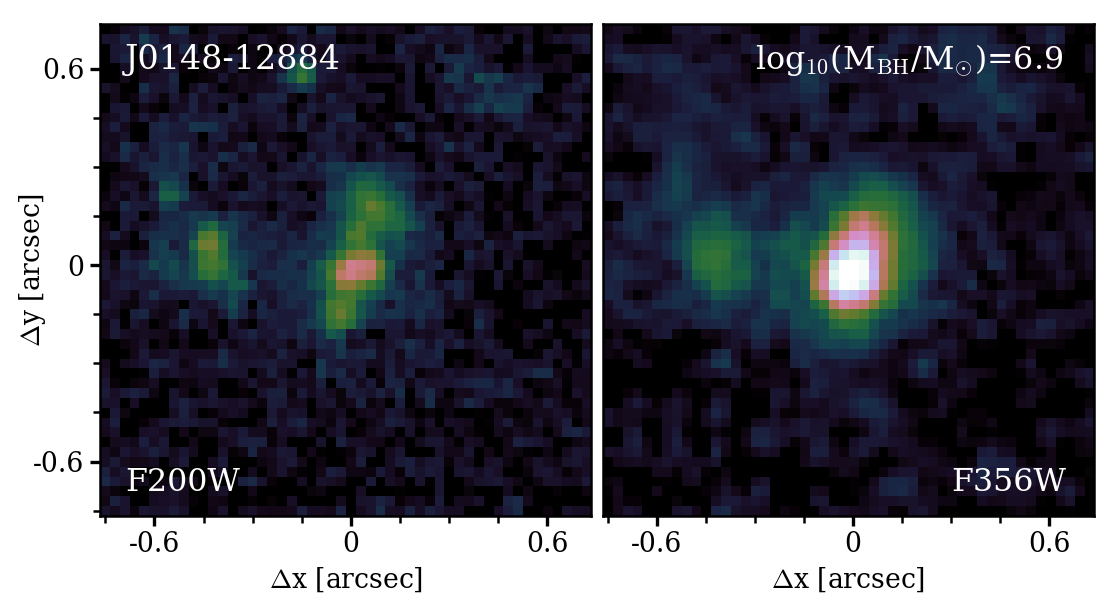} \\ \vspace{-0.6cm}
\includegraphics[width=8.1cm]{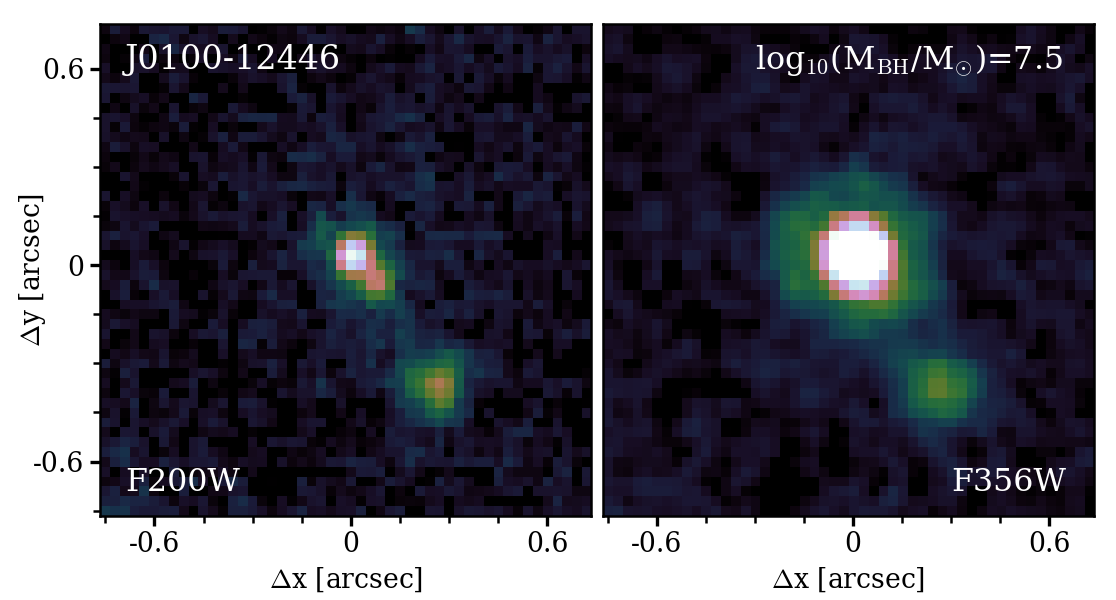} \\ \vspace{-0.6cm}
\includegraphics[width=8.1cm]{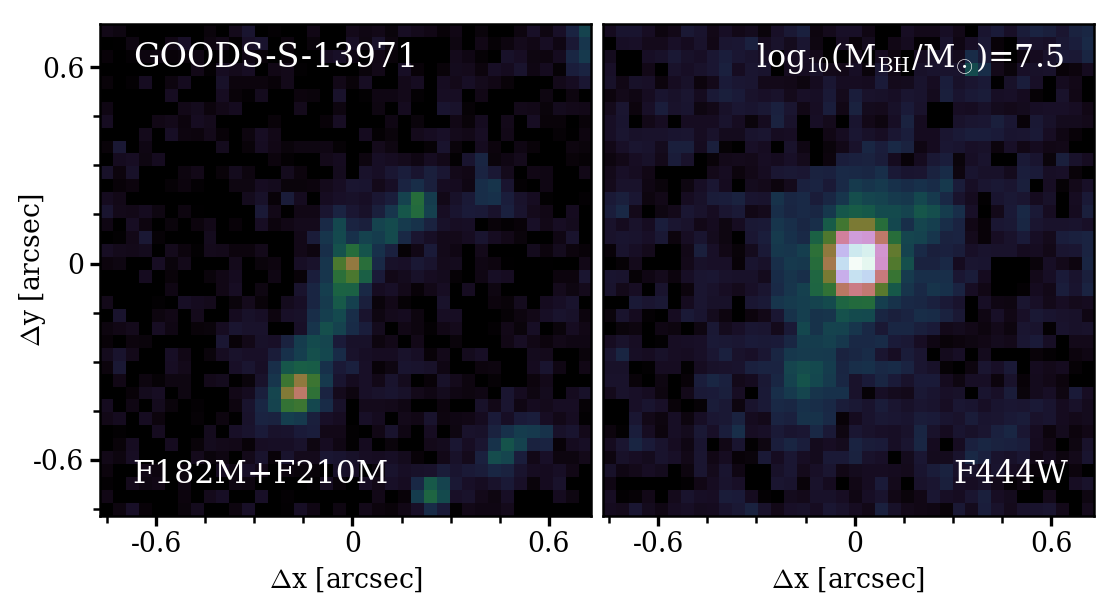} \\ \vspace{-0.6cm}
\includegraphics[width=8.1cm]{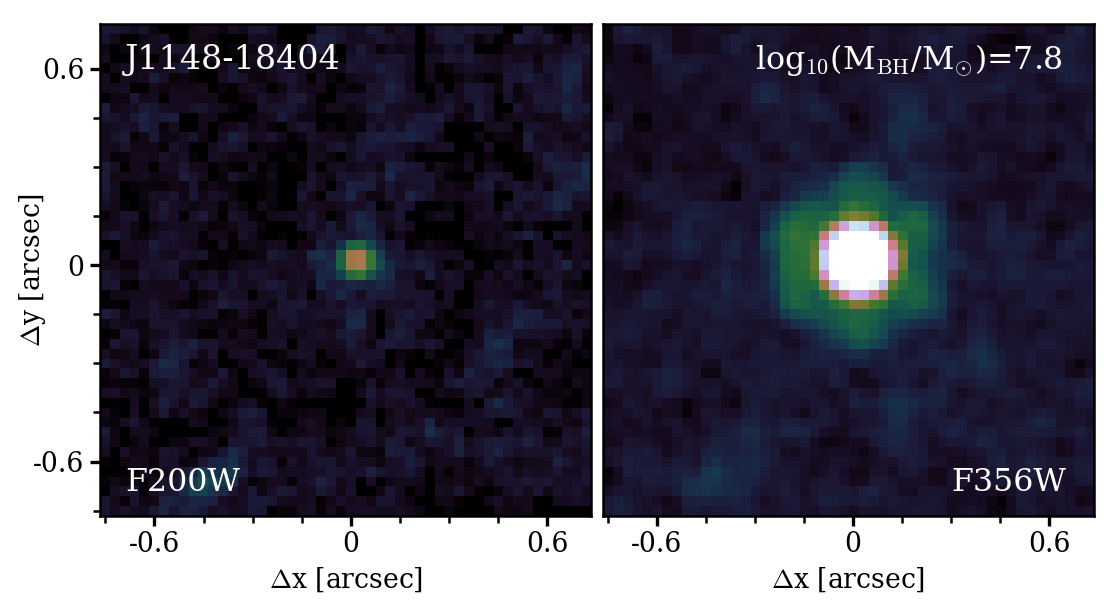} \\ \vspace{-0.6cm}
\includegraphics[width=8.1cm]{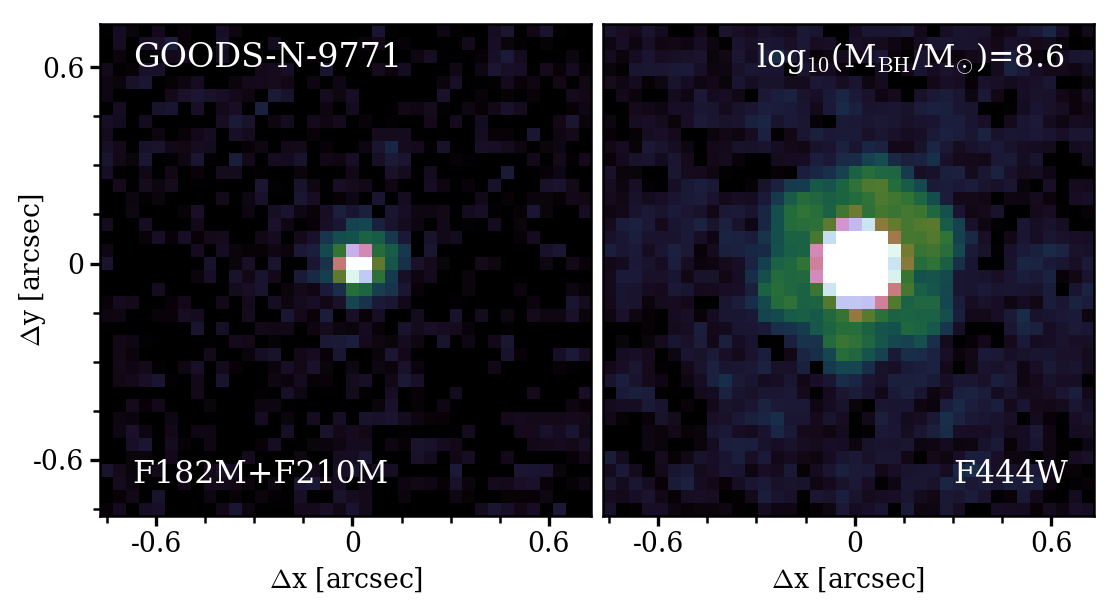} \\ \vspace{-0.23cm}
    \caption{{\bf Zoom-in stamps of 5 example BL H$\alpha$ emitters ordered by increasing BH mass from log$_{10}$(M$_{\rm BH}$/M$_{\odot}$)= 6.9, 7.4, 7.5, 7.8, 8.6.} The left column shows the F200W (EIGER) or F182M+F210M (FRESCO) image that has a PSF FWHM$\approx0.06''$. The right column shows the F356W or F444W image respectively (PSF FWHM $\approx0.12''$) that contains the H$\alpha$ line emission. The colour scaling follows a power-law with exponent $\gamma=0.6$ to highlight low surface brightness emission. All stamps are shown in Appendix $\ref{appendix:stamps}$.} 
    \label{fig:stamps}
\end{figure} 

As we lack extremely deep multi-wavelength imaging -- in particular in bands at $>2.5$ micron that span the Balmer break -- a full multi-wavelength modeling of the spectral energy distributions is beyond the scope of this paper. For the purpose of pinpointing the origin of the emission from the broad H$\alpha$ component, we measure the relative contribution of the red point-source to the total 2 micron flux of all components in the displayed stamps (Fig. $\ref{fig:stamps}$). We limit ourselves to the EIGER data as the short-wavelength imaging is significantly more sensitive due to the use of a wider filter. This is done by fitting the morphology in the F200W images with a combination of point sources and exponential profiles using \texttt{imfit} \citep{Erwin2015}. The PSF is modeled using the WebbPSF package \citep{Perrin12}. We base the number of included exponential components on the reduced $\chi^2_{\rm red}$. If adding one component does not decrease it by more than 0.5, we no longer add components. We typically fit one point source and two exponential components, but we find that in some cases a single point source suffices (for example J1148-18404), while J0148-12884 is the most complex system with five exponential components, see Fig. $\ref{fig:stamps}$.

\begin{figure}
    \centering
    \includegraphics[width=8.2cm]{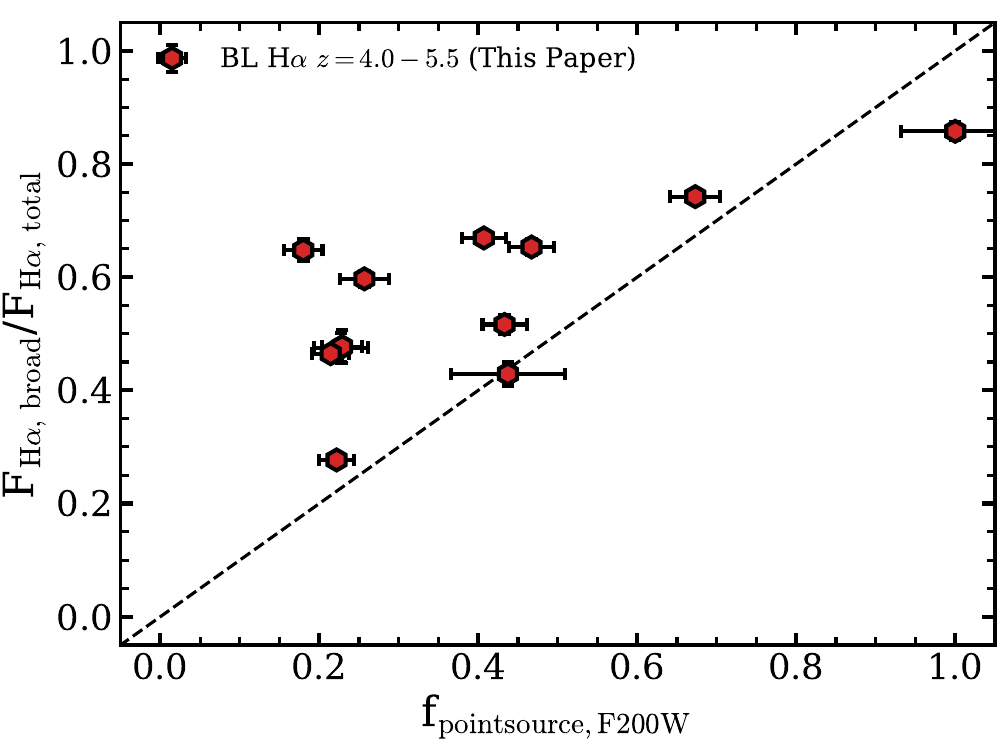}
    \caption{{\bf The fraction of the H$\alpha$ flux that is in the broad component versus the fraction of the flux in the point source in the F200W data}. Here only the deeper EIGER imaging is used. The dashed line shows the 1:1 relation. The correlation between the two fractions that are independently derived is further evidence that the broad component originates from a point-source.}
    \label{fig:corefrac}
\end{figure}

We find that typically $\approx$ 40 \%  (between 20 and 100 \%) of the flux in the F200W filter is due to the point source. As shown in Fig. $\ref{fig:corefrac}$, the fraction in the central component in the F200W filter correlates with the H$\alpha$ flux in the broad component. This is further independent evidence that the broad component predominantly originates from a point source, and not from e.g. a collection of surrounding clumps with high velocity dispersion.

\subsubsection{H$\alpha$ line-profiles} \label{sec:Haprofile}
In addition to emission from the broad line regions in AGN, broad components of the H$\alpha$ emission line have been studied in the context of galactic outflows either powered by star formation \citep[e.g.][]{Arribas14,Davies19,Freeman19,Swinbank19,FSWuyts20,Hogarth20,Mainali22}, supernovae \citep[e.g.][]{Filippenko97,Baldassare16} or by AGN \citep[e.g.][]{ForsterSchreiber19,Leung19}. Transient broad components from supernovae can appear as broad as 1000-4000 km s$^{-1}$, but those are typically shifted by $\gtrsim100$ km s$^{-1}$ from the narrow component \citep{Baldassare16}. In comparison to the broad components we identify, the line-profiles that are analysed in studies of star-forming galaxies typically have significantly narrower broad components ($\approx500-1000$ km s$^{-1}$), are blue-shifted with respect to the narrow component and have a much lower broad-to-total flux ratio. Based on stacks of star forming galaxies at $z\approx2$, \cite{Davies19} find a typical broad to narrow H$\alpha$ flux ratio of 50 \%, about two times lower than our typical value. Their broad components have FWHM $\approx940$ km s$^{-1}$. \cite{Llerena23} analyse broad emission-line profiles of low-mass galaxies at $z\approx2$ that are analogues of typical high-redshift galaxies and find typical FWHM $\approx350$ km s$^{-1}$. 

In the hypothetical case that our broad components would originate from outflows driven by star formation, their high line-width and their relatively high flux with respect to the narrow components would indicate a very high mass loading factor, $\eta=\frac{\rm \dot M_{\rm out}}{\rm SFR}$, as the numerator depends on the width and luminosity of the broad component, and the denominator on the luminosity of the narrow component. If we calculate the star formation rate (SFR) based on the luminosity of the narrow H$\alpha$ component, we obtain a typical SFR of 15 M$_{\odot}$ yr$^{-1}$ (from 2-90 M$_{\odot}$ yr$^{-1}$) following the \cite{KennicuttEvans12} conversion without attenuation correction. We calculate the outflow velocity following e.g. \cite{Genzel11} and \cite{Davies19}: $v_{\rm out}=\langle v_{\rm broad} \rangle + 2\sigma_{\rm broad}$, where $\langle v_{\rm broad}\rangle$ is the absolute velocity difference between the narrow and broad component, which we find to be consistent with zero in our fits. Typical values of the outflow velocity are $v_{\rm out}=1840$ kms $^{-1}$, with a minimum of 990 km s$^{-1}$. Then, scaling the relation between ${\rm \dot M_{\rm out}} \propto L_{\rm H\alpha, broad} \, v_{\rm out}$ from \cite{Ubler23}, following their (standard) assumptions on the geometry and electron density and assuming an outflow size $r_{\rm 1/2}=0.2$ kpc (i.e. unresolved in our grism data), we derive median mass outflow rates of $73$ M$_{\odot}$ yr$^{-1}$ (ranging from 23 to 1500 M$_{\odot}$ yr$^{-1}$), and median mass loading factors $\eta=6.1$ (from $1-27$). These mass loading factors are much higher than typically found in star forming galaxies at high-redshift ($\eta\approx0.1-0.5$; e.g. \citealt{Bordoloi16,Davies19,Llerena23}). Thus, unless the relative dust attenuation of the narrow component over the broad component is very high, which would shift $\eta$ down, this back-of-the-envelope calculation does not support an outflow origin of the broad components. 

AGN driven outflows have broad line widths that are similar to the widths measured here, and they also have relatively high broad-to-total flux ratios \citep{ForsterSchreiber19}. However, these outflows often have broad components that are blue-shifted with respect to the narrow component, are spatially extended, and often show relatively strong [NII] components suggestive of shock ionization \citep[e.g.][]{ForsterSchreiber14,Leung19}, all unlike the spectra that we measure in our sample. A particularly relevant comparison is the blue galaxy that contains an AGN at $z=5.55$ studied with {\it JWST}/NIRSpec by \citet{Ubler23} (see also \citealt{Vanzella10c}). This object has an H$\alpha$ line decomposed into a narrow, broad (AGN) and intermediate (outflowing) component. The broad component has a FWHM of 3300 km s$^{-1}$ which is similar to some of the objects in our sample (see Table $\ref{tab:linefits}$), whereas the outflowing component that they detect has a FWHM of 720 km s$^{-1}$, which is narrower than any broad component we detect.

In summary, the H$\alpha$ line profiles that we measure are significantly different from the profiles in star-forming galaxies, where much fainter and somewhat narrower broad components have been interpreted as signposting outflows. While the spectra are more similar to those identified in galaxies that contain galactic scale AGN-driven outflows, the spatial compactness of the broad component is strongly suggestive of an origin in the broad line region of the AGN. The typical broad to total H$\alpha$ flux ratios and the broad line-widths and equivalent widths (Table $\ref{tab:linefits}$) are somewhat lower than the average broad line H$\alpha$ selected sample of AGN in the Sloan Digital Sky Survey with similar luminosity (SDSS; \citealt{SternLaor12}; $\langle$L$_{\rm broad}$/L$_{\rm tot}\rangle=0.92$, $v_{\rm FWHM}=3700$ km s$^{-1}$), but well within the dynamic range probed by these Seyfert 1.8 galaxies.

\begin{table*}
    \centering
    \caption{{\bf BH and galaxy properties of the BL H$\alpha$ emitters.}}
    \begin{tabular}{cccccc}
    ID & log$_{10}$(M$_{\rm BH}$/M$_{\odot}$) & $L_{\rm bol}$/ $10^{44}$ erg s$^{-1}$ & M$_{\rm UV}$ & $\beta_{\rm UV}$ & $\beta_{\rm opt}$ \\ \hline
GOODS-N-4014 & $7.58\pm0.08$ & $ 9.3 \pm 0.5 $ & $-18.0\pm0.2$ & $ -2.04^{+0.70}_{-0.93} $ & $ 2.00^{+1.12}_{-0.82} $ \\
GOODS-N-9771 & $8.55\pm0.03$ & $ 65.8 \pm 1.6 $ & $-19.5\pm0.1$ & $ -0.61^{+0.13}_{-0.13} $ & $ 0.76^{+0.31}_{-0.36} $ \\
GOODS-N-12839 & $8.01\pm0.06$ & $ 31.2 \pm 1.2 $ & $-19.0\pm0.1$ & $ -1.64^{+0.19}_{-0.18} $ & $ 1.92^{+0.24}_{-0.22} $ \\
GOODS-N-13733 & $7.49\pm0.10$ & $ 5.2 \pm 0.3 $ & $-17.9\pm0.2$ & $ -1.59^{+0.55}_{-0.48} $ & $ 1.23^{+0.41}_{-0.32} $ \\
GOODS-N-14409 & $7.21\pm0.14$ & $ 7.4 \pm 0.9 $ & $-18.3\pm0.1$ & $ -2.26^{+0.37}_{-0.36} $ & $ 0.09^{+0.40}_{-0.37} $ \\
GOODS-N-15498 & $7.71\pm0.11$ & $ 10.4 \pm 1.9 $ & $-17.7\pm0.2$ & $ -1.63^{+0.50}_{-0.44} $ & $ 2.75^{+0.38}_{-0.34} $ \\ 
GOODS-N-16813 & $7.55\pm0.12$ & $ 9.1 \pm 1.0 $ & $-19.7\pm0.1$ & $ -2.19^{+0.12}_{-0.13} $ & $ -0.60^{+0.23}_{-0.23} $ \\
GOODS-S-13971 & $7.49\pm0.25$ & $ 5.5 \pm 1.2 $ & $-19.4\pm0.1$ & $ -2.07^{+0.09}_{-0.09} $ & $ 1.45^{+0.14}_{-0.14} $ \\
J1148-7111 & $7.92\pm0.10$ & $ 10.8 \pm 0.8 $ & $-18.6\pm0.1$ & $ -1.00^{+0.31}_{-0.28} $ & $ 0.25^{+0.18}_{-0.17} $ \\
J1148-18404 & $7.79\pm0.14$ & $ 6.9 \pm 1.4 $ & $-17.5\pm0.1$ & $ -0.20^{+0.43}_{-0.35} $ & $ 2.73^{+0.17}_{-0.15} $ \\
J1148-21787 & $7.59\pm0.18$ & $ 6.7 \pm 1.6 $ & $-19.2\pm0.1$ & $ -1.85^{+0.23}_{-0.23} $ & $ -0.42^{+0.17}_{-0.16} $ \\
J0100-2017 & $7.44\pm0.11$ & $ 6.7 \pm 0.8 $ & $-19.3\pm0.1$ & $ -1.57^{+0.19}_{-0.18} $ & $ -1.01^{+0.17}_{-0.17} $ \\
J0100-12446 & $7.46\pm0.06$ & $ 12.9 \pm 0.6 $ & $-19.0\pm0.1$ & $ -0.92^{+0.21}_{-0.20} $ & $ -0.20^{+0.14}_{-0.14} $ \\
J0100-15157 & $7.35\pm0.08$ & $ 6.5 \pm 0.6 $ & $-20.2\pm0.1$ & $ -1.87^{+0.14}_{-0.14} $ & $ -2.34^{+0.19}_{-0.24} $ \\
J0100-16221 & $7.53\pm0.07$ & $ 7.1 \pm 0.6 $ & $-18.9\pm0.1$ & $ -1.42^{+0.19}_{-0.19} $ & $ -0.50^{+0.14}_{-0.14} $ \\
J0148-976 & $7.11\pm0.18$ & $ 5.2 \pm 0.7 $ & $-19.1\pm0.1$ & $ -1.44^{+0.18}_{-0.18} $ & $ -1.52^{+0.19}_{-0.17} $ \\
J0148-4214 & $7.32\pm0.10$ & $ 5.9 \pm 0.4 $ & $-20.0\pm0.1$ & $ -1.71^{+0.14}_{-0.13} $ & $ -0.83^{+0.13}_{-0.11} $ \\
J0148-12884 & $6.91\pm0.15$ & $ 5.0 \pm 0.6 $ & $-19.5\pm0.1$ & $ -1.65^{+0.17}_{-0.16} $ & $ -0.59^{+0.12}_{-0.14} $ \\
J1120-7546 & $7.56\pm0.11$ & $ 13.8 \pm 1.6 $ & $-17.6\pm0.3$ & $ -0.41^{+0.94}_{-0.68} $ & $ 1.59^{+0.27}_{-0.27} $ \\
J1120-14389 & $7.65\pm0.07$ & $ 8.4 \pm 0.8 $ & $-19.1\pm0.1$ & $ -1.58^{+0.19}_{-0.19} $ & $ 0.62^{+0.13}_{-0.13} $ \\
\hline
    \end{tabular}
    \label{tab:properties}
\end{table*}

\subsection{Central black hole properties} \label{sec:BHprops}
Now, since we argued that the morphology and H$\alpha$ line-profiles of our sample are strongly suggestive that the broad component is powered by AGN activity, we derive the bolometric luminosity and black hole mass based on the virial relations and the fitted broad component following recent works \citep[e.g.][]{Ubler23,Kocevski23,Harikane23}.\footnote{We do not apply a dust correction to the broad H$\alpha$ luminosity, which means that our measurements are likely lower limits.} Following \cite{Reines13}, we estimate the BH mass using the equation:
\begin{equation}
\begin{split}
    {\rm log_{10}(M_{\rm BH}/M_{\odot})} = 6.57 + {\rm log}_{10}(\epsilon) + \\ 0.47 \, {\rm log}_{10}(L_{\rm H\alpha, broad}/10^{42} {\rm erg \,s}^{-1}) + \\ 2.06 \,  {\rm log}_{10}(v_{\rm FWHM, broad}/10^3 {\rm km \, s^{-1}}),
\end{split}
\end{equation}

where $\epsilon$ is a geometric correction factor related to the properties of the broad line region that we assume to be 1.075 following \cite{ReinesVolonteri15}. An estimate of the systematic uncertainty of BH mass measurements based on this virial relation is 0.5 dex \citep{ReinesVolonteri15}. The bolometric luminosity is challenging to measure as the observed photometry is likely significantly contaminated by emission from star forming regions around the AGN. We therefore follow the approach from \cite{Harikane23} by estimating the AGN continuum luminosity from the broad H$\alpha$ line \citep{GreeneHo2005} and applying the relevant bolometric correction from \cite{Richards08}:

\begin{equation}
  L_{\rm bol}/10^{44} {\rm erg\,s^{-1}}=  10.33(L_{\rm H\alpha, broad}/5.25\times10^{42})^{1/1.157}
  \label{eqn:lbol}
\end{equation}

\begin{figure}
    \includegraphics[width=8.5cm]{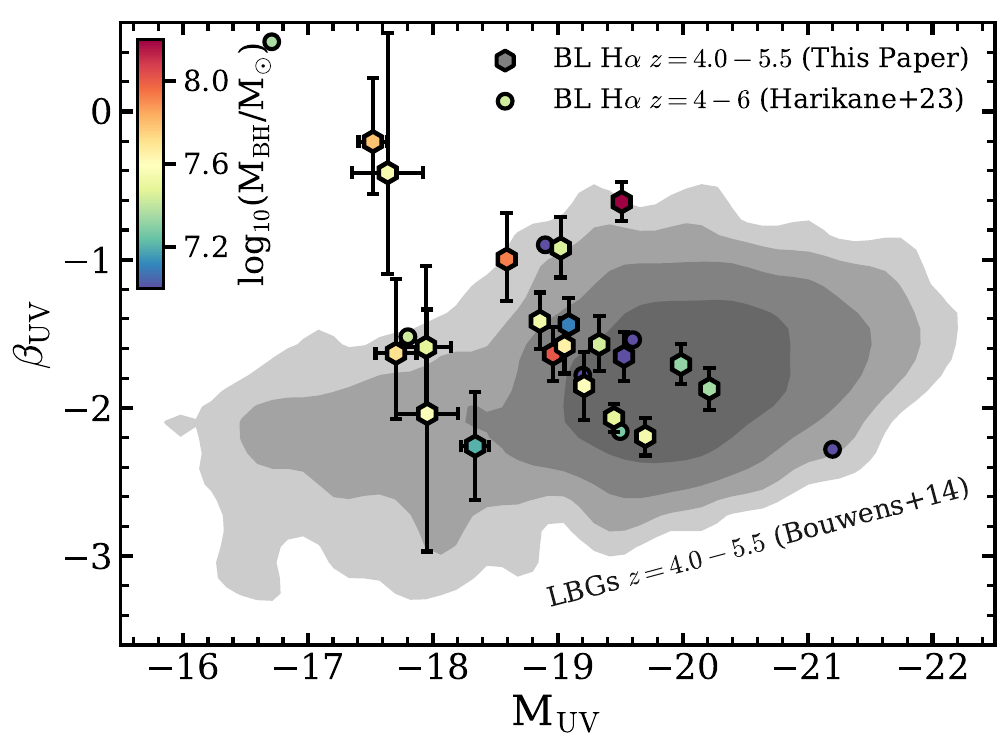}
    \caption{{\bf Most BL H$\alpha$ emitters have comparable UV colors as typical galaxies at $z\approx5$.} BL H$\alpha$ emitters identified in this paper are shown with hexagons, colored by BH mass increasing from blue to red. BL H$\alpha$ emitters identified by \citet{Harikane23} are shown in circles, with the same color coding. The shaded contours show the location of the UV-selected galaxy population at $z\approx5$ in this plane (from \citealt{Bouwens14}).} 
    \label{fig:MUVbeta}
\end{figure} 

Our sample is characterised by a typical black hole mass of M$_{\rm BH} = 3\times10^7$ M$_{\odot}$, ranging from $10^{6.9-8.6}$ M$_{\odot}$. These masses are a factor ten lower than samples at similar redshift drawn from ground-based surveys \citep[e.g.][]{Trakhtenbrot11,Matsuoka18} and up to 1000 times lower than the most massive BHs known at high redshift \citep[e.g.][]{Eilers22,Fan22,Farina22}. The bolometric luminosities -- typically $7\times10^{44}$ erg s$^{-1}$ -- are about 50 times lower than in typical samples of faint quasars. We list the BH masses and bolometric luminosites in Table $\ref{tab:properties}$. It is of interest to compare the estimate of the bolometric luminosity to the Eddington luminosity to derive the normalised accretion rate. The normalised accretion rate can be derived following \cite{Trakhtenbrot11}: $L/L_{\rm edd} = L_{\rm bol} / (1.5\times10^{38}$ M$_{\rm BH}$/M$_{\odot}$). We find $L/L_{\rm edd}$ typically 0.16 (in the range 0.07-0.4) which is somewhat lower than the more massive BHs ($L/L_{\rm edd}\approx0.6$; \citealt{Trakhtenbrot11}) suggesting that these faint AGN are not as efficiently accreting gas as more luminous quasars. However, we note that these estimates are subject to uncertainties in the attenuation corrections that may impact BH masses and bolometric luminosities. An attenuation as extreme as $A_V=4$ would yield a factor five underestimate in the BH mass and a factor 10 higher bolometric luminosity \citep[e.g.][for an AGN with such high $A_V$ and a $\beta_{\rm opt}\approx2$]{Kocevski23}, i.e. a factor two higher normalised accretion rate.

\begin{figure}
    \includegraphics[width=8.5cm]{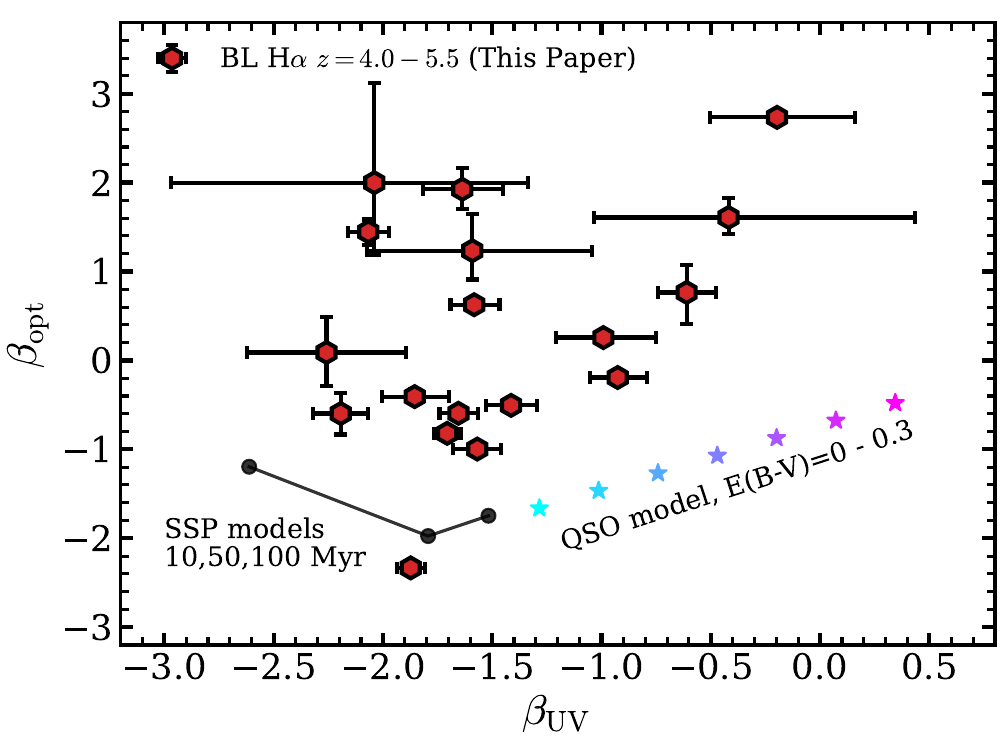}
    \caption{{\bf The rest-frame UV and optical colours of our sample of BL H$\alpha$ emitters at $z\approx5$.} Black circles show the colours for model single stellar populations with ages 10, 50 and 100 Myr from Starburst99 \citep{Leitherer1999}. Cyan to magenta stars show a quasar template \citep{Selsing16} that is increasingly reddened from E($B-V$)=0 to 0.3 assuming a \cite{Calzetti00} law. }
    \label{fig:betabeta}
\end{figure}

\begin{figure*}
    \centering
    \begin{tabular}{cc}
    \hspace{-1.5cm}
    \includegraphics[width=8.5cm]{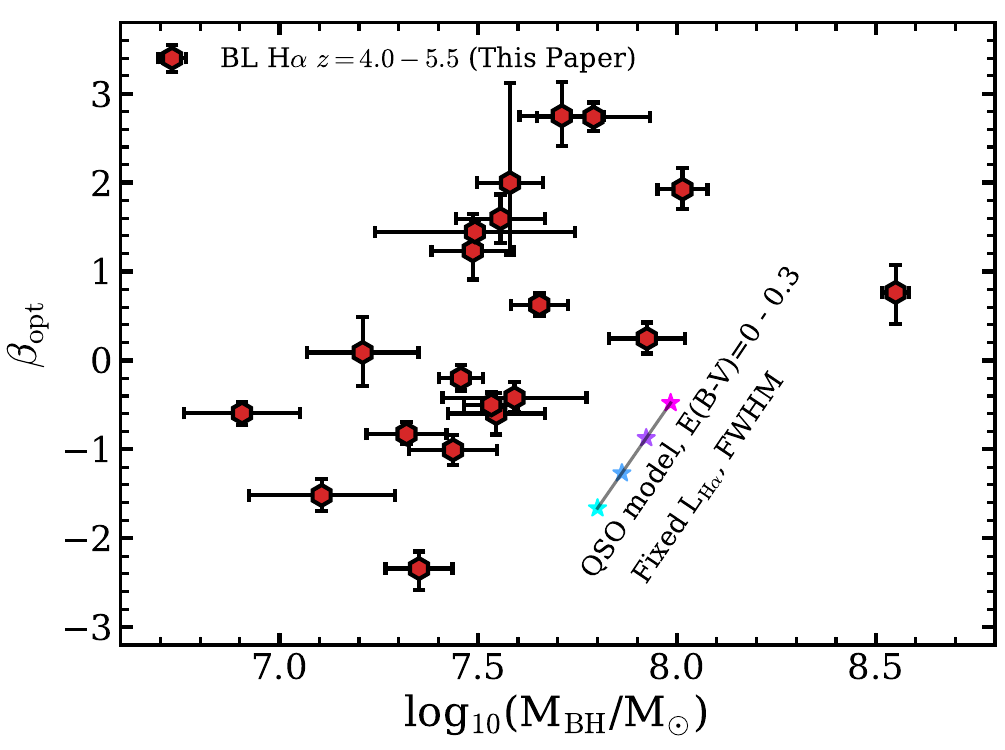}&
    \includegraphics[width=8.5cm]{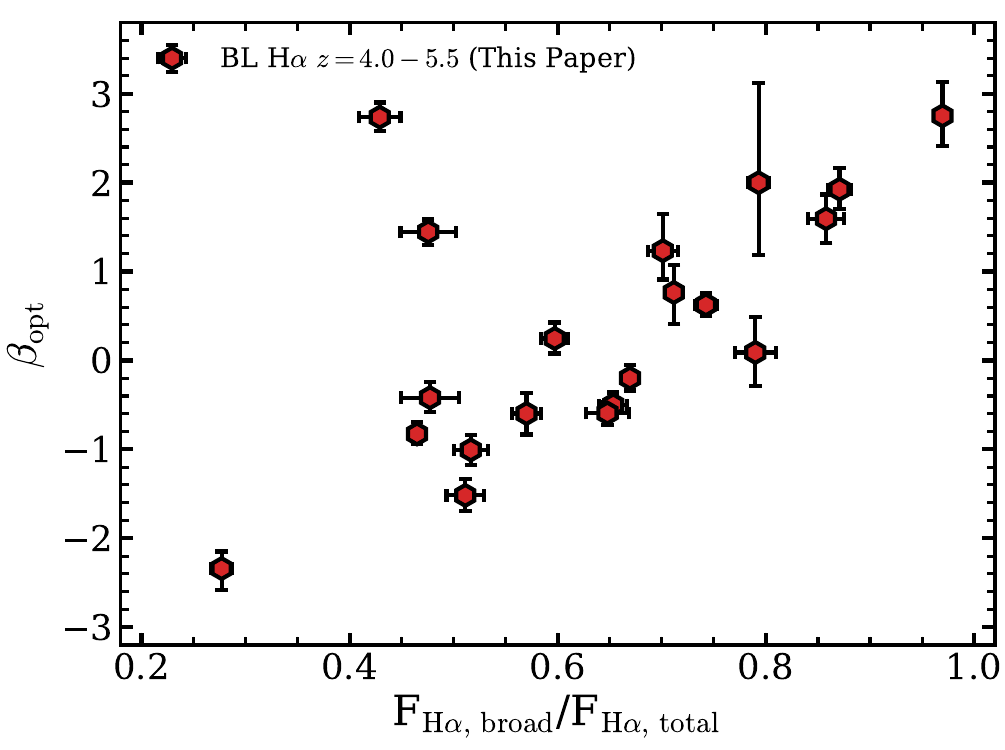}
    \end{tabular} 
    \caption{{\bf The optical redness of BL H$\alpha$ emitters correlates with BH mass and the fraction of broad to total H$\alpha$ flux, showing that the faint AGN are red.} In the {\it left} panel we show the relation between optical slope BH mass. To illustrate possible selection effects due to our broad line H$\alpha$ luminosity limit, we show a line of fixed H$\alpha$ luminosity along which a quasar template moves when varying the dust attenuation that reddens the spectrum. In the {\it right} panel we show the relation between the optical continuum slope and the fraction of the H$\alpha$ flux that is in the broad component.} 
    \label{fig:MBHbroadbeta}
\end{figure*}

\begin{figure*}
    \centering
    \includegraphics[width=0.9\linewidth]{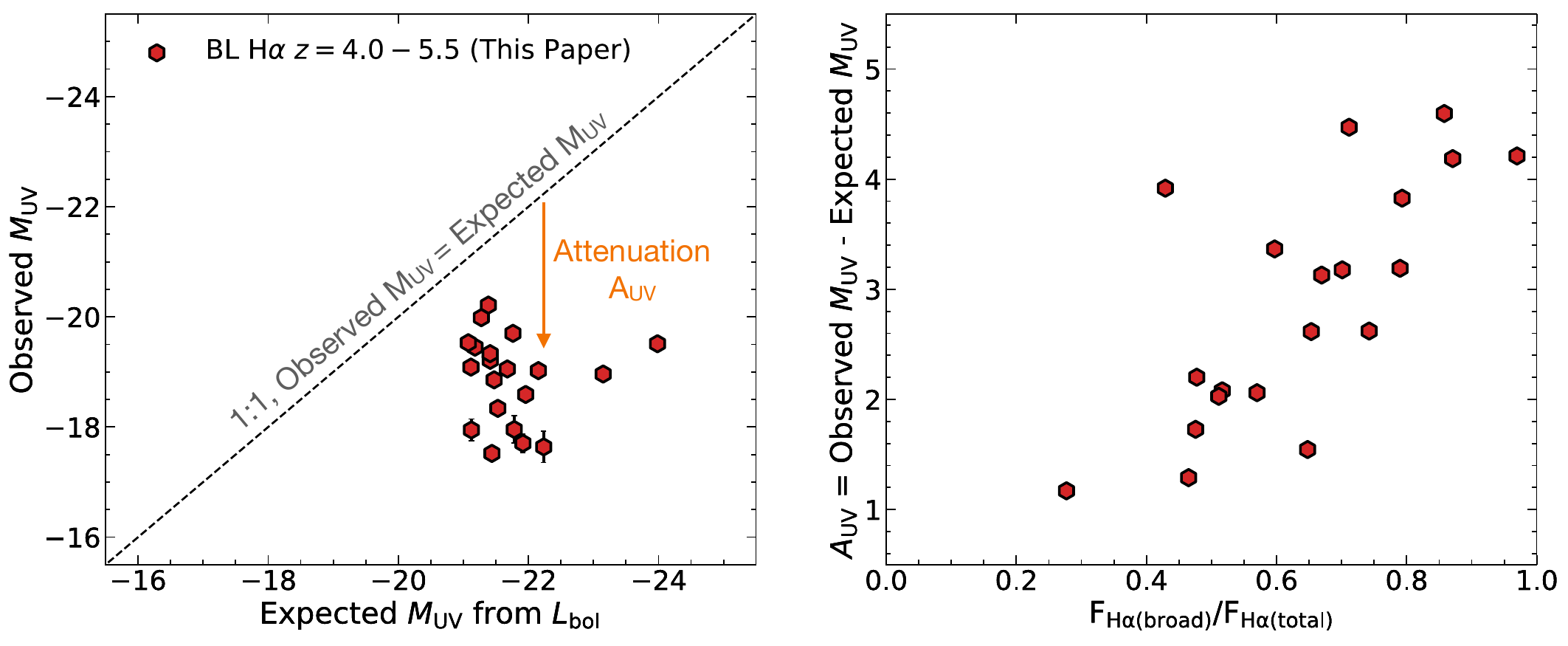}
    \caption{\textbf{The redness of BL H$\alpha$ emitters is due to extreme dust attenuation.} \textbf{Left:} Comparison of the observed $M_{\rm{UV}}$ of the BL H$\alpha$ emitters against the $M_{\rm{UV}}$ expected for their bolometric luminosity ($L_{\rm{bol}}$). The bolometric luminosity is computed from H$\alpha$ (Eq. \ref{eqn:lbol}), and the expected $M_{\rm{UV}}$ for this $L_{\rm{bol}}$ is as per the \citet{Shen20} scaling relation for typical AGN. The $M_{\rm{UV}}$ of our sample is much fainter than expected for its $L_{\rm{bol}}$. We interpret this difference as arising due to intense dust obscuration of growing black holes, with an $A_{\rm{UV}}$ ranging from 1 to 5 mags. \textbf{Right:} Sources in which the broad-line region dominates the H$\alpha$ flux (as quantified via $F_{\rm{H\alpha(broad)}}/F_{\rm{H\alpha(total)}}$) show a stronger deviation between their observed and expected $M_{\rm{UV}}$. This is likely because their highly obscured and reddened black holes outshine the star-forming regions in the galaxy, making the sources appear red overall. On the other hand, the galaxies with relatively weaker black holes still have significant representation from blue star-formation in their UV.} 
    \label{fig:MBH_AV}
\end{figure*}

\subsection{Galaxy properties} \label{sec:galprops}
In order to understand the nature of the BL H$\alpha$ emitters and compare them to the general galaxy population at $z\sim5$, it is of interest to compare the BH properties to properties of the (host) galaxies. As discussed in Section $\ref{sec:highresimaging}$, we lack the long-wavelength imaging data to perform detailed spatially resolved photometry and SED modeling required to decompose the stellar and AGN components. Here, we therefore focus on global quantities based on the total photometry. We derive the UV luminosity M$_{\rm UV}$ and the UV beta slope $\beta_{\rm UV}$, both normalised at a rest-frame wavelength of 1500 {\AA}, based on $\approx1-2$ micron data obtained from {\it JWST} photometry (EIGER data), and a combination of {\it JWST} and {\it HST}/WFC3 F105W and F125W photometry (FRESCO). This is done by simply fitting a power-law $f_\lambda \propto \lambda^{\beta}$ to filters that probe between rest-frame Lyman-$\alpha$ emission and the Balmer break. We also similarly measure the optical slope, $\beta_{\rm opt}$, between 2 and 4 micron using the emission-line subtracted photometry in the F356W and F444W filters, respectively. The measurements are listed in Table $\ref{tab:properties}$. 

The UV luminosities are on average M$_{\rm UV}=-18.9$, spanning from M$_{\rm UV}=-20.2$ to UV luminosities as faint as M$_{\rm UV}=-17.5$. These are several magnitudes fainter than the limiting magnitude from previous ground-based high-redshift AGN surveys (M$_{\rm UV}\approx-22$; e.g. \citealt{Matsuoka18}), and all fainter than the typical L$^{\star}$ of the galaxy luminosity function (M$_{\rm UV}^{\star}\approx-21$; e.g. \citealt{Bouwens21}). The UV slopes span a large range, from $\beta_{\rm UV}=-2.3$ to $\beta_{\rm UV}=-0.2$, but the median UV slope is relatively blue, $\beta_{\rm UV}=-1.6$. In Fig. $\ref{fig:MUVbeta}$ we compare the UV slope and luminosity to the distribution of the UV-selected galaxy sample at $z=4.0-5.5$ by \cite{Bouwens14}, color-coded by their BH mass. In the UV, the typical BL H$\alpha$ emitter is only somewhat redder than the typical galaxy at comparable UV luminosity. The BL H$\alpha$ emitters that have more massive BHs are tentatively redder in the UV. [NII] is not detected in any spectrum, with typical 3$\sigma$ upper limit of [NII]/H$\alpha < 0.1$. In the stacked spectrum, the upper limit is as low as [NII]/H$\alpha < 0.02 (0.3)$ for the narrow (broad) component. Such line-ratios are rare among broad line AGN at $z\sim0$ \citep[e.g.][]{Hviding22}, and suggest a low metallicity and high excitation conditions, similar to typical galaxies at $z\sim5$ \citep[e.g.][]{Sanders23}.

In Fig. $\ref{fig:betabeta}$ we compare the rest-frame UV colours to the optical colours of our sample of BL H$\alpha$ emitters. Similar to other samples of faint AGN at $z\sim5$ \citep[e.g.][]{Kocevski23,Greene23,Labbe23}, our sample shows a wide range in colours. Most surprisingly, very red rest-frame optical colours are found in systems with blue UV slopes. In Fig. $\ref{fig:betabeta}$, we also show the colours expected in simple models of single stellar populations and a (reddened) quasar template. It is clear that these simple models can not simultaneously account for the variation among the UV and optical colours, as extreme dust attenuations (A$_V\approx2-4$) are required to produce the reddest optical slopes. This suggests that the SEDs of the BL H$\alpha$ emitters are composed of hybrid models.

It is possible that the UV emission originates from star-formation of the host galaxy, but the UV emission could also originate from a (small fraction of) scattered light from the quasar \citep{Zakamska05,Glikman23,Kocevski23,Furtak23,Greene23}. For example, we could produce a $\beta_{\rm opt}\approx+3$ and $\beta_{UV}\approx-0.5$ for models with a heavily attenuated quasar ($A_V\approx4$) in combination with about 1 \% of un-attenuated quasar light (for the specific \citealt{Selsing16} template that we use). UV slopes that are bluer than $\beta_{UV}\lesssim-1.5$ can not be explained with such quasar template, and either require contribution from a host galaxy or a bluer quasar spectrum. It could be that different explanations apply to different sources within our sample. Further data, in particular sensitive spectra over the full rest-frame UV and optical range and imaging data in multiple filters are required to investigate the origin of the UV emission in more detail.

As discussed in detail in \cite{Barro23}, \cite{Noboriguchi23} and \cite{PerezGonzalez22}, these combinations of colours are not unique to the $z\sim5$ BL H$\alpha$ emitters. For example, hot dust obscured galaxies with blue excess \citep[e.g.][]{Assef15} and Type II AGN \citep[e.g.][]{Alexandroff13} have been reported at $z\sim2-4$ with similar colours, and in the case of the Type II AGN, with similar H$\alpha$ line profiles as well \citep[e.g.][]{Greene14}. On the other hand, by nature of the deep {\it JWST} data-sets used in this work, the BL H$\alpha$ emitters probe a significantly fainter luminosity range than these samples at $z\sim3$. Whether there is significant overlap in the different populations or whether the relatively faint BL H$\alpha$ at $z\sim5$ are the progenitors of Type II AGN or strongly obscured galaxies at $z\sim3$ requires a more detailed comparison and analogues broad H$\alpha$ line search at $z\sim3$ that is beyond the scope of this paper.

\subsection{The relation between reddening, BH mass and the H$\alpha$ line profile} \label{sec:BHtrends}
Within our sample, we identify trends between the rest-frame optical color, the BH mass and the relative fraction of the H$\alpha$ flux that is emitted in the broad component (Fig. $\ref{fig:MBHbroadbeta}$). While the lack of very red objects with low BH masses could be ascribed to selection effects due to our H$\alpha$ luminosity limit, the lack of bluer massive can not. These trends suggest a connection between optical redness and the relative importance of the AGN over the host galaxy emission (in case that dominates the narrow H$\alpha$ line emission), and it also adds further support that the AGN in these galaxies are dust obscured. In Figure \ref{fig:MBH_AV}, we explore an independent line of inquiry to probe how dust attenuates the AGN emission in our sample. We compare the observed $M_{\rm{UV}}$ of our sources to the $M_{\rm{UV}}$ expected for their bolometric luminosity as per typical AGN scaling relations. The bolometric luminosity is inferred from the broad H$\alpha$ line (Equation \ref{eqn:lbol}, Table \ref{tab:properties}), and converted to a UV luminosity as per \cite{Shen20}. We find that the BL H$\alpha$ emitters are UV-fainter than expected for typical AGN by a factor $\sim40$. We attribute this faintness to dust, and interpret the difference between the expected and observed $M_{\rm{UV}}$ as attenuation ($A_{\rm{UV}}$), with an average A$_{\rm UV}\approx2.9$ and a range spanning 1.2 -- 4.6. We note that this $A_{\rm{UV}}$ is likely underestimated as the observed $M_{\rm{UV}}$ also includes significant contributions from star-formation (as captured in the broad to narrow-line flux ratios). Accounting for any rest-frame UV emission due to star-formation or scattered AGN light, the chasm between observed and expected $M_{\rm{UV}}$ in Fig. \ref{fig:MBH_AV} would be even wider implying even higher $A_{\rm{UV}}$. Direct attenuation measurements, for example using broad H$\beta$ line measurements, are required.

Further, in the right panel of Fig. \ref{fig:MBH_AV}, we show that $A_{\rm{UV}}$ strongly correlates with the fraction of H$\alpha$ flux that is in the broad-line emission ($F_{\rm{H\alpha (broad)}}/F_{\rm{H\alpha (total)}}$). That is, as the obscured AGN begins to outshine star-formation (assuming the narrow H$\alpha$ emission is due to star-formation), the overall SED grows redder. Taken together with Figure \ref{fig:MBHbroadbeta}, these trends suggest that BH growth is accompanied by an increasing dust attenuation of the entire galaxy. We discuss the interpretation and implication of this result in Section $\ref{sec:discussion}$.

\begin{figure}
\includegraphics[width=8.5cm]{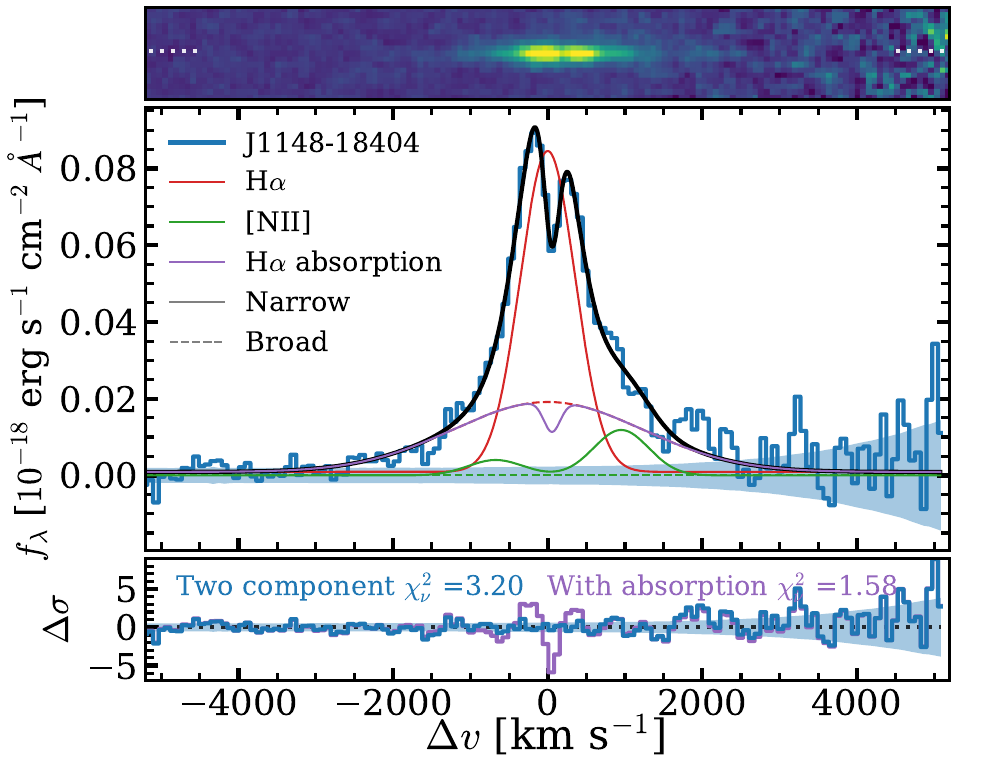} \\
\includegraphics[width=8.5cm]{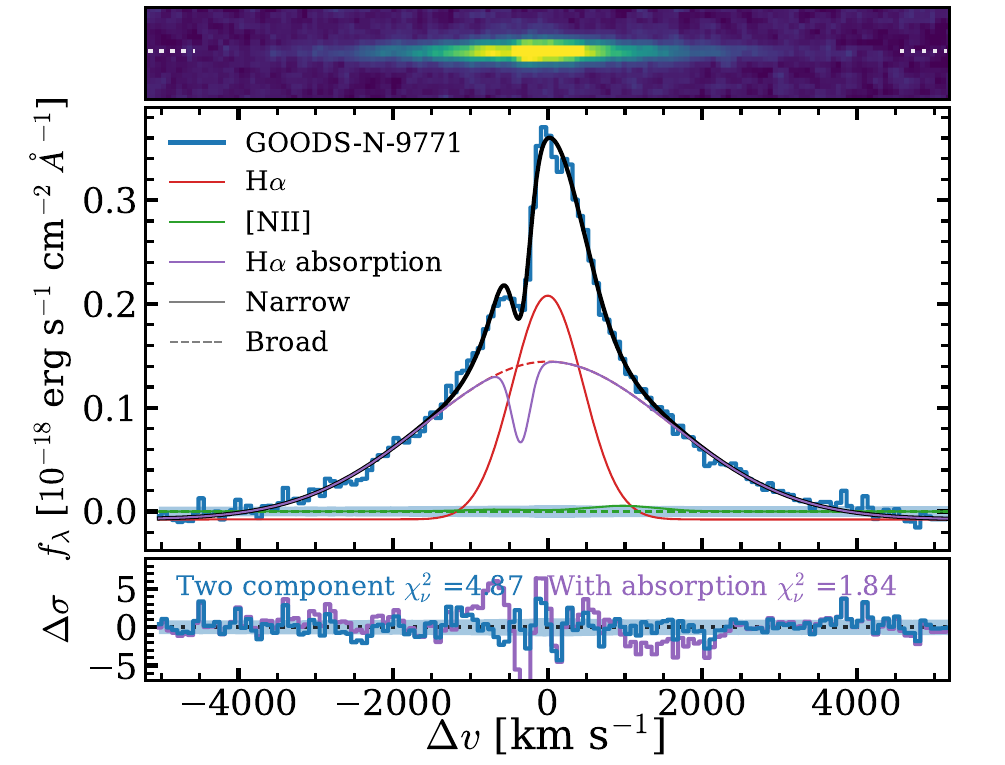}
    \caption{{\bf Detection of narrow H$\alpha$ absorption close to the systemic redshift in two BL H$\alpha$ emitters.} As Fig. $\ref{fig:1D_fit}$, we show 2D and 1D spectra, and residuals to the best-fit with an additional absorption component on the broad line. The inclusion of an absorption component in the fit decreases the reduced $\chi^2$ significantly, but in the case of J1148-18404 leads to strong degeneracies with the luminosity of the narrow component due to the small redshift of the absorption with respect to the systemic (+50 km s$^{-1}$). We note the [NII] emission is not detected at S/N$>3$. }
    \label{fig:absorption}
\end{figure}

\subsection{Detection of H$\alpha$ absorption} \label{sec:abs}
In two objects we measure a complex line profile that cannot be described by a simple combination of a narrow and a broad gaussian component. Their spectra are shown in Fig. $\ref{fig:absorption}$. Complex or double-peaked H$\alpha$ profiles have previously been detected and explained with Keplerian accretion disks \citep[e.g.][]{Eracleous2003,Luo09}, but they could also arise from closely separated galaxies and/or AGN \citep[e.g.][]{Maiolino23b}, or originate from Balmer absorption \citep[e.g.][]{Hall07}.

There is no indication of a secondary spatially resolved component in the spectra of the objects shown in Fig. $\ref{fig:absorption}$, nor in any of the deep imaging (Fig $\ref{fig:stamps}$). The spectrum of J1148-18404 is further covered by both of NIRCam's modules that disperse spectra in orthogonal directions. The observed line-profile is consistent in both spectra, demonstrating the complexity originates from processes below the spatial resolution scale. Typically, Keplerian disk profiles in quasar spectra have shown more symmetric double peaks with significantly larger separation than required to cause the narrow features we observe. These observations therefore indicate that we are seeing absorption. We explore this scenario and fit these line profiles similarly as in Section $\ref{sec:fit1D}$, but now add an absorption component on top of the broad component. We find that the fits are significantly improved, leading to $\chi^2_{\rm red}$ reductions of $\approx-2$. The absorption components have widths FWHM of 240-280 km s$^{-1}$. The absorption is blue-shifted with $-340$ km s$^{-1}$ with respect to the narrow component of GOODS-N-9771, while the absorption is redshifted by $+50$ km s$^{-1}$ with respect to the narrow component of J1148-18404. We note that due to this small redshift, the flux of the narrow component and the EW of the absorption are strongly degenerate. The rest-frame absorption EWs are $3.4\pm0.4$ {\AA} for GOODS-N-9771 and $3.6\pm0.8$ {\AA} for J1148-18404, respectively.

Balmer absorption lines have previously been detected in reddened quasars at $z\approx2$ with low-ionization broad absorption lines, although it is very rare \citep[e.g.][]{Aoki06,Shi16,Schulze18}. The absorption likely arises due to neutral hydrogen with column densities around $10^{19}$ cm$^{-2}$, where Lyman-$\alpha$ trapping significantly increases the number of hydrogen atoms with electrons in the $n=2$ shell more efficiently than collisional excitation \citep{Hall07}. We interpret the origin of the detected H$\alpha$ absorption therefore as high density gas in the broad line region that is outflowing / inflowing for GOODS-N-9771 / J1148-18404 \citep[see also][]{Shi16,Zhang18}. In contrast to our sample, other Balmer absorption lines are typically found in more massive BHs (M$_{\rm BH}\sim10^{9-10}$ M$_{\odot}$; \citealt{Schulze18}) with Balmer absorption that is often stronger and at (much) higher velocities \citep[e.g.][]{Hall13,Williams2017}. 

The detection of such rare absorption features that are relatively narrow and close to the systemic redshift opens a promising window towards studying the early stages of SMBH formation and feedback. These two detections can be further confirmed when the redshifts from the narrow and broad emission components can simultaneously be constrained with other emission lines such as H$\beta$ and [OIII]. Detections in other Balmer lines will improve the characterisation of the absorbing gas. Whether such dense gas clouds are common or rare could inform us whether they are short-lived phenomena or have low covering fractions. Among the full sample, the objects that show absorption have among the broadest H$\alpha$ lines and are relatively red (FWHMs $2800$ and $3700$ km s$^{-1}$, $\beta_{\rm opt}=+2.7$ and $+0.8$, respectively). GOODS-N-9771 is the brightest object in the sample with the highest BH mass and J1148-18404 is the object that has the brightest F356W magnitude within the EIGER sample (despite being the UV faintest within the sample; see Table $\ref{tab:properties}$). This means that we can not rule out that the detection of H$\alpha$ absorption in these particular objects is mainly due to their spectra having among the highest signal-to-noise (see for example Fig $\ref{fig:1Dgrid}$). Deep, high resolution spectroscopy is required to detect or rule out similar absorption features in other broad H$\alpha$ line samples \citep[e.g.][]{Kocevski23,Harikane23,Maiolino23b}.

\begin{figure}
    \centering
    \includegraphics[width=8.3cm]{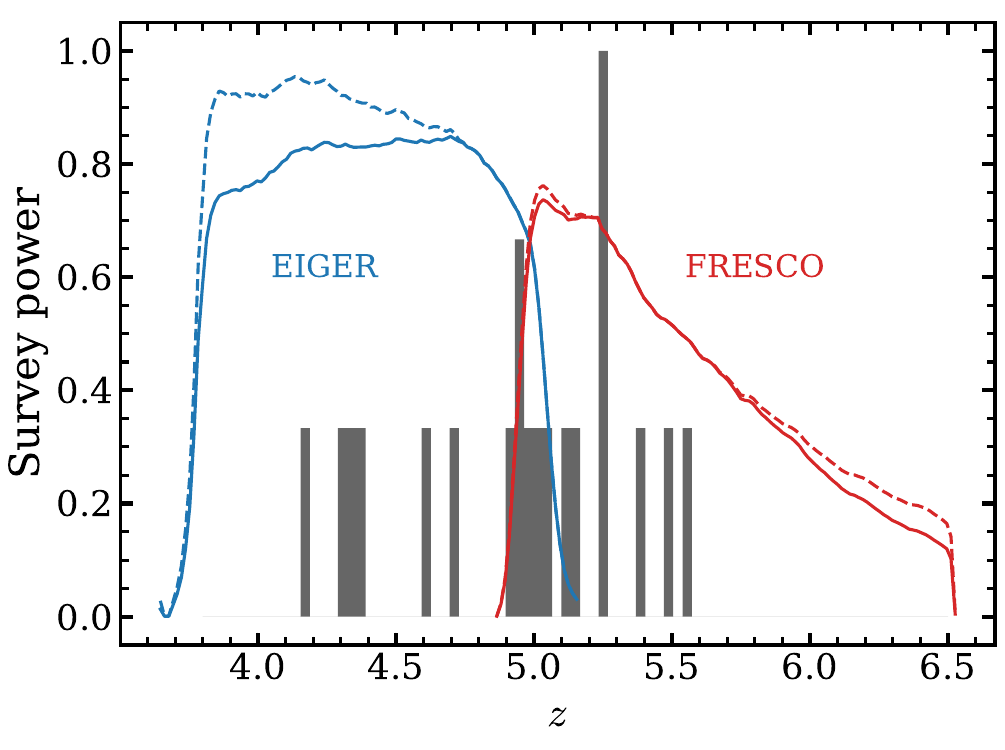}
    \caption{{\bf The redshift distribution of the BL H$\alpha$ sample}. We compare the redshift distribution to the expected distribution based on the so-called survey power in the EIGER and FRESCO surveys in blue and red, respectively. Survey power is defined as luminosity sensitivity (dashed) scaled by the redshift dependence of the covered volume (solid). }
    \label{fig:zdist} 
\end{figure}

\section{The number density of broad H$\alpha$ AGN} \label{sec:numberdensity}
One of the key motivations for our systematic search for broad line H$\alpha$ lines in NIRCam/WFSS data is the unbiased availability of spectra for objects in the field of view. This allows us to estimate the survey volume by modeling the wavelength dependent field of view using the grism trace models and the mosaic designs. 

In Fig. $\ref{fig:zdist}$ we show the redshift distribution of the BL H$\alpha$ emitters in comparison to the `survey power' of EIGER and FRESCO, which is the combination of the redshift dependence of the sensitivity and the volume. Despite covering $z=3.8-6.5$, the redshift distribution is confined to $z\approx4.2-5.5$. While shot noise with a $N=20$ sample is relatively high, this distribution is expected given that the NIRCam grism is most sensitive around 3.9 $\mu$m as this is the wavelength where the zodiacal background is the lowest. Due to the NIRCam grism design, 3.9 $\mu$m is covered by the full field of view, such that $z\approx5$ is the redshift where we are most sensitive to detect broad emission lines. As the area and sensitivity are significantly lower at the outer redshifts of the probed volume, we restrict our number density analysis to $z=4.0-6.0$.

In order to measure the number density of our sample as a function of UV magnitude $M$, we follow the standard $V_{\rm max}$ method \citep{Schmidt68}:

\begin{equation}
\Phi(M) = \frac{1}{\Delta M} \sum_i \frac{1}{V_{{\rm max}, i}}. 
\end{equation}

\begin{figure}
    \centering
    \includegraphics[width=8.3cm]{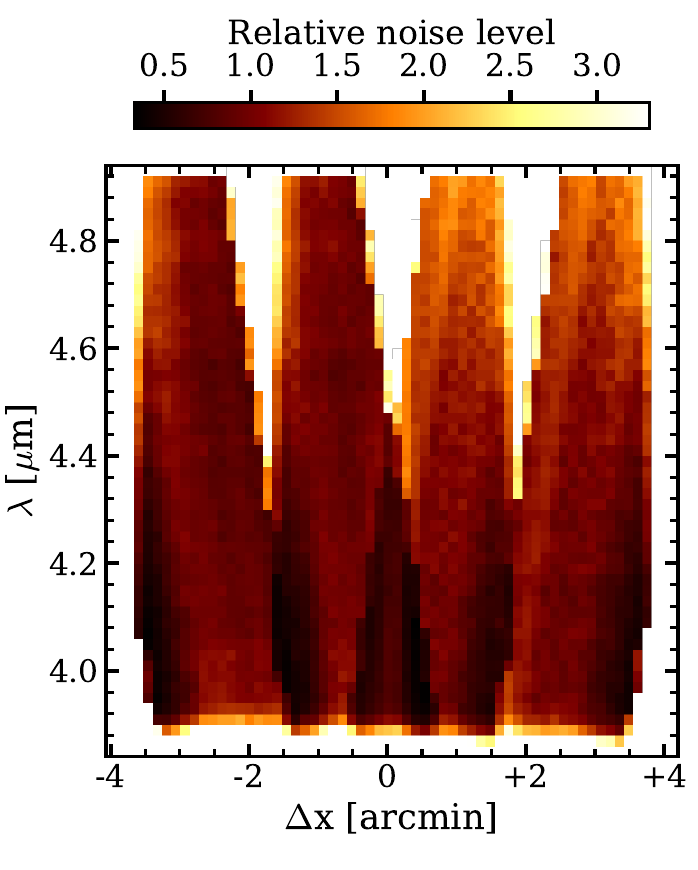} \vspace{-0.4cm}
    \caption{{\bf The wavelength and spatial dependence of the spectroscopic sensitivity in the GOODS-S field.} Darker colors correspond to a better sensitivity. We only show the dependence on the horizontal position in the mosaic (see \citealt{Oesch23}), as it almost does not vary with the vertical direction. The wavelength dependence of the areal coverage is taken into account in our number densities.}
    \label{fig:sens_wav}
\end{figure}

Here, $V_{{\rm max}, i}$ is the maximum volume over which a source could have been detected in the data. We measure the volume in each of the four EIGER fields and the two FRESCO fields by constructing a sensitivity cube (with dimensions x, y, $\lambda$) following the methodology outlined in \cite{Matthee23} (see also Mackenzie et al. in prep). We first generate a grid of spatial positions uniformly covering each of the mosaics with a separation of 12$''$. Then, for each spatial position, we extract a continuum-filtered spectrum from the grism data (see Section $\ref{sec:optextract}$) and we measure the noise level in the center of the 2D extraction at a range of wavelengths with 100 {\AA} intervals. These measurements are stored in a sensitivity cube. Fig $\ref{fig:sens_wav}$ shows an example of the spatial and wavelength dependent noise level in the GOODS-S field. Given that, with the FRESCO mosaic design, the noise level and wavelength coverage varies more in the horizontal than in the vertical direction, we here show a slice in the central y position of the mosaic. Based on the sensitivity cubes, we measure the maximum volume in which a BL H$\alpha$ emitter could have been detected given its redshift. The total volume that is covered is $2.7\times10^5$ cMpc$^{3}$ at $z=4.0-5.05$ in EIGER and $3.0\times10^5$ cMpc$^3$ at $z=5.05-6.00$ in FRESCO, i.e. $5.7\times10^5$ cMpc$^{3}$ over $z=4.0-6.0$ in total.

In our number density measurement we assume that we are fully complete. This is motivated by our conservative selection criteria, in particular the relatively high limiting luminosity (Section $\ref{sec:colors}$). Fully modeling the completeness of the detectability of broad components in galaxy spectra is not trivial as it depends on the luminosity and width of the broad component, which we find to vary from $\approx 1100 - 3740$ km s$^{-1}$. We have tested that our line-profile fitting (Section $\ref{sec:fit1D}$) recovers the broad component of the {\it median stacked} H$\alpha$ profile (Section $\ref{sec:resolvedspec}$) when we inject it in spectra at the 10 \% least sensitive regions at any redshift between $z=4.0-6.0$. 

\subsection{The faint AGN UV luminosity function at $z\approx5$} \label{sec:UVLF}
\begin{figure*}
    \centering
    \includegraphics[width=15.3cm]{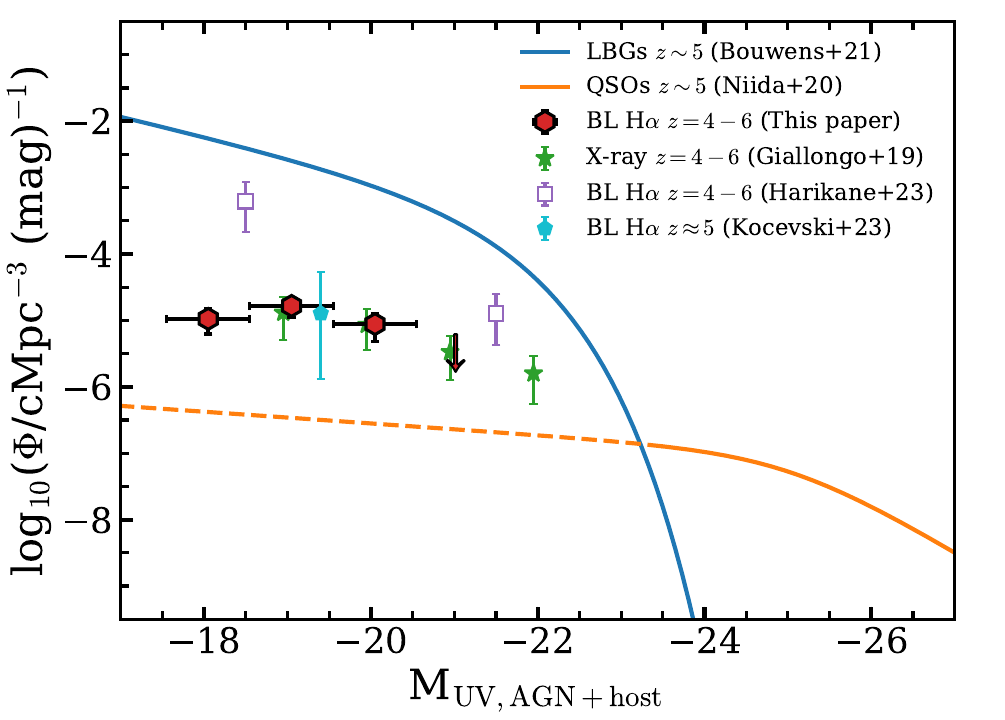}
    \caption{{\bf The UV luminosity function of the BL H$\alpha$ emitters (red hexagons; shifted horizontally by -0.05 dex for visualisation purposes) compared to the Lyman break galaxy population (blue line; \citealt{Bouwens21}), the extrapolated quasar luminosity function (orange line; dashed for extrapolated range; \citealt{Niida20}) and various faint AGN samples at $z\approx5$.} The red arrow shows the one-sided 1$\sigma$ upper limit following \citet{Gehrels86}. Green stars show the number density of galaxies with X-Ray detections at $z\approx5$ (\citealt{Giallongo19}; shifted horizontally by +0.05 dex for visualisation purposes). The blue pentagon and the open purple squares are both based on recent broad line H$\alpha$ measurements of selected galaxies observed with {\it JWST}/NIRSpec (\citealt{Kocevski23,Harikane23}). Note that the measured UV luminosities do not decompose the AGN from the host galaxy emission, and that the UV-faintest BL H$\alpha$ emitters typically are significantly obscured.}
    \label{fig:LF}
\end{figure*}

After deriving the probed volumes, we measure the number densities as a function of UV luminosity in bins of 1 magnitude and list the results in Table $\ref{tab:NumDens}$. The uncertainties are estimated from the poissonian errors on the counts. The number densities are fairly constant around $10^{-5}$ cMpc$^{-3}$. In Fig. $\ref{fig:LF}$ we show the number densities in comparison to the galaxy population at $z=5$ \citep{Bouwens21}, an extrapolation of the bright quasar UV luminosity function \citep{Niida20} and other published number densities of faint AGN samples at $z\approx5$, either selected through X-Ray emission \citep{Giallongo19} or broad line H$\alpha$ emitters (BL H$\alpha$) with {\it JWST} spectroscopy \citep{Kocevski23,Harikane23}.

\begin{table}
    \centering
    \caption{\bf The number densities of BL H$\alpha$ emitters at $\mathbf{z=4-6}$ as a function of UV luminosity. }
    \begin{tabular}{ccc}
    M$_{\rm UV, AGN+host}$ & $N$ & $\Phi$ / cMpc$^{-3}$ mag$^{-1}$ \\ \hline
$ -20.0 \pm0.5$ & 5 &$-5.06^{+0.16}_{-0.26}$\\
$ -19.0 \pm0.5$ & 9 & $-4.78^{+0.13}_{-0.18}$\\
$ -18.0 \pm0.5$ & 6 & $-4.98^{+0.15}_{-0.23}$\\ \hline
    \end{tabular}
    \label{tab:NumDens}
\end{table}

Compared to the UV-selected galaxy population at $z=5$, our sample of BL H$\alpha$ emitters are rare and they imply that only a very low fraction of the UV emission at these magnitudes is due to AGN emission: $\approx0.5$ \% in the range M$_{\rm UV} = -19$ to $-21$ and around 0.01-0.1 $\%$ at even fainter luminosities. These fractions are in line with estimates from \cite{Adams22} based on the joint fitting of star-forming and AGN UV luminosity functions at $z=5$. 

Our number density is about a factor ten higher than the extrapolated quasar UV LF at $z\sim5$ from \cite{Niida20}. The measured number density of BL H$\alpha$ emitters is similar to their estimated of the number density at $z\approx5$ by \cite{Kocevski23}, and with X-Ray sources with photometric redshifts $z\approx5$ from \cite{Giallongo19}. Our number densities are significantly lower than the number density estimates from \cite{Harikane23} that are based on the broad line H$\alpha$ emitter fraction within their studied sample of galaxies at $z=4-6$ that has been followed up with NIRSpec. We compare to the latter two samples in more detail in Section $\ref{sec:discuss_dens}$.

\subsection{The $z\approx5$ H$\alpha$ luminosity function} \label{sec:HaLF} 
\begin{figure}
    \centering
    \includegraphics[width=8.6cm]{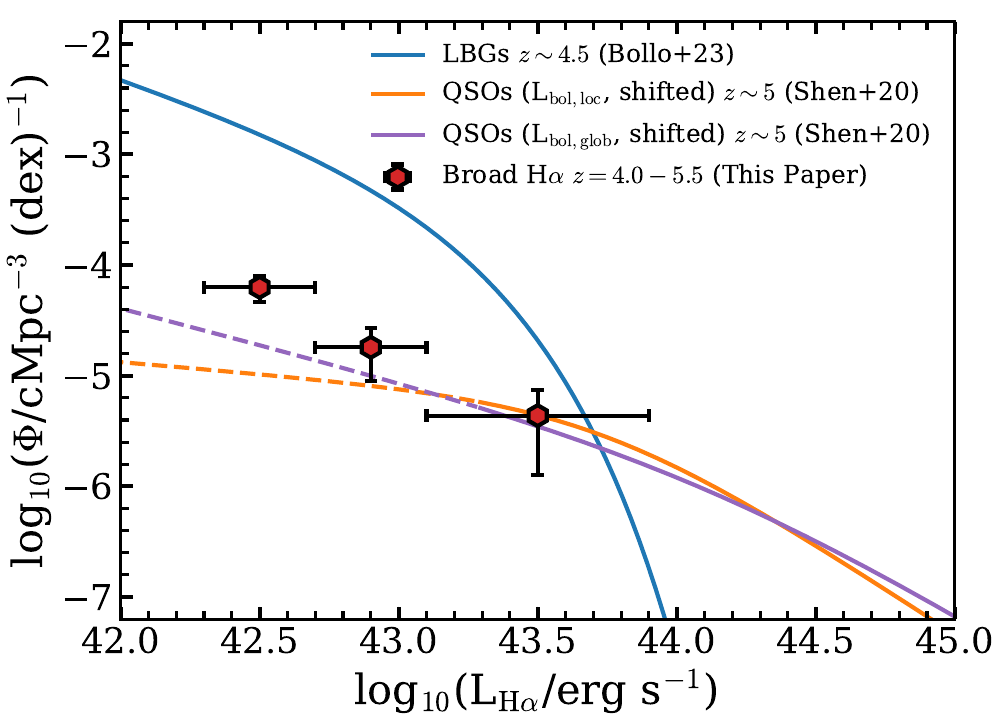}
    \caption{{\bf The broad H$\alpha$ luminosity function of our sample (red hexagons) compared to the total H$\alpha$ luminosity function of LBGs in blue \citep{Bollo23} and the broad line H$\alpha$ luminosity functions of quasars \citep{Shen20} in orange and purple.} The quasar LFs are derived by shifting the bolometric luminosity to the broad H$\alpha$ line LF following Equation $\ref{eqn:lbol}$. The orange LF uses so-called `local polished' LF at $z=5$ from \citealt{Shen20}, whereas the purple LF uses their `global' model for the evolution of the LF, that we evaluate at $z=5$. We illustrate the region in which the quasar luminosity functions are extrapolations with a dashed linestyle.}
    \label{fig:HaLF}
\end{figure}

Given that our sample is selected on broad H$\alpha$ line luminosity rather than UV luminosity and that a fraction of the UV light may originate from star formation, we also derive the broad H$\alpha$ line luminosity function of our sample with the same methods as described above, except that we bin our sample in broad H$\alpha$ line luminosity ranges. The measured number densities are listed in Table $\ref{tab:NumDensHa}$. Figure $\ref{fig:HaLF}$ shows the luminosity function compared to the recently measured H$\alpha$ luminosity function of Lyman break galaxies at $z\sim4.5$ \citep{Bollo23} and to the estimated quasar broad H$\alpha$ line luminosity at $z\sim5$ that we derive by shifting the bolometric luminosity function to H$\alpha$ assuming Equation $\ref{eqn:lbol}$. For the latter, we show the estimated LFs based both on the so-called `local polished' and the global bolometric LF at $z=5$ from \cite{Shen20}.

It is clear that at fixed H$\alpha$ luminosity, the typical broad line H$\alpha$ emitters that constitute our sample are significantly rarer than star-forming galaxies. At the luminous end probed by our sample, the number densities agree relatively well with the quasar LF and the LBG LF. This suggests that a significant fraction of the H$\alpha$ flux in such bright galaxies is due to AGN emission (consistent with $z\approx0-2$ results; e.g. \citealt{Sobral16}). The H$\alpha$ LF of our sample of faint AGN is steeper than the UV LF of our sample. As Fig. $\ref{fig:HaLF}$ illustrates, previous quasar luminosity functions (from \citealt{Shen20}) have a wide range in faint-end slopes. Our measured LF is significantly higher and steeper than the `local' model displayed in orange, but the slope is comparable to the `global A' model displayed in purple, where we only measure a higher number density in our faintest bin. These comparisons show that the number densities of faint AGN are among the highest extremes of (extrapolations of) previous estimates at $z\sim5$, highlighting the complimentarity of the {\it JWST} BL H$\alpha$ sample to previous studies. Extending the broad H$\alpha$ LF to fainter luminosities with more sensitive spectroscopic data would be particularly helpful constraining the faint end of the quasar LF.

\begin{table}
    \centering
    \caption{\bf The number densities of BL H$\alpha$ emitters at $\mathbf{z=4-6}$ as a function of broad H$\alpha$ line luminosity. }
    \begin{tabular}{ccc}
    log$_{10}$(L$_{\rm H\alpha, broad}$/erg s$^{-1}$) & $N$ & $\Phi$ / cMpc$^{-3}$ mag$^{-1}$ \\ \hline
$ 42.5 \pm0.2$ & 14 &  $-4.20^{+0.10}_{-0.14}$ \\
$ 42.9 \pm0.2$ & 4 &$-4.74^{+0.18}_{-0.30}$ \\
$ 43.5 \pm0.4$ & 2 & $-5.36^{+0.23}_{-0.53}$ \\ \hline
    \end{tabular}
    \label{tab:NumDensHa}
\end{table}

\subsection{The $z\approx5$ SMBH mass function} \label{sec:BHMF}

Our data allow us to derive a first measurement of the supermassive BH mass function at $z\approx5$ that probes number densities around $\sim10^{-5}$ cMpc$^{-3}$ and higher. This is unlike previous quasar surveys that only probed much rarer sources that are typically only found in survey volumes that are at least 100 times larger \citep[e.g.][]{Vestergaard09,Willott10,Fan22}. Number densities of $\sim10^{-5}$ cMpc$^{-3}$ imply that these systems should be captured by state-of-the-art large hydrodynamical simulations such as EAGLE, Horizon-AGN, IllustrisTNG and Simba \citep{Schaye15,Volonteri16,Pillepich18,Dave19} that simulate volumes of $10^6$ cMpc$^3$ (way too small for bright high-redshift quasars). Each of these simulations includes the seeding, growth and feedback of SMBHs with a different implementation. Measurements of the lower mass end of the SMBH mass function at high-redshift could potentially differentiate among models \citep[e.g.][]{Trinca22}. Most simulations seed SMBHs when halos obtain masses of a few times $\approx10^{10}$ M$_{\odot}$ with SMBH masses $\sim10^{4-6}$ M$_{\odot}$ (see \citealt{Habouzit21} for a detailed comparison).

\begin{table}
    \centering
    \caption{\bf The number densities of BL H$\alpha$ emitters at $\mathbf{z=4-6}$ as a function of the SMBH mass.}
    \begin{tabular}{ccc}
    log$_{10}$(M$_{\rm BH}$/M$_{\odot}$) & $N$ & $\Phi$ / cMpc$^{-3}$ dex$^{-1}$ \\ \hline
$ 7.1 \pm0.2$ & 3 & $-4.86^{+0.20}_{-0.37}$ \\
$ 7.5 \pm0.2$ & 12 & $-4.27^{+0.11}_{-0.15}$\\
$ 8.1 \pm0.4$ & 4 & $-5.05^{+0.18}_{-0.30}$ \\ \hline
    \end{tabular}
    \label{tab:NumDensHa}
\end{table}

\begin{figure}
    \centering
    \includegraphics[width=8.7cm]{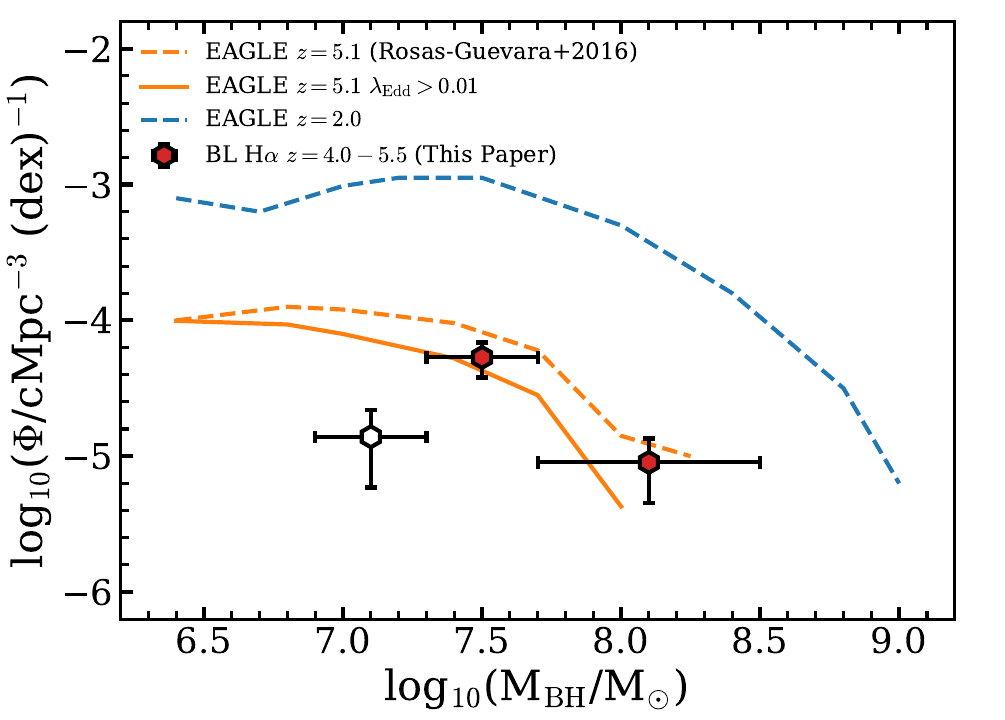}
    \caption{{\bf The supermassive black hole mass function in our sample of broad line H$\alpha$ emitters (open and red hexagons) compared to the EAGLE simulation (\citealt{Rosas2016}) at $z\approx2, 5$ (blue and orange lines, respectively).} For EAGLE we show both the total number density of all supermassive black holes (dashed) as well as those accreting at $\lambda_{\rm Edd}>0.01$ (solid). The measured number densities for SMBH masses $\approx10^{7.5-8}$ M$_{\odot}$ agree remarkably well with the simulations, while the downturn at lower masses suggests incompleteness (therefore shown as open symbol), likely because we only detect objects with higher accretion efficiencies.}
    \label{fig:MF}
\end{figure}

We derive the SMBH mass function at $z\approx5$ based on our data similar to the method described in Section $\ref{sec:UVLF}$, but now binning in SMBH mass instead of binning the sample in UV magnitude. The number densities are shown in Fig. $\ref{fig:MF}$, where we compare them to the SMBH mass function of all galaxies at $z=5$ in the 100 Mpc {\it reference} EAGLE simulation as published by \cite{Rosas2016}. Remarkably, our number densities for BH masses M$_{\rm BH}\gtrsim10^{7.5}$ M$_{\odot}$ agree well with those in EAGLE. \cite{Habouzit21} show that at $z=4$, the simulations listed above all agree with each other at the masses probed by our survey, whereas there are significant discrepancies around masses of $\sim10^{6.5}$ M$_{\odot}$ where IllustrisTNG and in particular Horizon-AGN have much higher number densities. 

We measure a significant decline in the number density for BH masses lower than $\sim10^7$ M$_{\odot}$. This is likely due to incompleteness effects as the required broad H$\alpha$ line luminosity ($>2\times10^{42}$ erg s$^{-1}$) requires them to have higher accretion efficiencies at fixed SMBH mass. We expect that at M$_{\rm BH}\approx10^{7.2}$ M$_{\odot}$, we can only identify AGN with $\lambda_{\rm Edd}>0.2$,  while this limit decreases to $>0.1$ and $>0.05$ at M$_{\rm BH}\approx10^{7.6, 8.0}$ M$_{\odot}$, respectively. Indeed, in our lower mass bin we find a median $\lambda_{\rm Edd}=0.3$, a factor 2-3 higher than in the other mass bins. We find that neither the luminosity or the width of the broad H$\alpha$ component depends on the UV luminosity, explaining why we do not see such sign of a strongly luminosity-dependent incompleteness in Section $\ref{sec:UVLF}$.

In EAGLE, SMBHs with masses of $1.5\times10^5$ M$_{\odot}$ are seeded in halos with mass $1.5\times10^{10}$ M$_{\odot}$. This suggests that our AGN sample contains SMBHs that have already grown their mass by at least two orders of magnitude through accretion. Therefore, the number densities of SMBHs with the masses that we currently probe are most sensitive to accretion and feedback physics instead of SMBH seeding. The EAGLE model has been tuned to reproduce the normalisation of the M$_{\rm BH}$-M$_{\rm star}$ relation and the galaxy stellar mass function at $z=0$ \citep{Schaye15}. By result it matches inferences of the $z=0$ SMBH mass function as well \citep{Rosas2016}. We note that since EAGLE does not model radiative transfer, the SMBH mass function constitutes the total SMBH function of obscured and unobscured, and active and inactive AGN. The fact that our number densities agree with EAGLE at M$_{\rm BH}\sim10^8$ M$_{\odot}$ intruigingly implies that our broad H$\alpha$ AGN selection contains the whole AGN population with such BH masses at $z\sim5$. In the case that there are substantial AGN populations at $z\sim5$ that are missed by our broad H$\alpha$ survey, or in the case that our SMBH masses are significantly underestimated, there would be a tension with the $z\sim5$ SMBH mass function in EAGLE. This could be the case if there are significant populations of heavily obscured Type II AGN with comparable BH masses, or if our BH mass estimates are strongly impacted by dust attenuation. How would models need to change? As shown in \cite{Rosas2016}, the M$_{\rm BH}$-M$_{\rm star}$ relation evolves little over cosmic time in their model. The EAGLE galaxy stellar mass function at $z=5$ roughly matches observational constraints \citep{Furlong15}. Therefore, an under-estimate in the SMBH mass function at $z\sim5$ would need to be balanced with a higher normalisation of the M$_{\rm BH}$-M$_{\rm star}$ relation. This would require substantial changes in the SMBH seed model (invoking much heavier seeds than $10^5$ M$_{\odot}$), or SMBHs should be able to grow even more quickly and/or in lower mass halos, possibly enabled by a less efficient self-regulation from their AGN feedback at high-redshift \citep[e.g.][]{Bower17,McAlpine18}. Importantly, one should note that such model changes would likely impact the (well-matched) simulated galaxy and BH properties in the present-day Universe. Finally, we note that \cite{Lyu23} and \cite{Scholtz23} recently reported such obscured and narrow-line AGN at similar redshifts as our sample. However, their bolometric luminosities are typically one or two orders of magnitude lower, suggestive of lower SMBH masses, and therefore not necessarily leading to large tensions with models as EAGLE.

\section{Discussion} \label{sec:discussion}
\subsection{Number density comparison} \label{sec:discuss_dens}

\subsubsection{Broad H$\alpha$ lines: do we only see the tip of an AGN iceberg?}
Fig. $\ref{fig:LF}$ shows that the number density of broad line H$\alpha$ emitters found in this work is $\approx10^{-5}$ cMpc$^{-3}$. This is about an order of magnitude higher than the extrapolated quasar UV LF \citep[e.g.][]{Niida20} and on the upper range in the \cite{FinkelsteinBagley22} model, but not as high as recent results from \cite{Harikane23} who searched for broad H$\alpha$ emission in {\ JWST}/NIRSpec data. While the sample from \cite{Harikane23} could have been biased due to the priorities given in the shutter allocation, we note that their sample contains several broad lines that are 5-10 times fainter than the broad lines in our sample due to the use of more sensitive spectroscopic data. In fact, applying our luminosity limits (Equation $\ref{eq:selection}$) to their data-set reduces their number density by a factor five, leading to somewhat more comparable number densities as those that we measure.

Additionally, relaxing our broad H$\alpha$ luminosity selection criterion (Section $\ref{sec:colors}$), we find that we could roughly double the number of sources with strong indications for broad H$\alpha$ emission. This suggests that the broad H$\alpha$ LF (see Fig. $\ref{fig:HaLF}$) does not significantly flatter below the luminosity range probed here. Besides being fainter, their broad components are also typically narrower ($\sim1000$ km s$^{-1}$) compared to the presented sample and their implied SMBH masses would be in the range $10^{6.7-7.2}$ M$_{\odot}$, M$_{\rm BH}=10^{6.9}$ M$_{\odot}$ on average. However, for these fainter sources our arguments regarding the broad component in favour of originating from a broad line region in an AGN activity are weaker: the fits to the line-profile are rather uncertain, we can not rule out velocity shifts between the narrow and broad component and the line-widths and relative broad to total H$\alpha$ fluxes are more similar to those powered by star-formation driven outflows (see Section $\ref{sec:Haprofile}$). Therefore, it is plausible that there exists a larger population of objects with lower BH masses, lower H$\alpha$ luminosities, and a relatively lower fraction of the UV flux that is due to AGN activity. However, for those it is more difficult to differentiate a broad line region from emission due to outflows or argue for AGN based on point-source morphologies. Indeed, the morphology of the majority of the \cite{Harikane23} sample that probes this lower mass regime is somewhat more complex than a point-source, suggesting an even more dominant light contribution from star formation in those systems.

\subsubsection{Comparison to X-Ray studies}
As shown in Fig. $\ref{fig:LF}$, our measured number densities of spectroscopically confirmed faint AGN at M$_{\rm UV}\sim-20$ at $z\approx5$ are in remarkable agreement with the results from \cite{Giallongo19} that are based on faint X-ray detections among galaxies with photometric redshifts in three CANDELS fields (including both GOODS fields). Seven objects from their sample are in the FRESCO coverage and have photometric redshift estimates that should lead to a detected H$\alpha$ line. However none of these seven shows an H$\alpha$ line in the FRESCO data. This suggests that the photometric redshifts (notoriously difficult for galaxies whose light originates from mixtures of AGN and star formation; e.g. \citealt{Parsa18}) of these seven sources are not very accurate. Alternatively, it could mean that the broad line H$\alpha$ objects probe a different AGN class than the X-Ray detections. None of our BL H$\alpha$ emitters are detected in the X-Rays, neither are recently published high-redshift AGN with broad H$\alpha$ lines similar to our sample \citep{Harikane23,Kocevski23,Ubler23}. We estimate the expected X-ray luminosity for our BL H$\alpha$ sample empirically based on data from the Sloan Digital Sky Survey. In particular, using data from \cite{SternLaor12}, we find that the luminosity of the broad H$\alpha$ line correlates well with the X-ray luminosity, following $\rm log_{10} (L_X$/$10^{42}$ erg s$^{-1}) = 0.52 + 0.74\, \rm log_{10}(L_{\rm H\alpha, broad}$/$10^{42}$ erg s$^{-1}$). Based on this calibration, we find that our sample should have a typical X-Ray luminosity of $10^{42.5}$ erg s$^{-1}$, which is a factor of about five below the X-Ray luminosity limit of our data.

\subsection{Implications for early BH growth scenarios}
\label{sec:cartoon}
\begin{figure*}
    \centering
    \includegraphics[width=14cm]{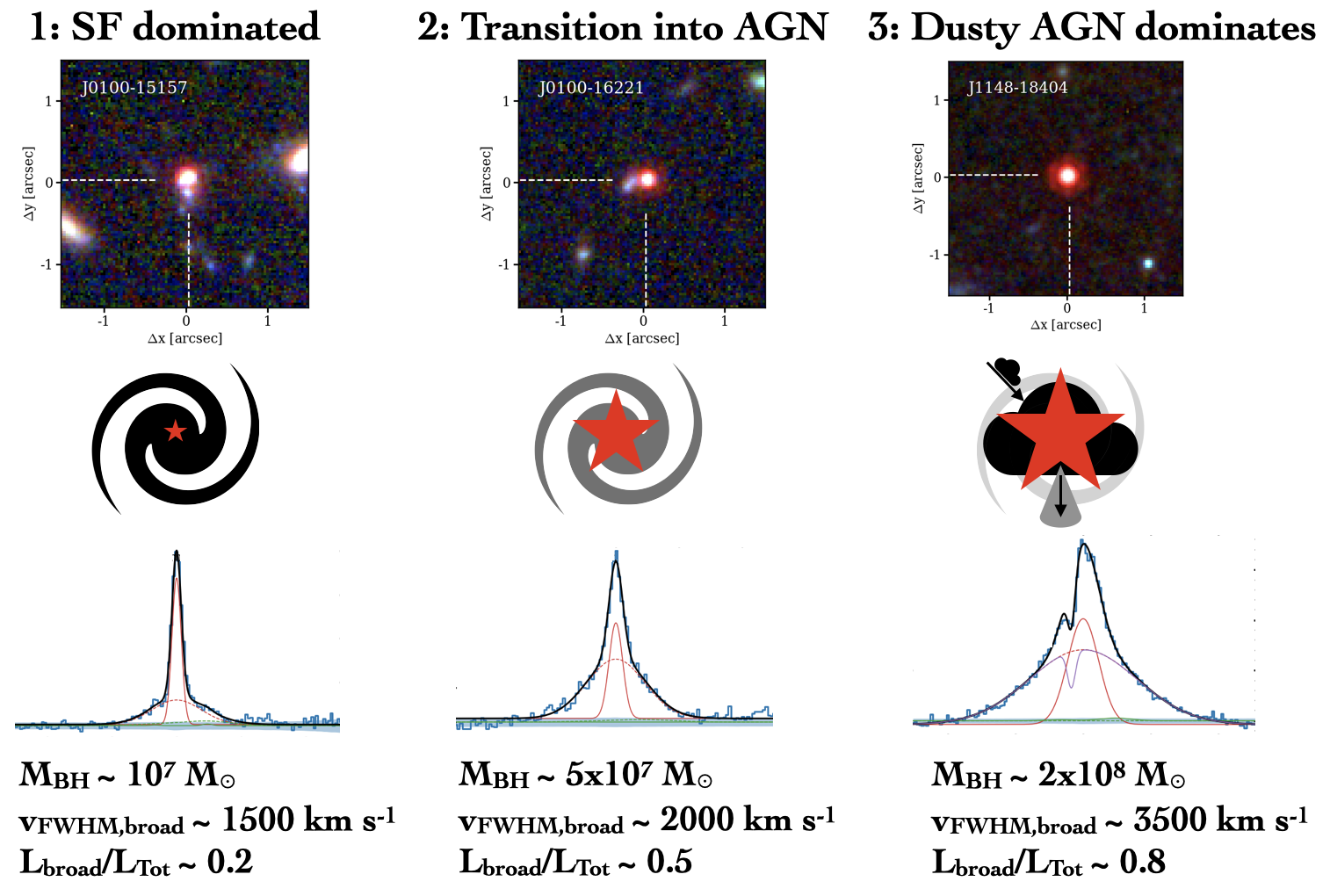}
\caption{{\bf Sketch of the three phases that describe our sample of faint AGN, illustrated by their apparent morphology and H$\alpha$ line profile.} In the top row, we show false-color {\it JWST}/NIRCam images of representative objects for each phase. We sketch the relative importance of star formation and AGN activity to the rest-frame UV-optical light in the middle row, where the red star shows the AGN, the black spiral the star-forming component and clouds illustrate dust content. In the bottom row we show the observed H$\alpha$ line profiles for the representative sources. We also list typical BH masses, FWHM of the broad components and the relative broad-to-total H$\alpha$ luminosity.} 
    \label{fig:sketch}
\end{figure*}

In this section we interpret our measurements in the context of SMBH growth scenarios in the early Universe. Our discussion is centered around  Fig. $\ref{fig:sketch}$, a sketch of the various kinds of objects that we identify in our sample and interpret in an evolutionary trend. The main motivation for this sketch are the comparison of the BL H$\alpha$ emitters to UV slopes and magnitudes of the general galaxy population at $z\sim5$ (Fig. $\ref{fig:MUVbeta}$) and the tentative trends identified between the optical color, the relative broad to total H$\alpha$ luminosity and BH mass (Figs. $\ref{fig:MBHbroadbeta}$ and $\ref{fig:MBH_AV}$).

We find that the galaxies with UV luminosities M$_{\rm UV}\approx-19$ have relatively similar UV colors as the star-forming galaxy population, where BL H$\alpha$ emitters with fainter UV luminosities are redder and have somewhat more massive SMBHs. We find particularly significant correlations between the SMBH mass and the relative broad to total H$\alpha$ luminosity and the optical color. This suggests that while the rest-frame UV flux is mainly dominated by star formation, especially for objects with relatively weak broad H$\alpha$ lines (phase I, SF dominated; Fig. $\ref{fig:sketch}$), the optical emission is increasingly dominated by emission from red and dusty AGN as the SMBH increases its mass. This is in line with models that suggest that the early formation phases of relatively numerous SMBHs are heavily obscured in the rest-frame UV \citep[e.g.][]{Hopkins05,Ni20,Peca23}.

Our deep imaging data reveals that the majority of BL H$\alpha$ emitters shows at least one spatially separated companion (see Figs. $\ref{fig:ap1}$ and $\ref{fig:ap2}$), for which the grism data in some cases detects narrow H$\alpha$ emission (Sections $\ref{sec:optextract}$ and $\ref{sec:resolvedspec}$). This suggests  that merging activity is common in galaxies that are experiencing AGN growth (see also \citealt{Trakhtenbrot17,Decarli18} for similar results in more luminous AGN at $z\gtrsim5$), as expected from simulations \citep[e.g.][]{McAlpine18}. 

More massive SMBHs in our sample have redder optical colors, and their (dust-reddened) AGN increasingly dominates the optical light (phase II, transition to AGN). This leads to broad components in the Balmer lines that increase their relative flux to the narrow component, and a UV continuum that becomes fainter and more obscured. Such obscuration of the UV light (Fig. $\ref{fig:MBH_AV}$) is similar to that seen in for example in the simulations by \cite{Trebitsch19}, and for example in the heavily obscured $z\sim7$ AGNs identified by \cite{Fujimoto22} and \cite{Endsley23}. In the likely case that redder optical colors are accompanied with dust attenuation of the broad line region and therefore an underestimate of the BH mass, this trend would even be stronger.

The most massive BHs in our sample are found among the galaxies that are the reddest and appear mostly as a point-source. The broad component is dominant in the phase where the red dusty AGN dominates the light (phase III). It is remarkable that in some of these objects we also find indications for Balmer absorption (Section $\ref{sec:abs}$), which we interpret to originate from dense inflowing and outflowing gas in the broad line region. These gas flows could reveal the fueling of the BH growth as well as the onset of feedback. We speculate that this phase predates the dust-reddened quasars that show a high fraction of broad absorption lines, in particular quasars that show broad absorption lines in low ionization states \citep{Urrutia09} --  these are also the kind of quasars for which Balmer absorption has previously been detected \citep{Schulze18}. Recently, \cite{Bischetti22} identified high velocity outflows in quasars at $z\approx6$ that are optically red, similar to our broad H$\alpha$ sample. While their outflow velocities are much higher than the velocities we measure in the Balmer absorption lines, these quasars are powered by a SMBH that is a factor $\sim100$ more massive. These more luminous quasars typically have bluer UV slopes, suggesting that the feedback associated with the AGN growth has cleared significant channels through the dust \citep[e.g.][]{Sanders88} that we infer to be present around the more common faint AGN at high-redshift. We measure a BH mass of $3.5\times10^8$ M$_{\odot}$ without an attenuation correction for GOODS-N-9771, which is by large the most luminous object in our sample. We also find indications for possible outflows in this object (Fig. $\ref{fig:absorption}$). It is interesting to note that an understimate of the BH mass of a factor 3 would already place this object in the BH mass regime occupied by bright quasars \citep[e.g.][]{Fan22}, suggesting that this object could be the obscured counterpart of these quasars.

\subsection{Implications for reionization} \label{sec:EoR}
The main drivers of cosmic reionization, which happened at $z\approx6-8$, remain elusive. A persisting question is whether the ionizing photons for this major phase transition arose from accreting black holes, from young stars, or some combination of both these channels. The most luminous quasars (M$_{\rm{{UV}}}<-23$) are too rare to have played an appreciable role (e.g., \citealt{Matsuoka18, Kulkarni19, Shen20, Jiang22, Schindler23}). However, fainter AGN may be substantial contributors provided they are (i) particularly numerous \textit{and} (ii) strongly ionizing (e.g., \citealt{Giallongo15, Giallongo19, Madau15, Finkelstein19}). For example, \citet{Madau15}, motivated by \citet{Giallongo15}, propose a purely AGN-driven reionization where faint AGN ($M_{\rm{UV}}\lesssim-18$) at $z\approx6-8$ are abundant ($\approx10^{-5}/\rm{cMpc}^{3}\rm{mag}^{-1}$), with their accretion disks effectively producing ionizing photons that escape with ease into the IGM (ionizing photon escape fractions, $f_{\rm{esc}}\approx100\%$). \cite{Grazian18,Grazian22} indeed find very high escape fractions of $\approx70$ \% for UV bright quasars ($-25<M_{\rm{UV}}<-23$) but our measurements question whether this still holds for much fainter AGN.

Our BL H$\alpha$ sample implies a high number-density for faint AGN at $z\approx5$ (Fig. \ref{fig:LF}), in agreement with the UV LF assumed in \citet{Madau15} for purely AGN-powered reionization. However, while numerous, it is unclear whether these AGN are strong ionizers. Most BL H$\alpha$ emitters have rest-UV colors that are almost indistinguishable from typical star-forming galaxies at these redshifts (Fig. \ref{fig:MUVbeta}) and the AGN are red. This means that the AGN are likely subdominant in the UV (as also suggested from the morphologies, see Fig. $\ref{fig:corefrac}$), as is expected for such faint AGN from simulations \citep[e.g.][]{Qin17,Trebitsch20}. Star-forming galaxies have been measured to have modest escape fractions ($\lesssim10\%$; e.g., \citealt{Steidel18,Pahl21}). In particular, galaxies with comparable UV slopes are suggested to have escape fractions in the range 0-7 \% (1.9 \% on average) following the calibration between the escape fraction and the UV slope based on low-redshift analogues \citep{Chisholm22}. 

The AGN in our sample seem to be heavily enshrouded in dust, presenting as red point-sources amidst blue star-forming clumps (see Figs. \ref{fig:RGBstamps} and $\ref{fig:MBH_AV}$). The rough estimate (lower limit) of the UV attenuation of the AGN emission obtained in Section $\ref{sec:galprops}$ is A$_{\rm UV}\approx2.8$. This suggests that there is a dearth of clear channels for ionizing photons around these AGN and that they hence have a low $f_{\rm{esc}}$. In the context of reionization, the BH growth scenario that we sketched in Section $\ref{sec:cartoon}$ could thus be considered as an analogy to the challenges possibly preventing ionizing photons that originate from short-lived massive stars to escape their dense and dusty natal clouds (e.g., Fig. 8 in \citealt{NaiduMatthee22}). A caveat is that it is possible that our broad H$\alpha$ line selection implicitly selects for dusty viewing angles, potentially allowing for a similarly high number density of faint AGN with unobscured lines of sight. While this can be addressed with alternative selection methods, we note that a high fraction of obscured SMBHs might explain the short UV luminous (unobscured) timescales inferred from small proximity zone sizes around high-redshift quasars \citep[e.g.][]{Eilers17,FDavies19,Zeltyn22,Satyavolu23}.

To summarize -- the faint AGN selected here are abundant, but they may be ineffective ionizing agents as their BH growth occurs in dust-reddened regimes. Therefore, our results so far indicate that star-forming galaxies remain leading candidates as the dominant sources of reionization (e.g., \citealt{Bouwens15c,Robertson15, Naidu20, Kashino22,MattheeNaidu22}). However, we also note that the current {\it JWST} surveys do not cover volumes that are large enough to identify significant numbers of AGN with intermediate UV luminosities around M$_{\rm UV} = -21$ to $M_{\rm UV}=-23$ (see Fig. $\ref{fig:LF}$). Whether these AGN are sufficiently abundant and whether they host the conditions favourable for ionizing photon escape (i.e. whether they are blue or red) remains an open question.

\subsection{Future directions} \label{sec:discuss_future}

The study of faint AGN with {\it JWST} is still in its early days. As the data-sets that were used in this work were not explicitly designed to study these objects, there are various avenues for improving our work that we list here:
{\bf i)} {\it a detailed characterisation of the connection between host galaxies and the AGN} would benefit significantly from deep NIRCam and MIRI imaging data over a suite of filters, in particular filters that probe beyond the Balmer break but are free of strong emission-lines. Additionally, deep NIRSpec IFU spectroscopy can benefit from the contrast enabled by spatially resolved line-profile fitting to decompose quasar and host spectra \citep[e.g.][]{Vayner23}, whereas mid infrared photometry and spectroscopy can help disentangling stellar light from dusty AGN emission;
{\bf ii)} {\it a clustering analysis to probe the host halo masses and duty cycles of the faint obscured AGN} can be performed with the existing WFSS data, but is beyond the scope of this paper; 
{\bf iii)} {\it more sensitive spectroscopy of fainter samples, for example using complete follow-up with NIRSpec} will be useful to confirm the AGN origin of galaxies that have fainter and somewhat narrower broad components through differential line-profile fitting of e.g. Balmer and forbidden lines as [OIII]; 
{\bf iv)} {\it larger samples, obtained with large area and wider band (slitless) spectroscopy extending from 2.7-5.0 micron} should enable us to investigate the evolution of the faint AGN luminosity function using broad H$\alpha$ lines over various redshift intervals in the $z\approx3-6.5$ range. The UV luminosity M$_{\rm UV}= -22$ to $-24$ is a critical regime that is currently poorly probed. Over this regime, the AGN fraction seemingly transition from $\approx 0-100$ \% at $z\approx5$ \citep{Sobral18,Adams22}. This is also the luminosity range where we could expect the transition from a majority of UV faint, red AGN, to a dominant AGN population that are UV bright, and plausibly have a high escape fraction that could contribute to the reionization of the Universe \citep[e.g.][]{Grazian22}. The AGN in this gap likely have number densities $\sim10^{-6}$ cMpc$^{-3}$, such that statistical samples require {\it JWST} surveys of $\approx0.6$ deg$^2$. Extending AGN identifications to higher redshifts $z>6.5$ will likely need to rely on the (much) fainter H$\beta$ line \citep[e.g.][]{Larson23}, or make use of X-ray data \citep{Bogdan23}, radio \citep{Endsley23}, rest-frame UV spectroscopy, or full SED modeling \citep{Fujimoto22}, which makes detailed estimates of their number density more difficult; 
{\bf v)} {\it a confirmation and more detailed study of the prevalence and properties of Balmer absorption in the faint AGN requires deep high resolution spectroscopy in both the H$\alpha$ and H$\beta$ lines}, whereas sub-mm spectroscopy targeting the bright [CII] line, for example with ALMA or NOEMA, could help to give strong priors on the systemic redshifts of the narrow components that could benefit the fitting process.

\section{Summary} \label{sec:summary}
Measuring the number density and properties of faint AGN in the early Universe is key for understanding the formation of SMBHs and determining their role in the reionization of the Universe. In this paper we exploit the new wide field IR slitless spectroscopic capabilities of NIRCam on the {\it JWST} to perform a systematic search for broad line H$\alpha$ emitters at $z\sim5$. We combine NIRCam imaging and grism data from the EIGER \citep{Kashino23} and FRESCO \citep{Oesch23} surveys. The data cover 220 arcmin$^2$ and constitute a total 70 hrs of observing time. A main benefit of these data is that they yield spectral coverage of virtually all sources in the field of view. Therefore, the availability of 3-5 micron spectra does not rely on pre-selected galaxies (in most Cycle 1 programs typically selected from {\it HST} or ground-based data).

Our main results are summarised as follows:

\begin{itemize} 
\item We identify 20 broad line H$\alpha$ emitters that have unambiguous broad components at $z=4.2-5.5$. The BL H$\alpha$ emitters occupy the rarest region in the NIRCam colour-colour space, having relatively flat/blue spectra from 1 - 2 micron, but a (very) red continuum from 2 - 4 micron. This is unlike extreme emission line galaxies at $z=5-9$ that are typically blue apart from the filters that contain strong [OIII] or H$\alpha$ emission, or dusty star forming galaxies that are typically red over the full wavelength range probed by NIRCam. [Section $\ref{sec:colors}$, Figs. $\ref{fig:RGBstamps}$, $\ref{fig:colselection}$]

\item We measure broad wings with widths FWHM$\approx2000$ km s$^{-1}$ (ranging from $1160-3700$ km s$^{-1}$). The broad components typically constitute 65 \% of the total H$\alpha$ flux. [Sections $\ref{sec:optextract}$, $\ref{sec:fit1D}$, Figs. $\ref{fig:2D_fits}$, $\ref{fig:1Dgrid}$, $\ref{fig:1D_fit}$ and Table $\ref{tab:linefits}$] 

\item Despite non-detections in the X-ray, as expected since our estimate suggests they are likely below current survey limits, we make the case that the broad components originate from the broad line region in an AGN as they are spatially unresolved sources and have no velocity offsets to the narrow components. The broad components are significantly broader and relatively more luminous compared to broad components typically associated to outflowing gas associated with star formation. We also detect spatially extended narrow H$\alpha$ emission due to star formation in the host galaxy and nearby companions. [Section $\ref{sec:AGN}$, Figs. $\ref{fig:2D_stack}$, $\ref{fig:stamps}$, $\ref{fig:corefrac}$]

\item Our sample has BH masses in the range $10^{7-8}$ M$_{\odot}$. The UV luminosities of the galaxies span a range M$_{\rm UV,  AGN + host}=-21$ to $-18$, significantly fainter than most previous high-redshift AGN samples. While the brighter objects in our sample have UV slopes that are comparable to those of the general galaxy population at $z\approx5$, the UV faint BL H$\alpha$ emitters tentatively have somewhat redder UV slopes. We find that the optical colors appear correlated with the mass of the BH and the broad to total H$\alpha$ flux, suggesting that dusty AGN (A$_{\rm UV} \gtrsim 2.8$) contribute significantly to rest-frame optical flux. [Sections $\ref{sec:BHprops}$, $\ref{sec:galprops}$, Table $\ref{tab:properties}$ and Figs. $\ref{fig:MUVbeta}$, $\ref{fig:MBHbroadbeta}$, $\ref{fig:MBH_AV}$]   

\item  We detect complex H$\alpha$ profiles in two red and bright BL H$\alpha$ emitters. We tentatively identify these as due to faint, narrow H$\alpha$ absorption. Until now, such absorption has only been detected in a handful of luminous red quasars at $z\approx2$. The absorption systems are shifted by +50 and -340 km s$^{-1}$ with respect to the line centre, respectively. We interpret them as tracing the dense gas inflows that are fueling the BH growth (the redshifted absorption), and outflowing gas within the broad line region that traces the onset of outflows driven by the AGN (the blueshifted absorption). [Section $\ref{sec:abs}$, Fig. $\ref{fig:absorption}$] 

\item The complete coverage allows us to measure the number densities of broad line H$\alpha$ emitters over a total volume $6\times10^5$ cMpc$^3$ over $z=4-6$. We measure a number density of $10^{-5}$ cMpc$^{-3}$, which is an order of magnitude higher than extrapolated quasar UV luminosity function, but similar to earlier faint AGN searches using X-Ray data. Despite the high number density, it implies that AGN only contribute $<$ 1 \% of the total UV emission at these luminosities. The H$\alpha$ luminosity function of the faint AGN is steeper than the UV luminosity function, but only slightly higher than previous (uncertain) extrapolations and in agreement with a low AGN fraction among the general galaxy population, except for the brightest H$\alpha$ luminosities. [Section $\ref{sec:numberdensity}$, Fig. $\ref{fig:LF}$, Table $\ref{tab:NumDens}$]

\item The BH mass function agrees with cosmological simulations over M$_{\rm BH}=10^{7.5-8}$ M$_{\odot}$, intruigingly suggesting that our AGN selection captures the full population of SMBHs with these masses, whereas at lower masses we are likely biased to detecting only more efficiently accreting BHs. [Section $\ref{sec:BHMF}$, Fig. $\ref{fig:MF}$]

\item We interpret the trends between BH masses, UV and optical colors and the H$\alpha$ profile, and the detected absorption components, in the context of an evolutionary sequence. For the lowest mass BHs, a significant fraction of the (UV) light originates from star formation yielding a blue color and a relatively faint broad component. As the BH grows, the red dust enshrouded AGN increasingly outshines the star forming regions, in particular in the optical, and the broad component starts to dominate. The H$\alpha$ absorption detected in the objects in which the red AGN dominates highlight the onset of feedback that clears the dust-free pathways through which more massive blue quasars are seen. [Section $\ref{sec:cartoon}$, Fig. $\ref{fig:sketch}$]

\item While the abundance of faint AGN at $z\approx5$ is higher than typical extrapolations of the quasar luminosity function, we find that these AGN likely have a minor contribution to the end stages of cosmic reionization. The galaxies -- particularly those where the AGN is dominant -- are relatively dust reddened, implying a low escape fraction of ionizing photons from these AGN. Moreover, a signification fraction of UV light is due to star formation in the bluer galaxies. [Section $\ref{sec:EoR}$] \vspace{0.4cm}
\end{itemize}

Our results highlight the potential of spectroscopy with {\it JWST} to improve our understanding of the abundance and properties of faint AGN in the high-redshift Universe. These will enable more detailed tests of models for the formation (seeding) of SMBHs and their models describing their growth. As discussed in Section $\ref{sec:discuss_future}$, there are various limitations to this work that can be addressed with dedicated observations of such AGN. While the BL H$\alpha$ emitters constitute a sample of faint AGN that complements the bright quasars, there remains a gap between the faint AGN samples with UV luminosites M$_{\rm UV}\geq-21$ identified with {\it JWST}, and quasars found with ground-based surveys, with M$_{\rm UV}\leq-23$. In this regime we may expect objects that are transitioning to being dominated by blue, UV bright AGN continuum emission, but future observations are required to identify this population. 


\facilities{\textit{JWST}, \textit{HST}}

\software{
    \texttt{Python},
    \texttt{matplotlib} \citep{matplotlib},
    \texttt{numpy} \citep{numpy}, 
    \texttt{scipy} \citep{scipy},
    \texttt{Astropy}
    \citep{astropy1, astropy2}, \texttt{Imfit} \citep{Erwin2015}. }
    
\acknowledgments{We thank an anonymous referee for their constructive comments that helped improving the manuscript.
This work is based on observations made with the NASA/ESA/CSA James Webb Space Telescope. The data were obtained from the Mikulski Archive for Space Telescopes at the Space Telescope Science Institute, which is operated by the Association of Universities for Research in Astronomy, Inc., under NASA contract NAS 5-03127 for \textit{JWST}. These observations are associated with programs \# 1243 and \# 1895. The specific observations analyzed can be accessed via \dataset[DOI: 10.17909/4xx0-zj76]{https://doi.org/10.17909/4xx0-zj76}. Funded by the European Union (ERC, AGENTS,  101076224). Views and opinions expressed are however those of the author(s) only and do not necessarily reflect those of the European Union or the European Research Council. Neither the European Union nor the granting authority can be held responsible for them. RPN acknowledges funding from {\it JWST} programs GO-1933 and GO-2279. Support for this work for RPN was provided by NASA through the NASA Hubble Fellowship grant HST-HF2-51515.001-A awarded by the Space Telescope Science Institute, which is operated by the Association of Universities for Research in Astronomy, Incorporated, under NASA contract NAS5-26555. Support for this work for GDI was provided by NASA through grant JWST-GO-01895 awarded by the Space Telescope Science Institute, which is operated by the Association of Universities for Research in Astronomy, Inc., under NASA contract NAS 5-26555. This work has received funding from the Swiss State Secretariat for Education, Research and Innovation (SERI) under contract number MB22.00072, as well as from the Swiss National Science Foundation (SNSF) through project grant 200020\_207349. The Cosmic Dawn Center (DAWN) is funded by the Danish National Research Foundation under grant No.\ 140. 

}

\bibliography{Biblio}
\bibliographystyle{apj}

\appendix 
\section{Photometry} \label{appendix:photometry}

In Tables $\ref{tab:photometry}$ and $\ref{tab:photometry_EIGER}$ we list the {\it HST} and {\it JWST} photometry of the BL H$\alpha$ emitters in the FRESCO and EIGER fields, respectively.

\begin{table*} 
    \centering
    \caption{{\bf Photometry of the broad H$\mathbf{\alpha}$ line emitters in the FRESCO data.} Fluxes are listed in nJy. Photometry is measured following the method described in Weibel et al. (in prep).}
    \begin{tabular}{cccccccccc}
ID & F606W &F775W & F814W &F850LP & F125W &F160W & F182M &F210M & F444W \\ \hline
GN-4014 & $ 3.3\pm3.4$ & $ 7.8\pm4.7$ & $ 10.4\pm2.1$ & $ 23.3\pm4.2$ & $ 22.9\pm5.3$ & $ 24.7\pm6.5$ & $ 13.0\pm9.5$ & $ 33.8\pm14.0$ & $ 389.3\pm13.4$ \\
GN-9771 & $ 2.9\pm3.9$ & $ 10.5\pm4.0$ & $ 13.3\pm2.6$ & $ 25.2\pm5.3$ & $ 65.6\pm5.4$ & $ 104.3\pm6.4$ & $ 158.9\pm7.9$ & $ 138.6\pm6.9$ & $ 2494.8\pm124.7$ \\
GN-12839 & $ -0.7\pm3.6$ & $ 15.4\pm4.2$ & $ 30.0\pm3.3$ & $ 38.8\pm5.5$ & $ 57.5\pm5.4$ & $ 63.3\pm6.3$ & $ 64.3\pm5.1$ & $ 61.1\pm6.3$ & $ 1250.3\pm62.5$ \\
GN-13733 & $ 2.6\pm3.9$ & $ 18.7\pm4.3$ & $ 17.2\pm3.4$ & $ 13.3\pm7.0$ & $ 16.1\pm6.8$ & $ 12.6\pm7.4$ & $ 21.2\pm4.3$ & $ 33.1\pm5.5$ & $ 245.8\pm12.3$ \\
GN-14409 & $ 1.1\pm3.2$ & $ 29.2\pm3.5$ & $ 29.3\pm2.5$ & $ 26.6\pm5.2$ & $ 32.2\pm7.0$ & $ 25.7\pm7.8$ & $ 35.5\pm5.5$ & $ 31.9\pm6.8$ & $ 224.3\pm11.2$ \\
GN-15498 & $ -2.9\pm3.5$ & $ 9.0\pm4.0$ & $ 16.2\pm3.2$ & $ 13.0\pm6.9$ & $ 16.5\pm5.2$ & $ 18.9\pm5.3$ & $ 21.4\pm3.4$ & $ 15.8\pm4.2$ & $ 488.7\pm24.4$ \\
GN-16813 & $ 6.5\pm3.5$ & $ 112.8\pm5.6$ & $ 117.3\pm5.9$ & $ 115.3\pm6.7$ & $ 93.9\pm5.7$ & $ 98.3\pm6.5$ & $ 119.4\pm7.2$ & $ 105.9\pm9.2$ & $ 439.2\pm22.0$ \\
GS-13971 & $ 1.5\pm0.9$ & $ 26.5\pm1.3$ & $ 32.6\pm3.8$ & $ 46.2\pm2.3$ & $ 53.7\pm2.7$ & $ 53.8\pm2.7$ & $ 61.2\pm3.1$ & $ 58.5\pm3.0$ & $ 673.2\pm33.7$ \\ \hline
    \end{tabular}
    \label{tab:photometry}
\end{table*}

\begin{table*} 
    \centering
    \caption{{\bf Photometry of the broad H$\mathbf{\alpha}$ line emitters in EIGER data.} Fluxes are listed in nJy. Photometry is measured following the method described in \cite{Kashino23}.}
    \begin{tabular}{cccc}
ID & F115W &F200W & F356W  \\ \hline
J1148-7111 & $ 48.5\pm6.5$ & $ 84.5\pm4.6$ & $ 592.7\pm12.6$ \\
J1148-18404 & $ 10.7\pm2.0$ & $ 29.0\pm1.8$ & $ 653.9\pm11.5$ \\
J1148-21787 & $ 121.8\pm10.9$ & $ 132.0\pm7.7$ & $ 535.5\pm13.0$ \\
J0100-2017 & $ 81.7\pm5.7$ & $ 103.6\pm4.3$ & $ 355.2\pm8.7$ \\
J0100-12446 & $ 58.5\pm5.0$ & $ 106.1\pm4.0$ & $ 590.7\pm11.3$ \\
J0100-15157 & $ 197.9\pm9.1$ & $ 212.4\pm7.3$ & $ 484.1\pm11.9$ \\
J0100-16221 & $ 71.4\pm5.4$ & $ 98.8\pm4.0$ & $ 391.1\pm8.6$ \\
J0148-976 & $ 100.3\pm6.4$ & $ 136.7\pm6.0$ & $ 325.7\pm11.1$ \\
J0148-4214 & $ 146.0\pm6.6$ & $ 171.8\pm5.3$ & $ 502.7\pm11.0$ \\
J0148-12884 & $ 122.6\pm7.1$ & $ 148.5\pm5.7$ & $ 435.0\pm11.0$ \\
J1120-7546 & $ 12.9\pm5.0$ & $ 30.9\pm4.2$ & $ 484.5\pm10.1$ \\
J1120-14389 & $ 65.0\pm4.5$ & $ 81.9\pm3.6$ & $ 528.2\pm10.8$ \\ \hline

    \end{tabular}
    \label{tab:photometry_EIGER}
\end{table*}

\section{All H$\alpha$ fits and stamps} \label{appendix:stamps}
\begin{figure*}
    \centering
    \begin{tabular}{ccccc}
    \hspace{-0.4cm}
 \includegraphics[height=3.6cm,trim={0 0 0.75cm 0},clip]{figs/profilefits/profile_fit_J0100_15157.pdf} &
 \hspace{-0.4cm}
 \includegraphics[height=3.6cm,trim={0.02cm 0 0.75cm 0},clip]{figs/profilefits/profile_fit_J1148_18404.pdf} &
 \hspace{-0.4cm}
 \includegraphics[height=3.6cm,trim={0.02cm 0 0.75cm 0},clip]{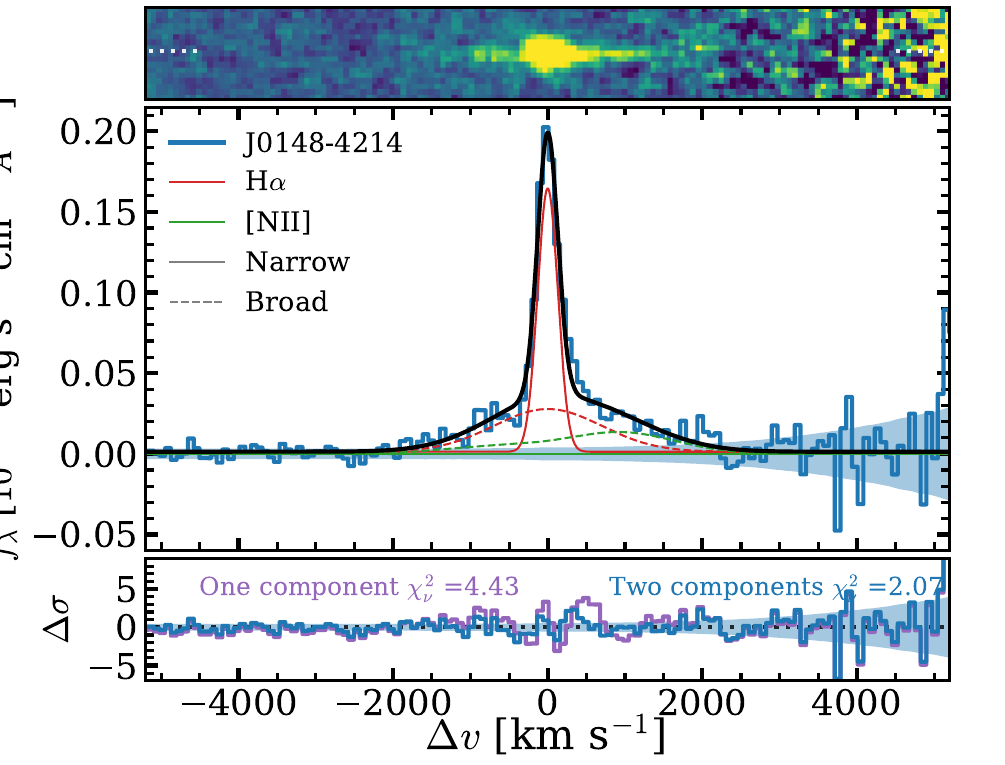} &
 \hspace{-0.4cm}
 \includegraphics[height=3.6cm,trim={0.02cm 0 0.75cm 0},clip]{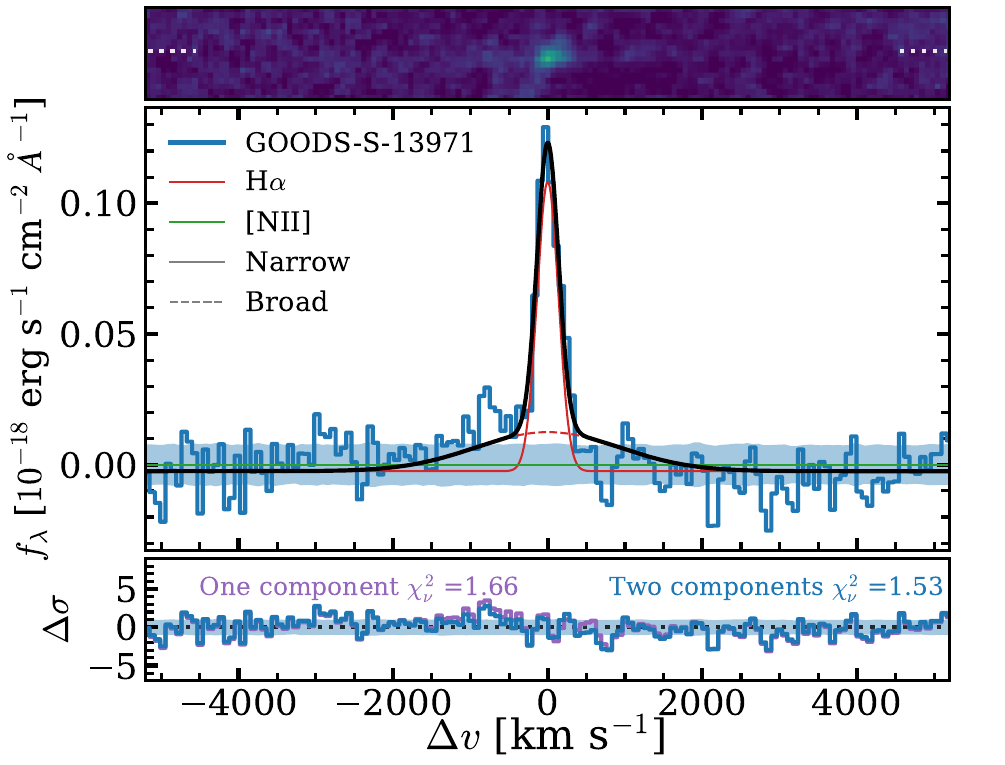} \\

 \hspace{-0.4cm}
 \includegraphics[height=3.6cm,trim={0 0 0.75cm 0},clip]{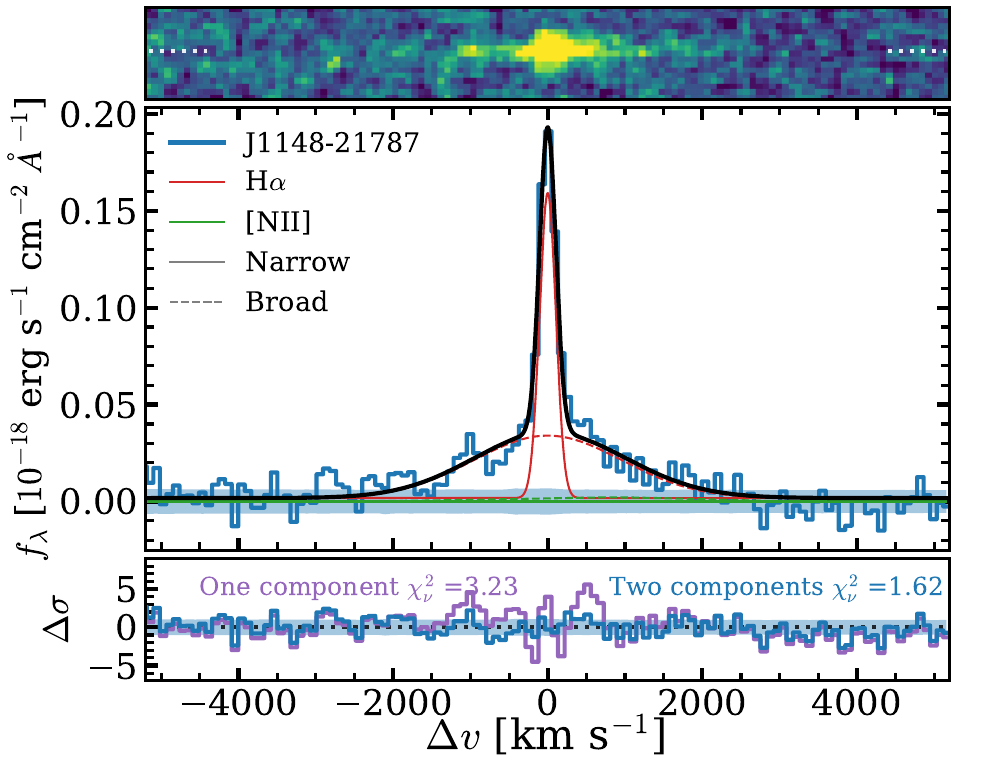} &
 \hspace{-0.4cm}
 \includegraphics[height=3.6cm,trim={0.35 0 0.75cm 0},clip]{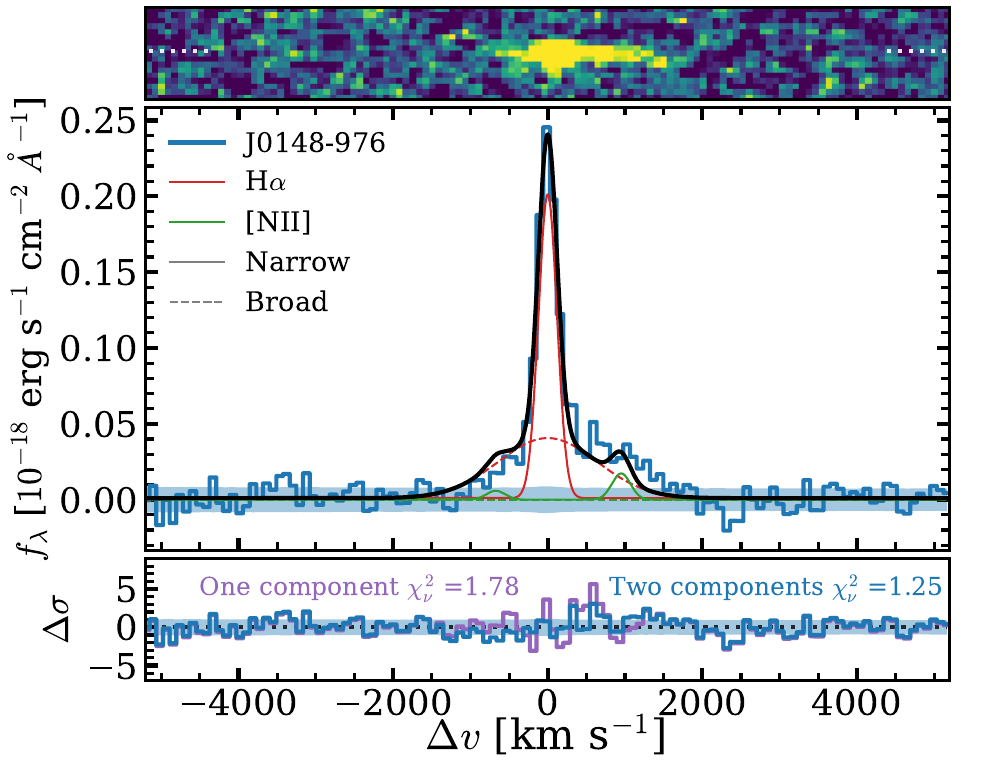} &
 \hspace{-0.4cm}
 \includegraphics[height=3.6cm,trim={0.02cm 0 0.75cm 0},clip]{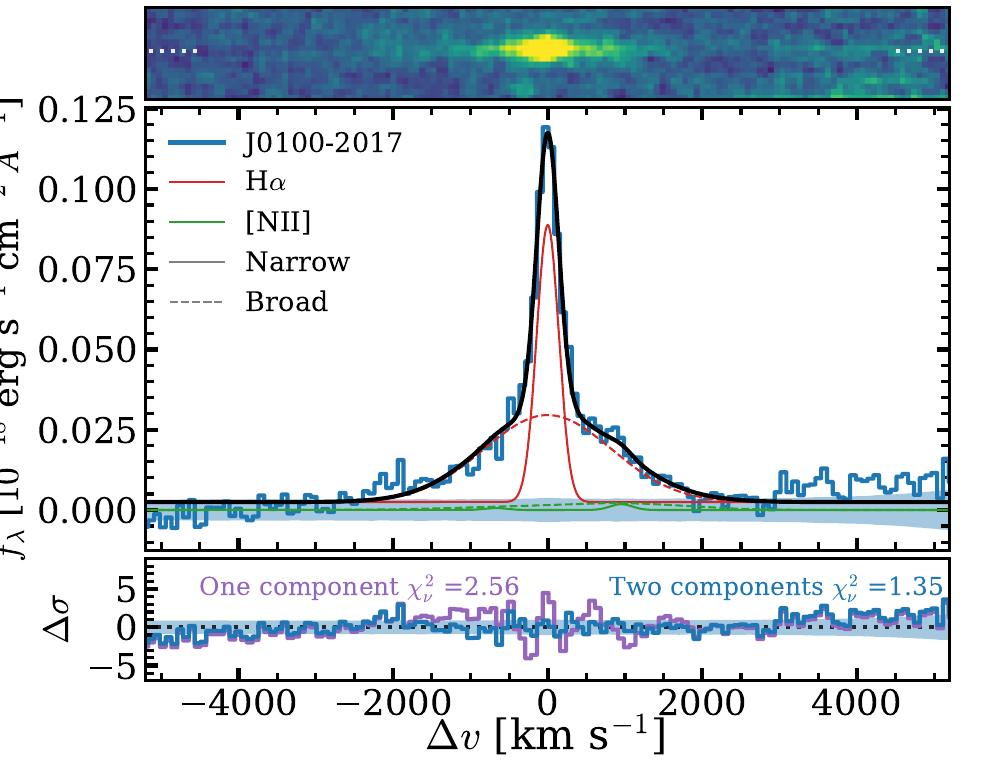} &
 \hspace{-0.4cm}
 \includegraphics[height=3.6cm,trim={0.02cm 0 0.75cm 0},clip]{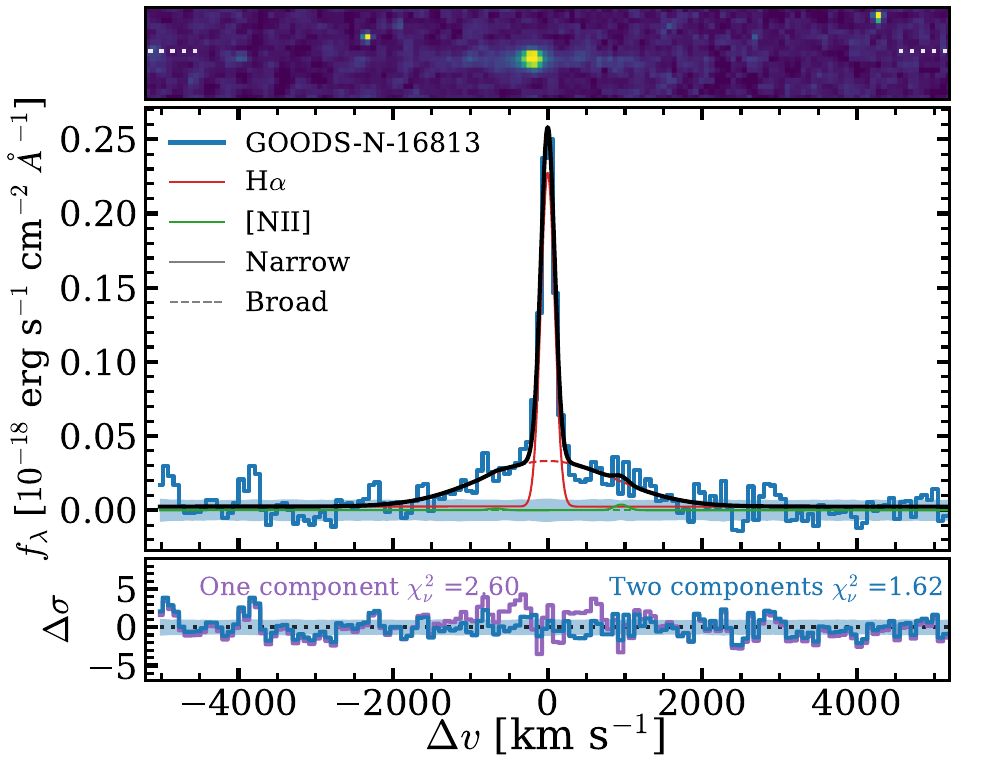} \\

 \hspace{-0.4cm}
 \includegraphics[height=3.6cm,trim={0.02cm 0 0.75cm 0},clip]{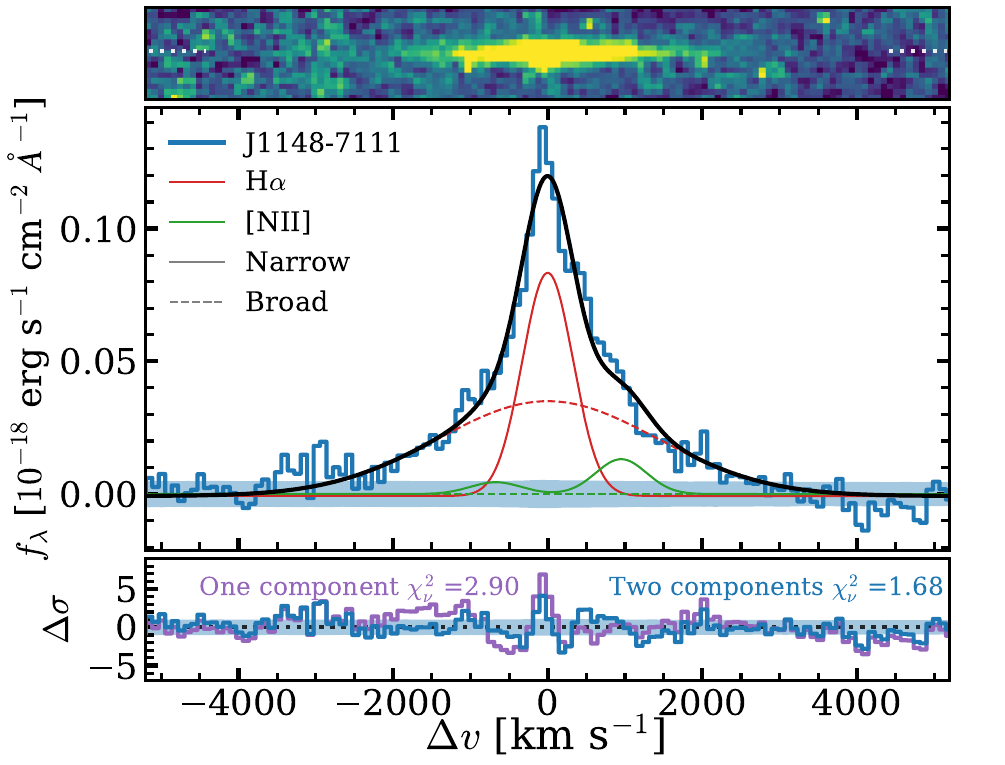} &
  \hspace{-0.4cm}
 \includegraphics[height=3.6cm,trim={0.02cm 0 0.75cm 0},clip]{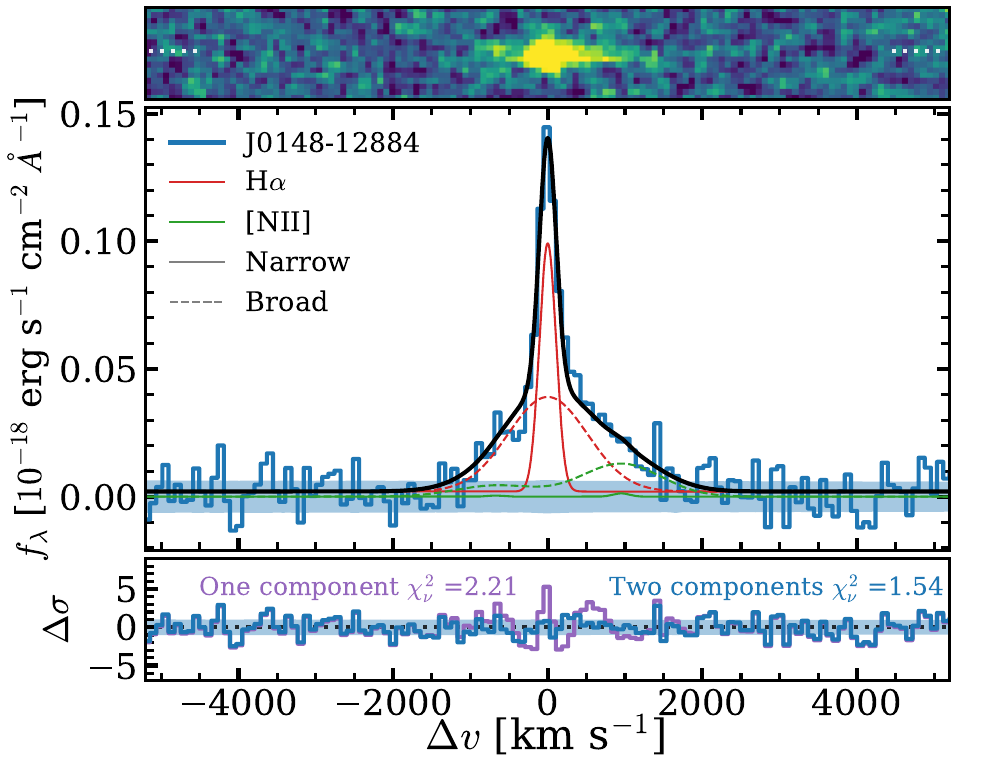} &
  \hspace{-0.4cm}
  \includegraphics[height=3.6cm,trim={0.02cm 0 0.75cm 0},clip]{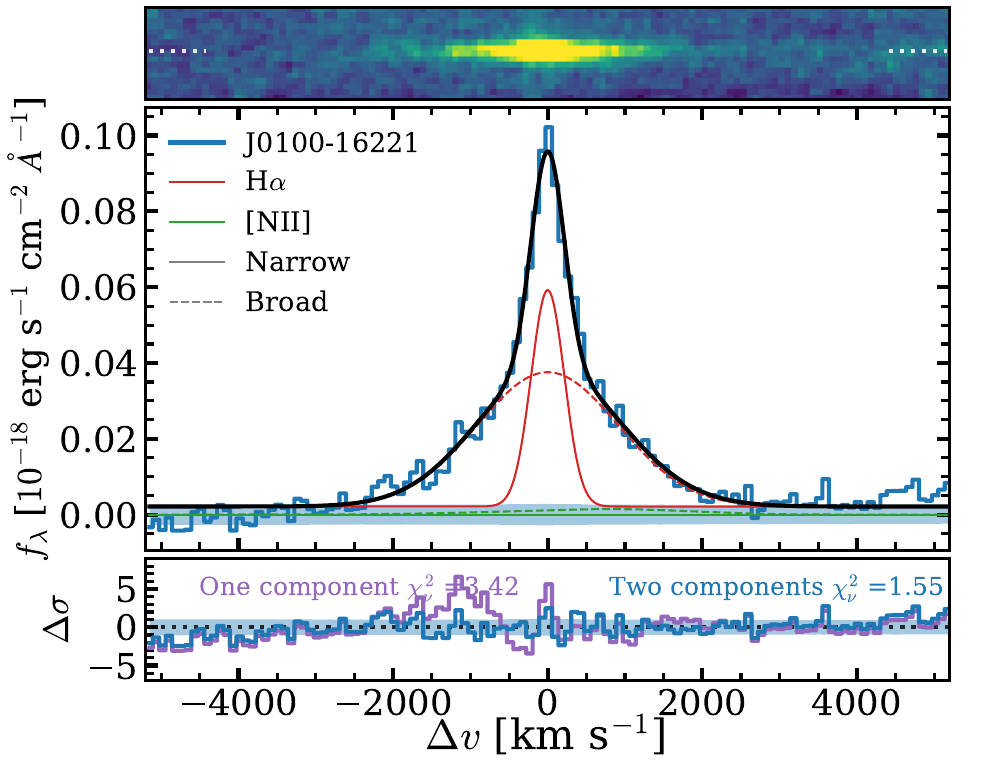} &
 \hspace{-0.4cm}
 \includegraphics[height=3.6cm,trim={0.02cm 0 0.75cm 0},clip]{figs/profilefits/profile_fit_J0100_12446.pdf} \\

 \hspace{-0.4cm}
 \includegraphics[height=3.6cm,trim={0.02cm 0 0.75cm 0},clip]{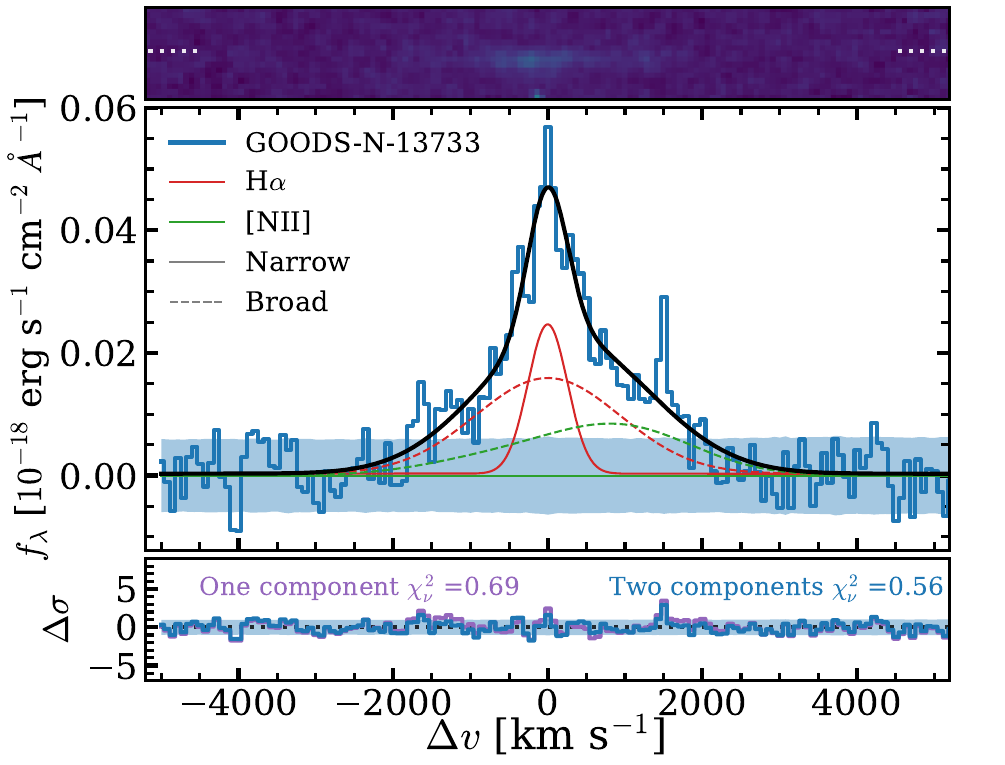} &
 \hspace{-0.4cm}
 \includegraphics[height=3.6cm,trim={0.02cm 0 0.75cm 0},clip]{figs/profilefits/profile_fit_GOODS-N_9771.pdf} &
 \hspace{-0.4cm}
 \includegraphics[height=3.6cm,trim={0.02cm 0 0.75cm 0},clip]{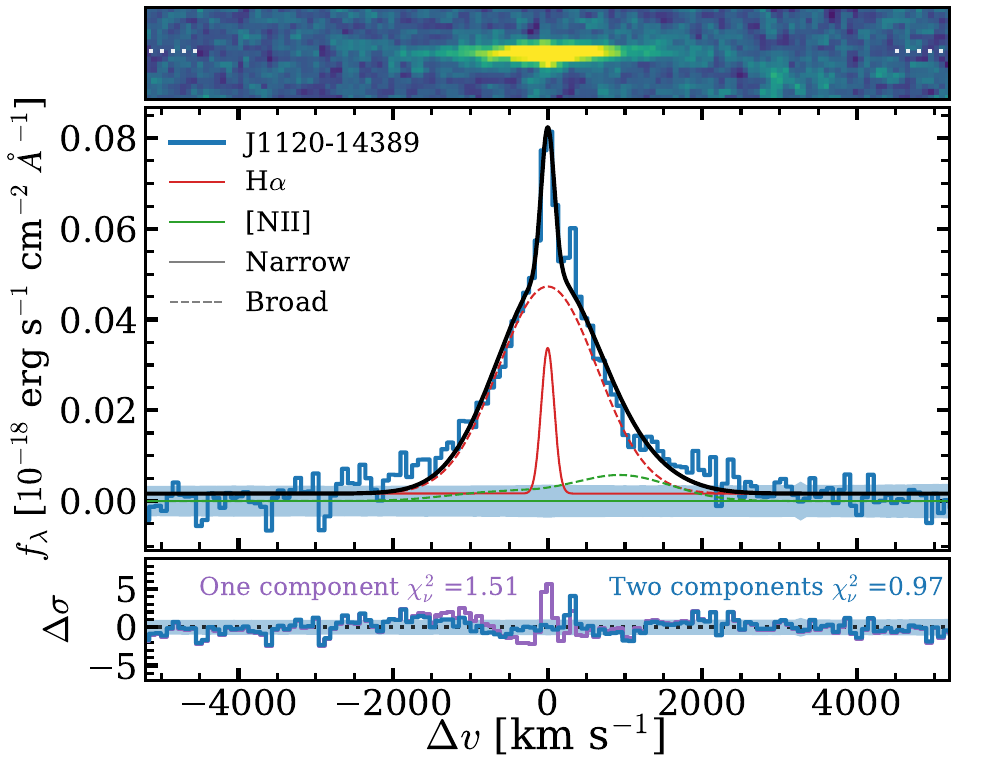} &
 \hspace{-0.4cm}
  \includegraphics[height=3.6cm,trim={0.02cm 0 0.75cm 0},clip]{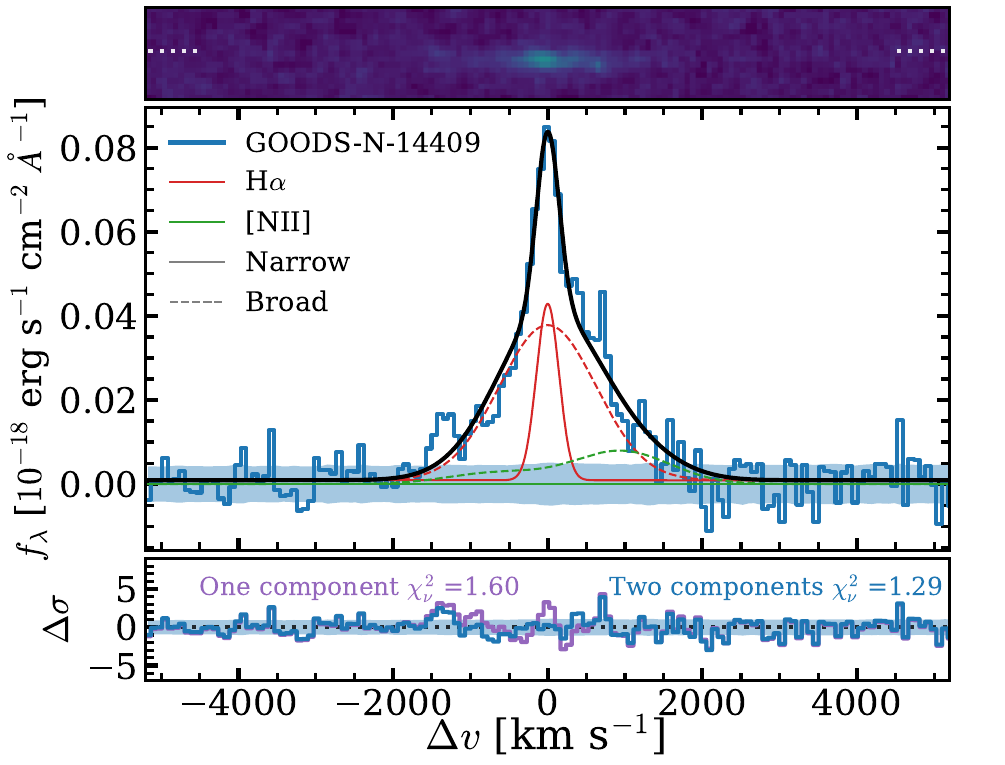} \\

 \hspace{-0.4cm}
 \includegraphics[height=3.6cm,trim={0.02cm 0 0.75cm 0},clip]{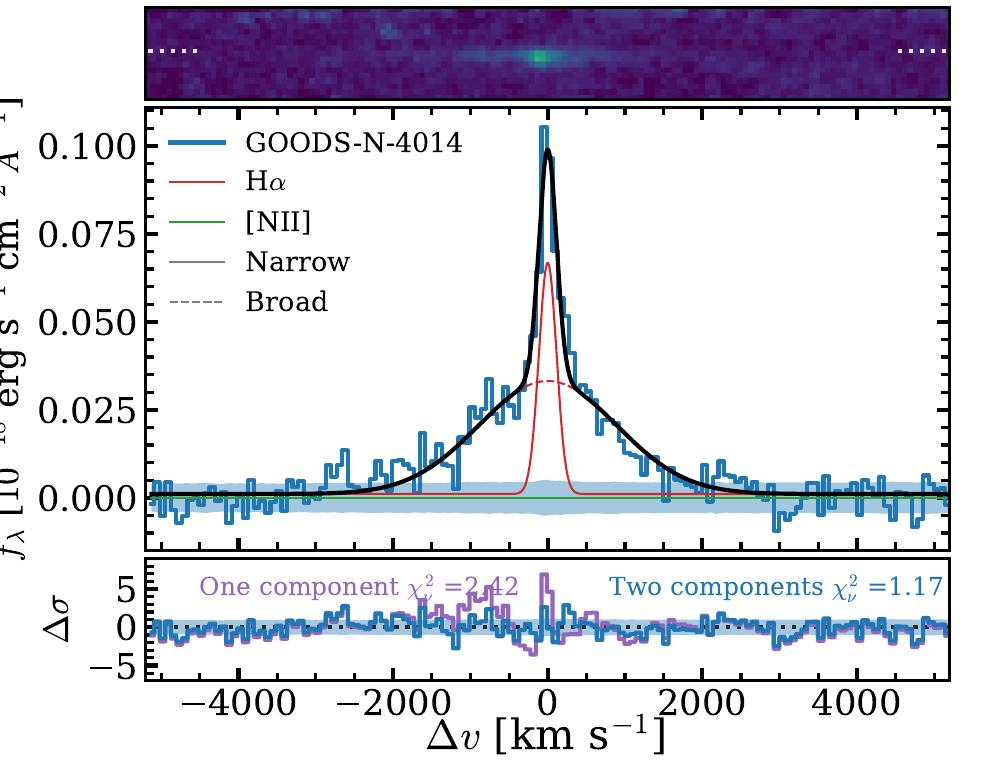} &
 \hspace{-0.4cm}
 \includegraphics[height=3.6cm,trim={0.02cm 0 0.75cm 0},clip]{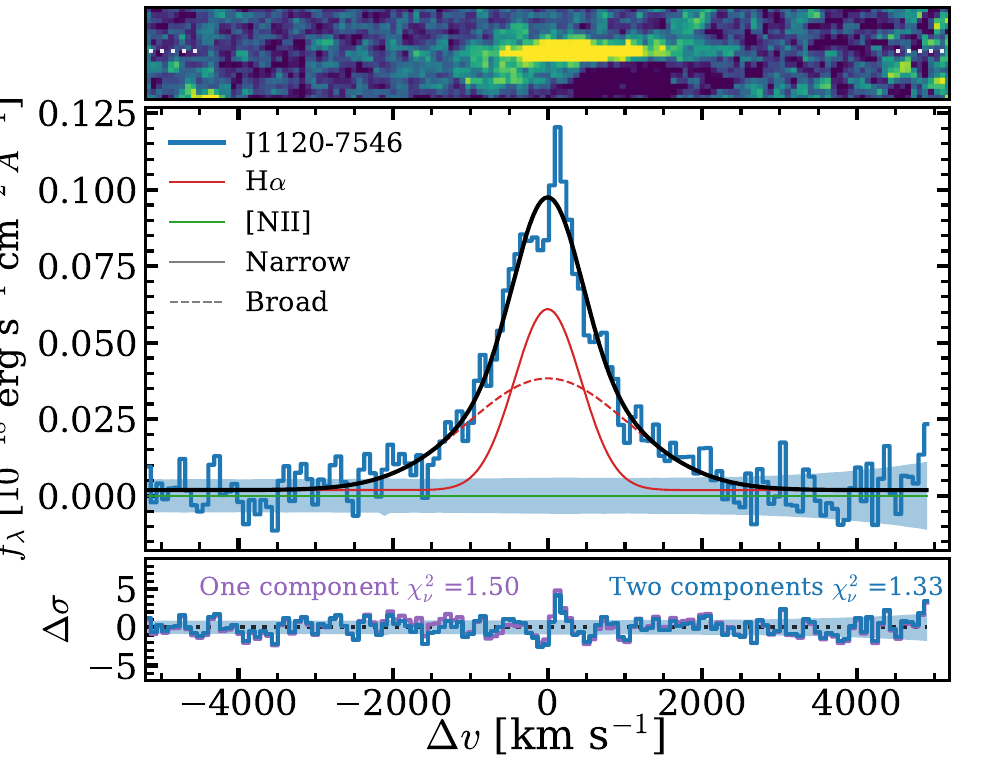} &
 \hspace{-0.4cm}
 \includegraphics[height=3.6cm,trim={0.02cm 0 0.75cm 0},clip]{figs/profilefits/profile_fit_GOODS-N_12839.pdf} &
 \hspace{-0.4cm}
 \includegraphics[height=3.6cm,trim={0.02cm 0 0.75cm 0},clip]{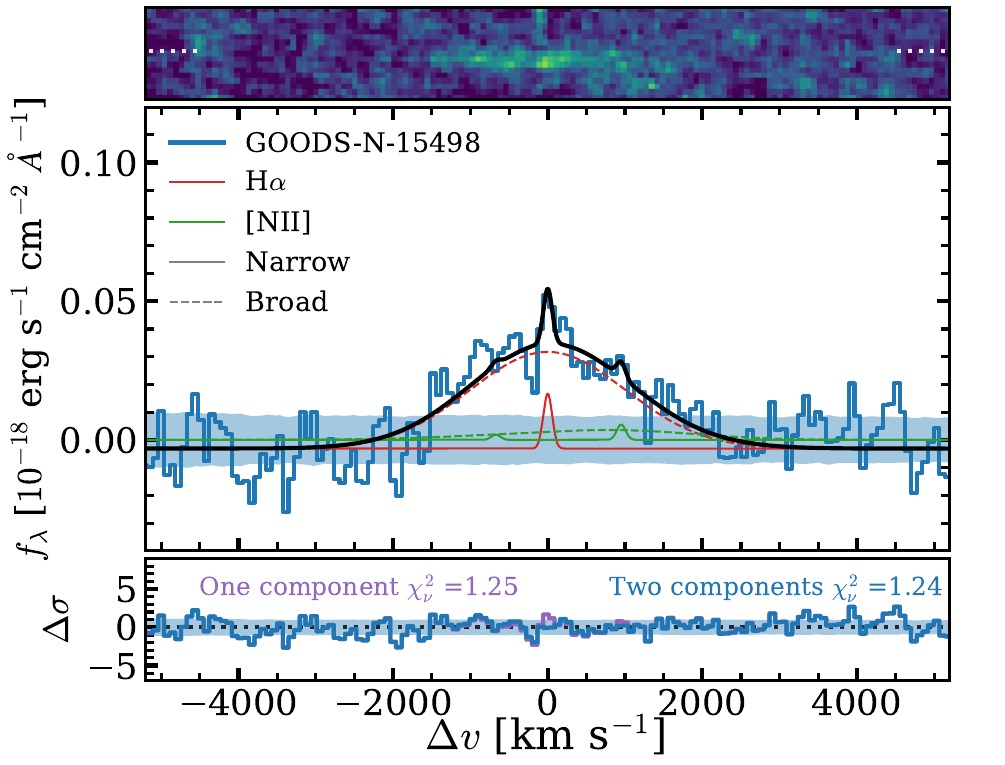} \\
        \end{tabular}
    \caption{{\bf Fits to all broad H$\mathbf{\alpha}$ line profiles, as in Fig. $\ref{fig:1D_fit}$.} In each of the three panels, we show the 2D emission-line spectrum, the optimally extracted 1D spectrum (blue, errors in blue shades) and the best-fit two component model (black solid line, where the solid red line shows the narrow H$\alpha$ component and the dashed red line the broad H$\alpha$ component, and green shows [NII]) and the residual of the shown two component and the best-fit single component model.}
    \label{fig:profilegrid}
\end{figure*}

\begin{figure*}
    \centering
    \includegraphics[width=16cm]{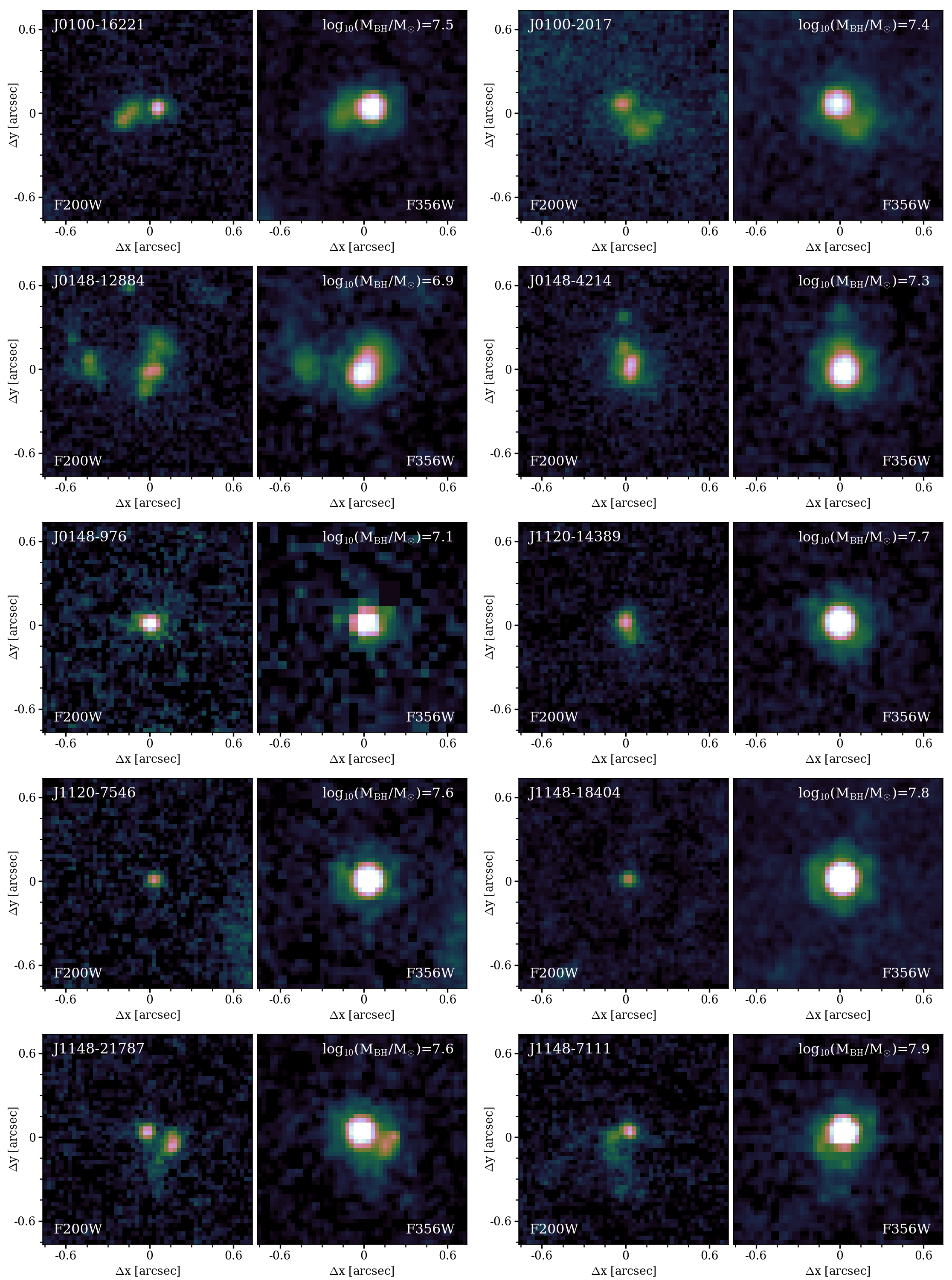}
    \caption{{\bf Zoom-in stamps of all BL H$\alpha$ emitters, as in Fig. $\ref{fig:stamps}$.} For each object we show the F200W (EIGER) or F182M+F210M (FRESCO) image that has a PSF FWHM$\approx0.06''$ and the F356W or F444W image (PSF FWHM $\approx0.12''$) that contains the H$\alpha$ line emission. The colour scaling follows a power-law with exponent $\gamma=0.6$ to highlight low surface brightness emission.}
    \label{fig:ap1}
\end{figure*}

\begin{figure*}
    \centering \includegraphics[width=16cm]{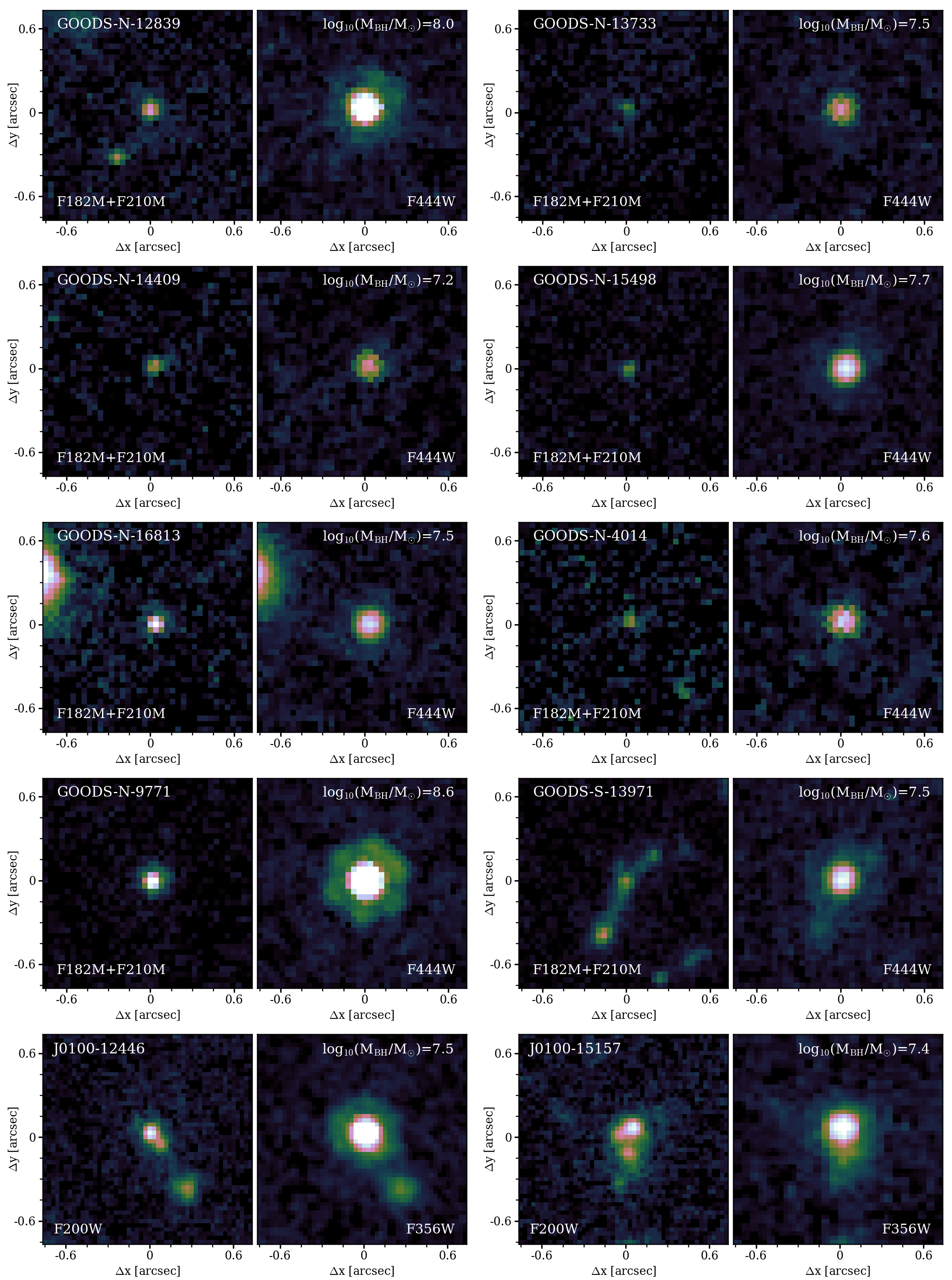}
    \caption{Fig. $\ref{fig:ap1}$, continued.}
    \label{fig:ap2}
\end{figure*}

In Figure $\ref{fig:profilegrid}$ we show the fitted H$\alpha$ profiles for all 20 BL H$\alpha$ emitters as described in \S $\ref{sec:fit1D}$.
In Figures $\ref{fig:ap1}$ and $\ref{fig:ap2}$ we show the F200W and F356W (for EIGER) and F182M+F210M and F444W (for FRESCO) stamps highlighting the differences in the morphology in the red filters that contain the broad H$\alpha$ emission, and the bluer filters that trace the rest-frame UV.

\end{CJK*}
\end{document}